\documentclass[12pt,a4paper]{article}
\pdfoutput=1
\usepackage{color}
\usepackage{amssymb,amsmath,bm,bbm}
\usepackage{epsf}
\usepackage{epsfig}
\usepackage{afterpage}
\usepackage{longtable}
\usepackage[dvipsnames]{xcolor}
\usepackage[linktoc=page,bookmarks=false,colorlinks=false,linkbordercolor=RoyalBlue,citebordercolor=ForestGreen,urlbordercolor=CornflowerBlue]{hyperref}
\usepackage{latexsym,mathrsfs,dsfont}
\usepackage[normalem]{ulem} 
\usepackage[compress]{cite}
\usepackage{graphicx}
\usepackage{url}
\usepackage{paralist}
\usepackage{bbold}
\usepackage{slashed}
\usepackage{multirow}

\usepackage[hypcap]{caption, subcaption}

\usepackage{fancyhdr}
\renewcommand{\sectionmark}[1]{
  \markboth{\thesection\ ~~#1}{\thesubsection\ #1}}   
\advance \headheight by 3.0truept       

\newcommand{\beq}{\begin{equation}}
\newcommand{\eeq}{\end{equation}}
\newcommand{\be}{\begin{equation}}
\newcommand{\ee}{\end{equation}}
\newcommand{\bi}{\begin{itemize}}
\newcommand{\ei}{\end{itemize}}
\newcommand{\ba}{\begin{array}}
\newcommand{\ea}{\end{array}}
\newcommand{\beqa}{\begin{eqnarray}}
\newcommand{\eeqa}{\end{eqnarray}}
\newcommand{\bea}{\begin{eqnarray}}
\newcommand{\eea}{\end{eqnarray}}
\newcommand{\beqn}{\begin{eqnarray}}
\newcommand{\eeqn}{\end{eqnarray}}

\definecolor{red}{cmyk}{0,1,1,0.4}

\definecolor{green1}{rgb}{0.06,0.66,0.06}
\definecolor{orange1}{rgb}{0.98,0.60,0.07}
\definecolor{darkgreen}{rgb}{0.0,0.6,0.0}
\definecolor{darkblue}{RGB}{12,13,115}
\definecolor{darkred}{RGB}{204,6,0}

\def \refeq#1{(\ref{#1})}
\def \refsec#1{Section~\ref{#1}}

\def \refapp#1{Appendix~\ref{#1}}
\def \reffig#1{Fig.~\ref{#1}}
\def \reftab#1{Table~\ref{#1}}

\newcommand{\Br}{\mathrm{Br}}

\newcommand{\tev}{\, {\rm TeV}}

\def\epe{\varepsilon'/\varepsilon}

\newcommand{\epsK}{\varepsilon_K}

\def\kpn{K^+\rightarrow\pi^+ \nu\bar\nu}
\def\klpn{K_{L}\rightarrow\pi^0 \nu\bar\nu}

\newcommand{\kepe}{\kappa_{\varepsilon^\prime}}

\newcommand{\GSM}{{\mathrm{G_{SM}}}}

\newcommand{\SUthreeC}{{\mathrm{SU(3)_c}}}
\newcommand{\SUtwoL}{{\mathrm{SU(2)_L}}}
\newcommand{\UoneY}{{\mathrm{U(1)_Y}}}
\newcommand{\UoneEM}{{\mathrm{U(1)_{\rm em}}}}

\def \One{\leavevmode\hbox{\small1\kern-3.6pt\normalsize1}} 

\newcommand{\wc}[3][{}]{[{\cal C}_{#2}^{#1}]_{#3}}
\newcommand{\dotwc}[3][{}]{[\dot{\cal C}_{#2}^{#1}]_{#3}}

\newcommand{\Wc}[2][{}]{{\cal C}_{#2}^{#1}}

\newcommand{\dotWc}[2][{}]{\dot{\cal C}_{#2}^{#1}}

\newcommand{\Op}[2][{}]{{\cal O}_{#2}^{#1}}
\newcommand{\op}[3][{}]{[{\cal O}_{#2}^{#1}]_{#3}}

\newcommand{\Yuk}[1]{Y_{#1}^{}}
\newcommand{\YukD}[1]{Y_{#1}^\dagger}

\newcommand{\muNP}{{\mu_{\Lambda}}}
\newcommand{\muEW}{{\mu_{\rm ew}}}
\newcommand{\muLow}{{\mu_{\rm low}}}

\begin{document}


\vspace{-14mm}
\begin{flushright}
   LMU-ASC 11/17 \\
  TUM-HEP-1077/17
\end{flushright}

\vspace{8mm}

\begin{center}

{\Large\bf\boldmath 
  Yukawa enhancement of $Z$-mediated New Physics \\[3mm]
  in $\Delta S = 2$ and $\Delta B = 2$ Processes 
} \\[8mm]

{\bf 
  Christoph Bobeth,${}^{1,2,4}$
  Andrzej~J.~Buras,${}^{1,2}$ 
  Alejandro Celis, ${}^3$ 
  Martin Jung ${}^{4}$ 
}\\[1cm]

{\small
${}^1$TUM Institute for Advanced Study, 
  Lichtenbergstr.~2a, D-85748 Garching, Germany \\[2mm]

${}^2$Physik Department, TU M\"unchen, 
  James-Franck-Stra{\ss}e, D-85748 Garching, Germany \\[2mm]

${}^3$Ludwig-Maximilians-Universit\"at M\"unchen, Fakult\"at f\"ur Physik,\\
  Arnold Sommerfeld Center for Theoretical Physics, 80333 M\"unchen, Germany \\[2mm]

${}^4$Excellence Cluster Universe, Technische Universit\"at M\"unchen,
  Boltzmannstr. 2, D-85748 Garching, Germany}

\end{center}

\vspace{3mm}

\begin{abstract}
\noindent
We discuss Yukawa-enhanced contributions from $Z$-mediated new physics to
down-type quark $\Delta F=2$ processes in the framework of the standard model
gauge-invariant effective theory (SMEFT). Besides the renormalization group (RG)
mixing of the $Z$-mediating $\psi^2 H^2 D$ operators into $\Delta F = 2$
operators, we include at the electroweak scale one-loop (NLO) matching
corrections consistently, necessary for the removal of the matching scale
dependence.  We point out that the right-handed $Z$-mediated interactions
generate through Yukawa RG mixing $\Delta F=2$ left-right operators, which are
further enhanced through QCD RG effects and chirally enhanced hadronic matrix
elements. We investigate the impact of these new effects on the known
correlations between $\Delta F=2$ and $\Delta F=1$ transitions in the SMEFT
framework and point out qualitative differences to previous parameterizations of
$Z$-mediated new physics that arise for the left-handed case. We illustrate how
specific models fit into our model-independent framework by using four models
with vector-like quarks.  We carry out model-independent analyses of scenarios
with purely left-handed and purely right-handed new-physics $Z$ couplings for
each of the three sectors $s\to d$, $b\to s$ and $b\to d$. Specifically we
discuss the correlations between $\epe$, $\varepsilon_K$, $K_L\to \mu^+\mu^-$
$\kpn$ and $\klpn$ in the Kaon sector, and $\phi_s$, $B_s\to\mu^+\mu^-$ and
$B\to K^{(*)} (\mu^+\mu^-, \nu\bar\nu)$ in the $b\to s$ sector and
$B_d\to\mu^+\mu^-$ in the $b\to d$ sector.

\end{abstract}

\setcounter{page}{0}
\thispagestyle{empty}
\newpage
\setcounter{tocdepth}{2}
\tableofcontents

\newpage

%
%
%
\section{
  Introduction
  \label{sec:intro}
}

In the Standard Model (SM) tree-level flavour-changing (FC) couplings of the $Z$
boson to quarks are forbidden by the Glashow-Iliopoulos-Maiani (GIM) mechanism
\cite{Glashow:1970gm}. At one-loop level the GIM mechanism is broken by the
disparity of quark masses and such couplings are generated, for instance through
the so-called $Z$-penguin diagrams. These play an important role
specifically in rare $K$ and $B_{s,d}$ decays ($\Delta F=1$) in which the GIM
mechanism is strongly broken because of the large top-quark mass. However, the
related one-loop function $C(x_t)$ ($x_t=m_t^2/M_W^2$), one of the Inami-Lim
functions \cite{Inami:1980fz}, is gauge dependent and additional diagrams have
to be included to cancel this gauge dependence. Only the resulting
gauge-independent one-loop functions have physical meaning and consequently
enter formulae for various rare decay observables \cite{Buchalla:1990qz}, see
\cite{Buras:2013ooa} for a recent review.

In models beyond the SM (BSM) FC quark couplings of the $Z$ can be present
already at tree-level and hence contribute to rare $K$ and $B_{s,d}$ decays as
well as neutral meson mixing, that is, $\Delta F=2$ processes. The relative size
of these contributions can be very large, given the loop- and GIM-suppression
present in the SM. Such scenarios can model-independently be treated in an
effective theory invariant under the SM gauge group
$\GSM = \SUthreeC \otimes \SUtwoL \otimes \UoneY$ (SMEFT)
\cite{Buchmuller:1985jz} if the new physics (NP) responsible for these couplings
is weakly coupled, can be decoupled at some high scale $\muNP \gg \muEW$, much
larger than the scale $\muEW$ of electroweak symmetry breaking (EWSB), and no
additional degrees of freedom are present besides the ones of the SM. FC quark
couplings of the $Z$ are then given at leading order (dimension six) by four
operators of the non-redundant ``Warsaw'' basis \cite{Grzadkowski:2010es}. Only
three of these are relevant for down-quark $\Delta F = 1,2$ flavour-changing
neutral current (FCNC) phenomenology: two operators with a left-handed (LH)
quark current, $\Op[(1,3)]{Hq}$, and one with a right-handed (RH) quark current,
$\Op{Hd}$. They belong to the class $\psi^2H^2D$ of operators with two fermion
fields, two Higgs-doublet fields and one covariant derivative. We do not consider
the fourth operator $\Op{Hu}$ of this class which is relevant for up-quark
processes, only. Operators of the class $\psi^2 H X$ of dipole operators, where
$X$ denotes the field strength tensors of $\GSM$, are suppressed by an
additional factor of a light fermion mass for the processes considered below and
hence also neglected.\footnote{Furthermore, these operators are generated only
  at loop level in weakly coupled gauge theories \cite{Arzt:1994gp}.}

The importance of $Z$-mediated FCNC processes has increased recently in view of
the absence of direct NP signals at the LHC and given that the neutral $Z$ is
particularly suited to be a messenger of possible NP even at scales far beyond
the reach of the LHC, see \cite{Buras:2012jb, Buras:2015yca, Buras:2015jaq,
  Endo:2016tnu} for recent analyses. Here we analyze $Z$-mediated NP in the
framework of SMEFT. The analysis yields several surprises concerning down-quark
$\Delta F = 2$ processes, which to our knowledge have not been noticed so far in
the literature:

\begin{enumerate}
\item[\bf 1.] 
In the presence of right-handed FC $Z$ couplings, \emph{i.e.} $\mathcal C_{H_d}
\neq 0$ for a FC transition, inspection of the renormalization group (RG) equations
due to Yukawa couplings in \cite{Jenkins:2013wua} yields that at $\muEW$ the
left-right $\Delta F=2$ operators $O_{{\rm LR},1}^{ij}$ in \refeq{eq:DF2-LR} are generated
and are enhanced by the large leading logarithm $\ln \muNP/\muEW$. Such operators
are known to provide very important contributions to $\Delta F=2$ observables
because of their enhanced hadronic matrix elements and an additional enhancement
from QCD RG effects below $\muEW$, in particular in the $K$-meson system.
As a result \emph{these} operators -- and not $O_{{\rm VRR}}^{ij}$ in \refeq{eq:DF2-VLL},
as used in \cite{Buras:2012jb, Buras:2015yca, Buras:2015jaq} -- dominate 
$\Delta F=2$ processes. The results in \cite{Jenkins:2013wua} allow the
calculation of this dominant contribution including only leading logarithms.

\item[\bf 2.] Because of the usual scale ambiguity present at leading order (LO)
we calculate the next-to-leading order (NLO) matching corrections of $\Op{Hd}$
to $\Delta F = 2$ processes at $\muEW$ within SMEFT. One NLO contribution is
obtained by replacing the flavour-diagonal lepton vertex in the SM $Z$-penguin
diagram by $\wc{Hd}{ij}$ (see \reffig{fig:psi2H2D-DF2-me-1HPR}), which again
generates the operator $O_{{\rm LR},1}^{ij}$ simply because the flavour-changing
part of the SM penguin diagram is LH. In fact this contribution has been recently
pointed out in \cite{Endo:2016tnu} and used for phenomenology. Unfortunately,
such contributions are by themselves gauge dependent, simply because the function
$C(x_t)$ present in the SM vertex is gauge dependent. Hence, while the observation
made in \cite{Endo:2016tnu} is important, the analysis of these new contributions
presented there is incomplete.\footnote{After the appearance of our paper
the authors of \cite{Endo:2016tnu} included the remaining contributions obtaining
a gauge independent NLO correction to $\Wc{Hd}$ which agrees with ours. 
Unfortunately they did not include  RG Yukawa effects above  $\muEW$, discussed 
in point 1. above, so that their result depends very strongly on $\muEW$ 
as we will demonstrate in \refsec{sec:COMPL}.} In the present paper we calculate the missing
contributions using SMEFT, obtaining a new gauge-independent function.
This contribution itself is by about a factor of two smaller than the one
found in \cite{Endo:2016tnu} (with the same sign), however, the LO contribution
is not only more important due to the large logarithm $\ln\muNP/\muEW$, but has
also opposite sign, allowing to remove the LO scale dependence. The total new
contribution that includes LO and NLO terms in SMEFT is larger and has opposite
sign to the one found in \cite{Endo:2016tnu}. Moreover being strongly enhanced
with respect to the contributions considered in \cite{Buras:2012jb, Buras:2015yca,
Buras:2015jaq}, it  has a very large impact on the phenomenology; in 
particular the correlations between $\Delta F=2$ and $\Delta F=1$ observables are
drastically changed.

\item[\bf 3.] The situation for LH FC $Z$ couplings is different from the RH
case both qualitatively and quantitatively: inspecting again the RG equations
in \cite{Jenkins:2013wua} we find that the two operators $\Op[(1)]{Hq}$ and
$\Op[(3)]{Hq}$ in SMEFT generate only the $\Delta F=2$-operator $O_{{\rm VLL}}$
dominant already in the SM, which is the same as generated by the contributions
considered in \cite{Buras:2012jb}. The resulting NP effects in $\Delta F=2$
transitions are then much smaller
than in the RH case, because no LR operators are present. Importantly, the
correlations between $\Delta F=1$ and $\Delta F=2$ processes are changed
dramatically: while $\Delta F=1$ transition amplitudes are proportional to the 
sum $\Wc[(1)]{Hq}{} + \Wc[(3)]{Hq}{}$, the leading RG contribution to $\Delta F=2$
processes is proportional to the difference of these couplings. Correlations
between $\Delta F=1$ and $\Delta F=2$ processes are hence only present in specific
models. This is in
stark contrast to the contributions considered in \cite{Buras:2012jb} and also
used in \cite{Endo:2016tnu}, where the same couplings enter both classes of
processes.\footnote{In particular in  \cite{Endo:2016tnu} the choice 
$\Wc[(3)]{Hq}=0$ has been made, which can only be true at a single renormalization 
scale.} Of course correlations remain in each sector separately, since both
are governed by two parameters, only. Interestingly this allows to access
\emph{both} coefficients \emph{separately} from  $\Delta F=2$ and $\Delta F=1$
observables, which did not seem possible before. In models where $\Delta F=2$
and $\Delta F=1$ observables are correlated, the constraints become weaker,
allowing for larger NP effects in rare decays.

\item[\bf 4.] Also for the operators $\Op[(1,3)]{Hq}$ the NLO contributions 
to $\Delta F=2$ transitions corresponding to the replacement of the
flavour-diagonal lepton vertex in the SM $Z$-penguin diagram 
\reffig{fig:psi2H2D-DF2-me-1HPR} by $\Wc[(1,3)]{Hq}$ are gauge dependent. 
We include the remaining contributions to remove this gauge dependence and find
a second gauge-independent function. Since the NLO contributions are different for
$\Wc[(1)]{Hq}$ and $\Wc[(3)]{Hq}$, it is not just their difference contributing
to $O_{{\rm VLL}}$ anymore, but also their sum. In fact, $\SUtwoL$
gauge invariance in SMEFT imposes relations between down-type and up-type
quark FCNCs that are governed by these operators, for example between
$B_{d,s}$-mixing and $t \to (u,c)$ processes.

\item[\bf 5.] At NLO also new gauge-independent contributions are generated which
are unrelated to tree-level $Z$ exchanges and only proportional to $\Wc[(3)]{Hq}$, 
analogous to the usual box diagrams with $W^\pm$ and quark exchanges. They turn
out to be important for gauge-independence and depend not only on the coefficients
for the quark transition under consideration, but also on additional ones
corresponding to the possible intermediate quarks in the box diagrams.

\end{enumerate}

It should be stressed in this context that the contributions to $\Delta F=2$
transitions from FC quark couplings of the $Z$ could be less relevant in NP
scenarios with other sources of $\Delta F=2$ contributions. Most importantly,
$\Delta F=2$ operators could receive a direct contribution at tree-level at the
scale $\muNP$, but also in models where this does not happen $Z$ contributions
could be subdominant. Examples are models in which the only new particles are
vector-like quarks (VLQs), where box diagrams with VLQ and Higgs exchanges
generate $\Delta F=2$ operators at one-loop level \cite{Ishiwata:2015cga,
  Bobeth:2016llm}, which were found in these papers to be larger than the $Z$
contributions at tree-level. However, in \cite{Ishiwata:2015cga} and in the
first version of our analysis in \cite{Bobeth:2016llm} the effects listed above
have not been included. As we will see below, for right-handed FC $Z$ couplings
these box contributions are dwarfed by the LR operator contributions mentioned
at the beginning of our list in Kaon mixing, whereas in $B$-mixing they are
comparable.

The outline of our paper is as follows: In \refsec{sec:SMEFT:DF2-EFT} we
establish notation, recall the parts of SMEFT and the effective Hamiltonian for
$\Delta F=2$ transitions most important for our work, and summarise the
tree-level matching between the two at $\muEW$. In \refsec{sec:LLA:RGE} we
summarise the results of the RG evolution in SMEFT. In \refsec{sec:NLO-matching}
we present the results of the NLO contributions to the matching at $\muEW$ with
some details relegated to \refapp{app:SMEFT}. The main outcome of this section
are gauge-independent functions $ H_1(x_t,\muEW)$ and $H_2(x_t,\muEW)$ that are
analogous to the $\Delta F=1$ loop functions of \cite{Buchalla:1990qz}. Their
dependence on the electroweak scale $\muEW$ cancels the scale dependence of the
Wilson coefficients resulting from the RG evolution. We illustrate the size of
NLO effects model-independently considering LH and RH couplings individually and
in the context of models with vector-like quarks (VLQs) using the results from
\cite{delAguila:2000rc, Ishiwata:2015cga, Bobeth:2016llm}. In \refsec{sec:COMPL}
we compare the framework of SMEFT used here with simplified models of FC quark
couplings of the $Z$ considered previously in \cite{Buras:2012jb, Buras:2015yca,
  Buras:2015jaq, Endo:2016tnu}, stressing significant limitations of these
models as far as FCNCs mediated by $Z$ exchanges are concerned.\footnote{Our
  critical analysis does not apply to $Z^\prime$ models considered in these
  papers.} In particular we compare our results to those of \cite{Endo:2016tnu}
who included some of the NLO contributions considered by us in the framework of
simplified models, reaching conclusions rather different from our findings. In
\refsec{sec:implications} we study model-independently the impact of the
presence of $\psi^2H^2D$ operators on the correlation between $\Delta F=1$ and
$\Delta F=2$ processes, like the $s\to d$ transitions
$\epe$, $\varepsilon_K$, $\kpn$ and $\klpn$ and the $b\to d,s$ transitions
$\Delta m_{d,s}$, $B_{d,s}\to\mu^+\mu^-$ or $B\to K^{(*)}(\mu^+\mu^-, \nu\bar\nu)$.  
We conclude in \refsec{sec:summary}.

%
%
%
\section
{\boldmath SMEFT, $\Delta F\!=\!2$--EFT and Tree-level Matching
  \label{sec:SMEFT:DF2-EFT}
}
\sectionmark{SMEFT, $\Delta F=2$--EFT and Tree-level Matching}

Throughout we assume that NP interactions have been integrated out at some scale
$\muNP \gg \muEW$, giving rise to the SMEFT framework. The field content of the
SMEFT-Lagrangian are the SM fields and the interactions are invariant under the
SM gauge group $\GSM$; the corresponding Lagrangian can be written as
\begin{align}
  \label{eq:GSM:EFT}
  {\cal L}_{{\rm SMEFT}} & 
  = {\cal L}_{{\rm dim}-4} + \sum_a \Wc{a} \Op{a}\,. 
\end{align}
Here ${\cal L}_{{\rm dim}-4}$ coincides with the SM Lagrangian and a
non-redundant set of operators of dimension six (dim-6), $\Op{a}$, has been
classified in \cite{Grzadkowski:2010es}. The anomalous dimensions (ADM)
necessary for the RG evolution from $\muNP$ to $\muEW$ of the SM couplings and
the Wilson coefficients $\Wc{a}$ are known at one-loop \cite{Jenkins:2013zja,
Jenkins:2013wua, Alonso:2013hga}.  Given some initial coefficients
$\Wc{a}(\muNP)$, they can be evolved down to $\muEW$, thereby resumming leading
logarithmic (LLA) effects due to the quartic Higgs, gauge and Yukawa couplings
into $\Wc{a}(\muEW)$. Far above $\muEW$ it is convenient to work in the unbroken
$\SUtwoL \otimes \UoneY$ phase, however close to $\muEW$ EWSB is taking place
and it is more convenient to transform gauge bosons and fermions from the weak
to their mass eigenstates.

For the purpose of down-quark $\Delta F = 1,2$ phenomenology a second decoupling
of heavy SM degrees of freedom ($W^\pm, Z, H, t$) takes place at the electroweak
scale $\muEW$. It gives rise to the $\Delta F=1,2$ effective Hamiltonians (EFT)
with the gauge symmetry $\SUthreeC \otimes \UoneEM$ and number of active quark
flavours $N_f = 5,4,3$ when going below the $b$- and $c$-quark thresholds
\cite{Buchalla:1995vs}.

We use the following notation for Wilson coefficients and operators in the
corresponding effective theories:
\begin{equation}
\begin{aligned}
  \mbox{SMEFT:} & &             {\cal L}_{\rm SMEFT}  & \sim \Wc{a} \Op{a}\,,
\\
  \Delta F = 2\mbox{--EFT:} & & {\cal H}_{\Delta F=2} & \sim C_{a} O_{a}\,.
\end{aligned}
\end{equation}
Note the use of the Lagrangian ${\cal L}$ for SMEFT, but the Hamiltonian 
${\cal H}$ for the $\Delta F=2$--EFT.\footnote{Note the relative sign 
${\cal L} = - {\cal H}$.}

%
%
\subsection{\boldmath
  SMEFT
  \label{sec:SMEFT}
}

In this work we are concerned with operators that induce FC quark couplings of
the $Z$ and their impact on the four-fermion ($\psi^4$) operators that mediate
$\Delta F = 2$ down-type quark transitions. The relevant operators in the quark
sector belong to the class $\psi^2 H^2 D$. The ones with LH quark currents are
\footnote{In order to simplify notations we suppress flavour indices on the
  operators.}
\begin{align}
  \label{eq:LH13}
  \Op[(1)]{Hq} & = (H^\dagger i \overleftrightarrow{\cal D}_{\!\!\!\mu} H) 
                   [\bar{q}_L^i \gamma^\mu q_L^j]\,, &
  \Op[(3)]{Hq} & = (H^\dagger i \overleftrightarrow{\cal D}^a_{\!\!\!\mu} H) 
                   [\bar{q}_L^i \sigma^a \gamma^\mu q_L^j]\,,
\end{align}
including also modified LH $W^\pm$ couplings. The ones with RH quark currents are
\begin{align}
  \label{eq:RH1}
  \Op{Hd} & = (H^\dagger i \overleftrightarrow{\cal D}_{\!\!\!\mu} H) 
              [\bar{d}_R^i \gamma^\mu d_R^j], &
  \Op{Hu} & = (H^\dagger i \overleftrightarrow{\cal D}_{\!\!\!\mu} H)
              [\bar{u}_R^i \gamma^\mu u_R^j]\,.
\end{align}
Finally there is one operator with charged RH quark currents:
\begin{align}
  \label{eq:O-Hud}
  \Op{Hud} & = (\widetilde{H}^\dagger i {\cal D}_{\!\mu} H) 
              [\bar{u}_R^i \gamma^\mu d_R^j]\,.
\end{align} 
Here $\widetilde{H} \equiv 
i\sigma_2 H^*$ and more details on conventions are given in \refapp{app:SMEFT}.
The complex-valued coefficients of these operators are denoted by
\begin{align}
  \label{eq:SMEFT-wilson-coeffs}
  \wc[(1)]{Hq}{ij}, && \wc[(3)]{Hq}{ij}, && 
  \wc{Hd}{ij}, && \wc{Hu}{ij}, && \wc{Hud}{ij}\,,
\end{align}
where the indices $ij=1,2,3$ denote the different generations of up- and
down-type quarks.

After EWSB, the transition from a weak to the mass eigenbasis takes place for
gauge and quark fields. The quark fields are rotated by $3\times 3$ unitary
rotations in flavour space
\begin{align}
  \psi_L & \to V^\psi_L \psi_L\,, & \psi_R & \to V^\psi_R \psi_R\,,
\end{align}
for $\psi = u,d$, such that
\begin{align}
  V^{\psi\dagger}_L m_\psi V^\psi_R & = m_\psi^{\rm diag}, &
  V & \equiv (V^{u}_L)^\dagger V^d_L \,,
\end{align}
with diagonal up- and down-quark mass matrices $m_\psi^{\rm diag}$. The
non-diagonal mass matrices $m_\psi$ include the contributions of dim-6
operators. The quark-mixing matrix $V$ is unitary, similar to the CKM matrix of
the SM; however, in the presence of dim-6 contributions the numerical values are
different from those obtained in usual SM CKM-fits. Throughout we will take the
freedom to choose the weak basis such that down quarks are already mass
eigenstates, which fixes $V_{L,R}^d = \One$, and assume without loss of
generality $V_R^u = \One$, yielding $q_L = (V^\dagger u_L, d_L)^T$.

The $\psi^2 H^2 D$ operators lead to modifications of the couplings of quarks to
the weak gauge bosons ($V=W^\pm, Z$ and $g_Z\equiv \sqrt{g_1^2 + g_2^2}$): 
\begin{equation}
  \label{eq:SMEFT-qqV-dim6}
\begin{aligned}
  {\cal L}_{\psi\bar\psi V}^{\rm dim-6} = -\frac{g_Z}{2} v^2 Z_\mu & 
  \Big(
    \left[V_L^{d\dagger} (\Wc[(1)]{Hq} + \Wc[(3)]{Hq}) V_L^d \right]_{ij} 
    \big[\bar{d}_i \gamma^\mu P_L d_j \big] 
  + \left[V_R^{d\dagger} \Wc{Hd} V_R^d \right]_{ij} 
    \big[\bar{d}_i \gamma^\mu P_R d_j \big]  
\\ & \!\!\!
  + \left[V_L^{u\dagger} (\Wc[(1)]{Hq} - \Wc[(3)]{Hq}) V_L^u \right]_{ij} 
    \big[\bar{u}_i \gamma^\mu P_L u_j \big] 
  + \left[V_R^{u\dagger} \Wc{Hu} V_R^u \right]_{ij} 
    \big[\bar{u}_i \gamma^\mu P_R u_j \big]  
  \Big)
\\
 + \frac{g_2}{\sqrt{2}} v^2
 \Big( 
   \Big[V_L^{u\dagger} \Wc[(3)]{Hq} & V_L^d \Big]_{ij}
   \big[\bar{u}_i \gamma^\mu P_L d_j \big] W^+_\mu
 + \left[V_R^{u\dagger} \frac{\Wc{Hud}}{2} V_R^d \right]_{ij} 
   \big[\bar{u}_i \gamma^\mu P_R d_j \big] W^+_\mu
 + \mbox{h.c.} \Big)\,,
\end{aligned}
\end{equation} 
where we display all rotation matrices for completeness. Note that in our notation
fermion fields with an index for their handedness correspond to weak eigenstates,
whereas mass eigenstates -- like in this equation -- do not carry this index. 
The values for $v$ and the gauge couplings $g_{1,2}$ differ from the SM ones by
dim-6 contributions. However, since the couplings in \refeq{eq:SMEFT-qqV-dim6}
are already at the level of dim-6, such corrections would count as dim-8 
contributions, which is of higher order than considered here. From this equation
it is apparent that our definition of the Wilson coefficients $\wc{a}{ij}$ 
(at $\muEW$) for our special choice of the weak quark basis -- $V_{L,R}^d = 
\One$ and $V_R^u = \One$ -- is particularly convenient for the study of down-type
quark $\Delta F=1,2$ transitions, see also \cite{Aebischer:2015fzz}, because
additional CKM factors appear only in couplings of operators involving
left-handed up-type quarks. Thus the associated Wilson coefficients
$\Wc[(1,3)]{Hq}$ enter down- \emph{and} up-type-quark processes, leading to correlations
between the affected processes that depend on the appearing CKM factors. As an example
let us consider the top-quark FCNC coupling $t\to c Z$ ($i = c$, $j = t$),
\begin{align}
  \label{eq:corr-Bsmix-tcZ}
  {\cal L}_{t\to cZ}^{\rm dim-6} & 
  \; \propto \; \sum_{k,l} V_{ck} [\Wc[(1)]{Hq} - \Wc[(3)]{Hq}]_{kl} V^*_{tl} 
  \; \approx \; [\Wc[(1)]{Hq} - \Wc[(3)]{Hq}]_{sb} + {\cal O}(\lambda_C) \, ,
\end{align}
where we have neglected in the second step contributions suppressed by the Cabibbo
angle $\lambda_C$ and assumed that $V_{cs} \sim V_{tb} \sim {\cal O}(1)$ in 
SMEFT. Moreover, we have assumed that the Wilson coefficients themselves are 
all of the same size and hence do not upset the hierarchy of the CKM factors. 
In this case in fact $B_s$-mixing depends on above linear combination -- see the 
result \refeq{eq:1stLLA-VLL} -- and hence is directly correlated to $t\to cZ$ FCNC
decays. Under the same assumptions, also $B_d$ and Kaon mixing are related to 
$t\to uZ$ and $c\to uZ$ FCNC decays, respectively.

We will omit $\Op{Hu}$ and $\Op{Hud}$ in the phenomenological part of our
work,\footnote{This is mainly justified because their RG flow does not induce
  leading logarithmic contributions to down-type quark $\Delta F=1,2$
  processes.} where we deal with down-type quark $\Delta F=1,2$ processes. The
Wilson coefficients $\Wc[(1,3)]{Hq}$, $\Wc{Hd}$ (and $\Wc{Hu}$) are
complex-valued matrices in flavour space with a symmetric real part and
antisymmetric imaginary part, such that each contains $6+3=9$ real degrees of
freedom.

It is customary to parameterize FC quark couplings of the $Z$ as \cite{Buras:2012jb}
\begin{align}  
  \label{eq:Zcouplings}
  \mathcal{L}_{\psi\bar\psi Z}^{\rm NP} & 
  = Z_{\mu} \sum_{\psi = u,d} \bar \psi_i \, \gamma^{\mu} \left( 
        [\Delta_L^{\psi}]_{ij} \, P_L 
  \,+\, [\Delta_R^{\psi}]_{ij} \, P_R \right) \psi_j \,.
\end{align}
This parameterization introduces also complex-valued couplings, whose relation
to the dim-6 SMEFT tree-level contributions can be read off as
\begin{equation}
  \label{eq:Z-Deltas:dim-6-WC}
\begin{aligned}
  \phantom{x}[\Delta^u_L]_{ij} & 
  = -\frac{g_Z}{2} v^2 \left[\Wc[(1)]{Hq} - \Wc[(3)]{Hq}\right]_{ij} , \qquad &
  [\Delta^u_R]_{ij} & 
  = -\frac{g_Z}{2} v^2 \wc{Hu}{ij} ,
\\
  [\Delta^d_L]_{ij} & 
  = -\frac{g_Z}{2} v^2 \left[\Wc[(1)]{Hq} + \Wc[(3)]{Hq}\right]_{ij} , \qquad &
  [\Delta^d_R]_{ij} &
  = -\frac{g_Z}{2} v^2 \wc{Hd}{ij} ,
\end{aligned}
\end{equation}
in the special weak basis advocated before for the Wilson coefficients. We
will comment in detail in \refsec{sec:COMPL} on the validity of this approach
that does not incorporate the SM-gauge invariance and can only be justified 
when neglecting the Yukawa RG effects and hence adapted only at tree-level.
 
The $\psi^4$ operators that mediate $\Delta F=2$ transitions
are in SMEFT the $(\overline{L}L)(\overline{L}L)$ operators
\begin{align}
  \label{eq:SMEFT:DF2-LLLL}
  \op[(1)]{qq}{ijkl} & 
  = [\bar{q}_L^i \gamma_\mu q_L^j] [\bar{q}_L^k \gamma^\mu q_L^l] , &
  \op[(3)]{qq}{ijkl} & 
  = [\bar{q}_L^i \gamma_\mu \sigma^a q_L^j] [\bar{q}_L^k \gamma^\mu \sigma^a q_L^l] ,
\end{align}
the $(\overline{L}L)(\overline{R}R)$ operators
\begin{align}
  \label{eq:SMEFT:DF2-LLRR}
  \op[(1)]{qd}{ijkl} & 
  = [\bar{q}_L^i \gamma_\mu q_L^j] [\bar{d}_R^k \gamma^\mu d_R^l] , &
  \op[(8)]{qd}{ijkl} & 
  = [\bar{q}_L^i \gamma_\mu T^A q_L^j] [\bar{d}_R^k \gamma^\mu T^A d_R^l] ,
\end{align}
and the $(\overline{R}R)(\overline{R}R)$ operator
\begin{align}
  \label{eq:SMEFT:DF2-RRRR}
  \op{dd}{ijkl} & 
  = [\bar{d}_R^i \gamma_\mu d_R^j] [\bar{d}_R^k \gamma^\mu d_R^l] ,
\end{align}
with $kl = ij$ for $\Delta F = 2$ processes. The $T^A$ denote $\SUthreeC$ colour
generators.

%
%
\subsection[Effective Hamiltonian for $\Delta F = 2$]
{ \boldmath
   Effective Hamiltonian for $\Delta F = 2$
  \label{sec:DF2-EFT}
}

The decoupling of heavy SM degrees of freedom at $\muEW$ gives rise to the
$\Delta F=2$ effective Hamiltonian \cite{Ciuchini:1997bw, Buras:2000if}
\begin{align}
  \label{eq:DF2-hamiltonian}
  {\cal H}_{\Delta F = 2}^{ij} &
  = {\cal N}_{ij} \sum_a  C_a^{ij} O_a^{ij} + \mbox{h.c.} ,
\end{align}
where the normalisation factor and the CKM combinations are
\begin{align}
  \label{eq:DF2:norm-factor}
  {\cal N}_{ij} & 
  = \frac{G_F^2}{4 \pi^2} M_W^2 \left(\lambda^{ij}_t\right)^2 , \qquad 
\lambda^{ij}_{t}=V_{ti}^*V_{tj}\,,
\end{align}
with $ij = sd$ for kaon mixing and $ij = bd,bs$ for $B_d$ and $B_s$
mixing, respectively. The important operators for our discussion are in the 
operator basis of \cite{Buras:2000if}
\begin{align}
  \label{eq:DF2-VLL}
  O_{{\rm VLL}}^{ij} & 
  = [\bar{d}_i \gamma_\mu P_L d_j][\bar{d}_i \gamma^\mu P_L d_j] \,, &
  O_{{\rm VRR}}^{ij} & 
  = [\bar{d}_i \gamma_\mu P_R d_j][\bar{d}_i \gamma^\mu P_R d_j] \,, 
\\
  \label{eq:DF2-LR}
  O_{{\rm LR},1}^{ij} & 
  = [\bar{d}_i \gamma_\mu P_L d_j][\bar{d}_i \gamma^\mu P_R d_j] \,, &
  O_{{\rm LR},2}^{ij} & 
  = [\bar{d}_i P_L d_j][\bar{d}_i P_R d_j] \,,
\end{align}
and the complete set can be found in \cite{Ciuchini:1997bw, Buras:2000if}, where
also the ADM matrices have been calculated up to NLO in QCD. We use these
results in the QCD RG evolution from the electroweak scale $\muEW$ to low-energy
scales.

In the SM only
\begin{align}
  \label{eq:DF2:S0}
  C_{{\rm VLL}}^{ij}(\muEW)|_{\rm SM} & = S_0(x_t), &
  S_0(x) & 
  = \frac{x (4 - 11 x + x^2)}{4\, (x-1)^2} + \frac{3 x^3 \ln x}{2\, (x-1)^3}   
\end{align}
is non-zero at the scale $\muEW$. 

%
%
\subsection[$\Delta F=2$ Tree-level matching]
{ \boldmath
  $\Delta F=2$ Tree-level matching  
  \label{sec:SMEFT:DF2-EFT:matching}
}

The SMEFT is matched to the $\Delta F=2$--EFT at $\muEW$ and at tree-level one
finds the following modifications of the Wilson coefficients
\cite{Aebischer:2015fzz}:
\begin{equation}
  \label{eq:GSM-DF2-matching}
\begin{aligned}
  \Delta C_{\rm VLL}^{ij} & =
     - {\cal N}_{ij}^{-1} \left( \wc[(1)]{qq}{ijij} + \wc[(3)]{qq}{ijij} \right) , \qquad &
  \Delta C_{\rm VRR}^{ij} & = - {\cal N}_{ij}^{-1} \wc{dd}{ijij} ,
\\
  \Delta C_{{\rm LR}, 1}^{ij} & =
   -{\cal N}_{ij}^{-1} \left( \wc[(1)]{qd}{ijij} 
            - \frac{\wc[(8)]{qd}{ijij}}{2 N_c} \right) , \qquad &
  \Delta C_{{\rm LR}, 2}^{ij} & = {\cal N}_{ij}^{-1} \wc[(8)]{qd}{ijij} ,
\end{aligned}
\end{equation}
where ${\cal N}_{ij}$ is given in \refeq{eq:DF2:norm-factor}. As will be seen in
the next section, the $\Delta F = 2$ Wilson coefficients at $\muEW$ receive
leading logarithmic contributions via up-type Yukawa-induced mixing from
$\psi^2 H^2 D$ operators. The minus signs in the case of VLL and VRR operators
reflect the fact that $\wc[(1,3)]{qq}{ijij}$ and $\wc{dd}{ijij}$ are the
coefficients in the Lagrangian and the coefficients $C_{\rm VLL}^{ij}$ and
$C_{\rm VRR}^{ij}$ in the Hamiltonian. In the case of LR operators additional
Fierz transformations have to be made.

%
%
%
\section{
  Leading RG Effects in SMEFT
  \label{sec:LLA:RGE}
}

The scale dependence of Wilson coefficients is governed by the RG equation
\begin{align}
  \dotWc{a} & \equiv (4\pi)^2 \mu \frac{{\rm d} \Wc{a}}{{\rm d}\mu} 
  = \gamma_{ab}\, \Wc{b}
\end{align}
and determined by the ADM $\gamma_{ab}$. The ADM is known for SMEFT at one-loop
and depends on 1) the quartic Higgs coupling $\lambda$ \cite{Jenkins:2013zja},
2) the fermion Yukawa couplings to the Higgs doublet \cite{Jenkins:2013wua} and
3) the three gauge couplings $g_{1,2,s}$ \cite{Alonso:2013hga}. For small
$\gamma_{ab}/(4\pi)^2 \ll 1$ the approximate solution retains only the first
leading logarithm (1stLLA)
\begin{align}
  \label{eq:SMEFT-RGE}
  \Wc{a} (\muEW) &
  =  \left[ \delta_{ab} 
     - \frac{\gamma_{ab}}{(4 \pi)^2} \ln \frac{\muNP}{\muEW} \right]
     \Wc{b} (\muNP)\,,
\end{align}
which is sufficient as long as the logarithm is not too large, so that also
$\gamma_{ab}/(4\pi)^2\ln \frac{\muNP}{\muEW} \ll 1$ holds. Numerically one
expects the largest enhancements when the ADM $\gamma_{ab}$ is proportional to
the strong coupling $4\pi \alpha_s \sim 1.4$ or the top-Yukawa coupling
squared $y^2_t \sim 1$. The QCD mixing is flavour-diagonal
and hence cannot give rise to new genuine phenomenological effects in
$\Delta F=1,2$ observables. On the other hand, Yukawa couplings are the main
source of flavour-off-diagonal interactions responsible for the phenomenology
discussed here. The $\SUtwoL$ gauge interactions induce via ADMs
$\gamma_{ab} \propto g_2^2$ and are parametrically suppressed compared to
$y_t$-induced effects, such that we do not consider them here. The suppression
is even stronger for $\UoneY$.

Note that the RG equations of SMEFT are in principle a set of coupled
differential equations of the RG equations of SM couplings (quartic Higgs, 
gauge and Yukawa) and the ones of dim-6 Wilson coefficients. The solution of 
this system in full generality requires the application of numerical methods
and the imposition of boundary conditions might be highly nontrivial.
The 1stLLA neglects all these ``secondary mixing'' effects that would be 
present in the general leading logarithmic approximation (LLA), which would
also resum large logarithms to all orders in couplings. With ``secondary 
mixing'' we refer to the situation where an operator ${\cal O}_A$ might not 
have an ADM entry with operator ${\cal O}_B$ (no ``direct mixing''), but still
contributes to the Wilson coefficient $C_{B}(\muEW$), via a direct mixing with 
some operator ${\cal O}_C$ that in turn mixes directly into ${\cal O}_B$.

The most relevant mixing of the $\psi^2 H^2 D$ operators $\Op[(1,3)]{Hq}$ and
$\Op{Hd}$ into $\Delta F=2$-mediating $\psi^4$ operators
\refeq{eq:SMEFT:DF2-LLLL}-\refeq{eq:SMEFT:DF2-RRRR} proceeds via up-type Yukawa
couplings \cite{Jenkins:2013wua}, yielding for $(\bar{L}L)(\bar{L}L)$ operators
\begin{align}
  \label{eq:RGE:qq1}
  \dotwc[(1)]{qq}{ijij} & 
  = + [\Yuk{u} \YukD{u}]_{ij} \wc[(1)]{Hq}{ij} + \ldots ,
\\
  \label{eq:RGE:qq3}
  \dotwc[(3)]{qq}{ijij} &
  = - [\Yuk{u} \YukD{u}]_{ij} \wc[(3)]{Hq}{ij} + \ldots ,
\intertext{and for $(\bar{L}L)(\bar{R}R)$ operators}
  \label{eq:RGE:qd1}
  \dotwc[(1)]{qd}{ijij} & 
  = [\Yuk{u} \YukD{u}]_{ij} \wc{Hd}{ij} + \ldots \, .
\end{align}
The dots indicate other terms $\propto\Wc{\psi^2H^2D}$ that are not proportional
to $\Yuk{u}$ \footnote{Note that here $\Yuk{u}$ is defined
  \refeq{eq:SM-dim-4-Yuk} as the hermitean conjugate w.r.t
  \cite{Jenkins:2013wua}.} as well as terms $\propto \Wc{\psi^4}$, which become
relevant in scenarios where these Wilson coefficients are generated at $\muNP$.
As can be seen, the two $\Delta F=2$ operators $\Op{dd}$ and $\Op[(8)]{qd}$ do
not receive direct leading logarithmic contributions. It is well known though
that $\Wc[(8)]{qd}$ would be generated via secondary QCD mixing from
$\Wc[(1)]{qd}$ \cite{Buras:2000if}.

Under the transformation from weak to mass eigenstates for up-type quarks
\begin{align}
  \Yuk{u} &
  \; \stackrel{\rm dim-4}{\approx} \;
   \frac{\sqrt{2}}{v} V_L^u m_U^{\rm diag} V_R^{u\dagger}
  \; = \; \frac{\sqrt{2}}{v} V_{\rm CKM}^\dagger m_U^{\rm diag} ,
\end{align}
the ADMs transform as
\begin{align}
  \label{eq:ADM-basis-change}
  [\Yuk{u}\YukD{u}]_{ij} & 
  \; = \;\frac{2}{v^2} \sum_{k=u,c,t} m_k^2 V_{ki}^* V_{kj}^{}
  \; \approx \; \frac{2}{v^2} m_t^2 \lambda_t^{ij}\,,
\end{align}
with up-type quark masses $m_k$. Since the ADMs are needed here for the 
evolution of dim-6 Wilson coefficients themselves, we have used tree-level
relations derived from the dim-4 part of the Lagrangian only, thereby 
neglecting dim-6 contributions, which would constitute dim-8 corrections
in this context. In the sum over $k$ only the top-quark contribution
is relevant ($m_{u,c} \ll m_t$), if one assumes that the unitary 
matrix $V$ is equal to the CKM matrix up to dim-6 corrections.\footnote{
We expect only tiny contributions from $k=c$ in case that $ij = sd$,
for $ij = bd, bs$ such contributions are entirely negligible.}

From \refeq{eq:GSM-DF2-matching} and \refeq{eq:RGE:qd1} we find that the
presence of the RH operator ${\cal O}_{Hd}$ at a short distance scale $\muNP$,
i.e. $\wc{Hd}{ij}(\muNP) \neq 0$, generates through Yukawa RG effects a
leading-logarithmic contribution to the LR operator $O_{{\rm LR},1}^{ij}$ at the
electroweak scale $\muEW$, given by
\begin{align}
  \label{eq:1stLLA-LR1}
  \Delta_{\rm 1stLLA} C_{\rm LR,1}^{ij} (\muEW) & =
  v^2 \frac{\wc{Hd}{ij}(\muNP)}{\lambda_t^{ij}}
  x_t  \ln\frac{\muNP}{\muEW} ,
\end{align}
with $v \approx (\sqrt{2} G_F)^{-1/2}$. Similarly, the
presence of two operators ${\cal O}_{Hq}^{(1)}$ and ${\cal O}_{Hq}^{(3)}$
generates via \refeq{eq:GSM-DF2-matching}, \refeq{eq:RGE:qq1} and
\refeq{eq:RGE:qq3}
\begin{align}
  \label{eq:1stLLA-VLL}
  \Delta_{\rm 1stLLA} C_{\rm VLL}^{ij} (\muEW) & = 
  \frac{v^2}{\lambda_t^{ij}} 
  \left(\wc[(1)]{Hq}{ij}(\muNP) - \wc[(3)]{Hq}{ij}(\muNP) \right) 
   x_t \ln\frac{\muNP}{\muEW} .
\end{align}
In summary
\begin{itemize}
\item $\Op{Hd}$, corresponding to RH FC quark couplings of the $Z$, generates 
      the $\Delta F=2$ left-right operator $O_{\rm LR,1}$, which has
      numerically enhanced QCD running below $\muEW$ and chirally enhanced
      hadronic matrix elements;
\item $\Op[(1,3)]{Hq}$, corresponding to LH FC quark couplings of the $Z$,
      generate the $\Delta F=2$ left-left operator $O_{\rm VLL}$; 
\item in the LH case the contribution is $\sim (\Wc[(1)]{Hq} - \Wc[(3)]{Hq})
      \propto \Delta_L^u$, contrary to the linear combination $\Delta_L^d \propto 
      (\Wc[(1)]{Hq} + \Wc[(3)]{Hq})$ appearing in the coupling of down-type
      quarks to the $Z$ in \refeq{eq:SMEFT-qqV-dim6}.
\end{itemize}
The latter point implies that the result in \refeq{eq:1stLLA-VLL} and
consequently the contributions to $\Delta F=2$ processes cannot be presented
solely in terms of $\Delta_L^d$ but must involve due to $\SUtwoL$ gauge
invariance also $\Delta_L^u$. We will return to the phenomenological
implications in \refsec{sec:implications}.

%
%
%
\section{
  NLO Contributions in SMEFT
  \label{sec:NLO-matching}
}

In this section we present the results of the calculation of one-loop (NLO)
corrections of matrix elements of $\psi^2 H^2 D$ operators to $\Delta F = 2$
transitions in SMEFT, arising in the matching to the $\Delta F=2$--EFT.  The
divergent parts of these matrix elements determine the ADM given in
\cite{Jenkins:2013wua}, which have been used in \refsec{sec:LLA:RGE} for the
leading RG evolution.  Here we calculate the finite parts that also scale with
the top-quark Yukawa coupling. While they are not enhanced by a large logarithm,
they cancel the $\muEW$ dependence of the leading RG results in
\refeq{eq:1stLLA-LR1} and \refeq{eq:1stLLA-VLL}. As mentioned in
\refsec{sec:intro}, these finite parts involve novel gauge-independent
functions. Similar NLO matching corrections in the context of matching SMEFT
onto low-energy EFTs have been also calculated for operators entering
$\mu\to e\gamma$ \cite{Pruna:2014asa}, $\mu\to e\nu_\mu\bar{\nu}_e$
\cite{Gauld:2015lmb}, which is used to determine $G_F$, anomalous triple gauge
couplings in rare decays $d_j \to d_i + (\gamma,\, \ell^+\ell^-, \nu\bar\nu$)
\cite{Bobeth:2015zqa} and extensively for many $\Delta B = 1,2$ processes in
\cite{Aebischer:2015fzz}.  Regarding $\Delta F = 2$ transitions, the latter work
has considered only NLO matrix elements of $\psi^4$ operators ($\Op[(1,8)]{qu}$
and $\Op[(1,8)]{ud}$) to the matching of SMEFT and $\Delta F=2$--EFT.
Very recently, the NLO matching corrections of $\Op{Hud}$ have been
calculated for $\Delta F = 1, 2$ processes in \cite{Alioli:2017ces}.

%
%
\subsection[$\Delta F=2$ NLO matching] 
{
  \boldmath
  $\Delta F=2$ NLO matching  
  \label{sec:SMEFT:DF2-EFT:NLO:matching}
}

\begin{figure}
  \centering
  \begin{subfigure}[t]{0.24\textwidth}
    \centering
    \includegraphics[width=0.9\textwidth]{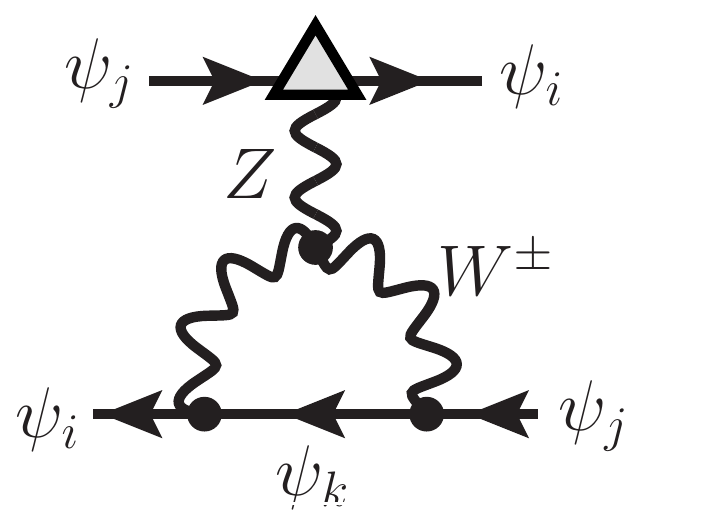}
    \caption{}
    \label{fig:psi2H2D-DF2-me-1HPR}
  \end{subfigure}
  \begin{subfigure}[t]{0.24\textwidth}
    \centering
    \includegraphics[width=0.9\textwidth]{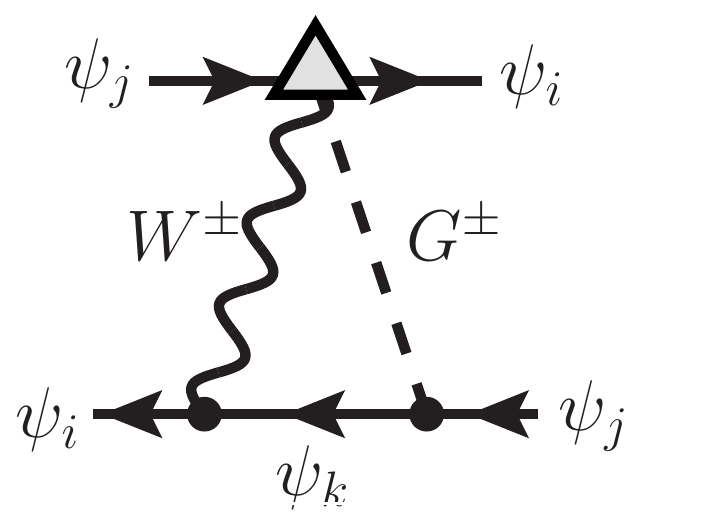}
    \caption{}
    \label{fig:psi2H2D-DF2-me-1PI-WG}
  \end{subfigure}
  \begin{subfigure}[t]{0.24\textwidth}
    \includegraphics[width=0.9\textwidth]{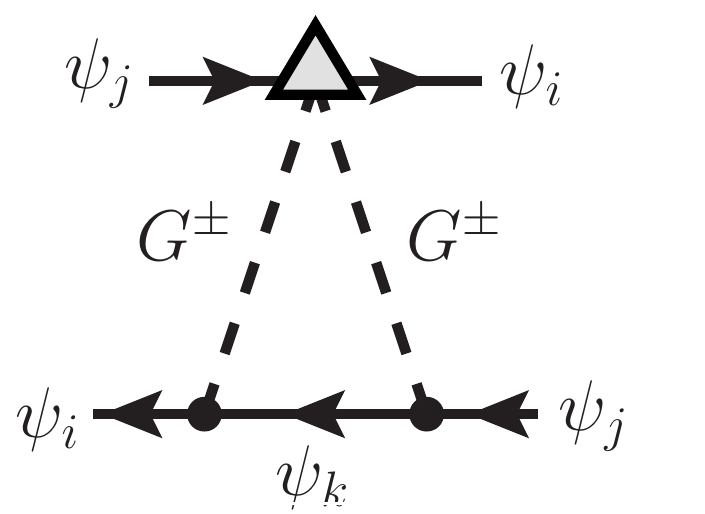}
    \caption{}
    \label{fig:psi2H2D-DF2-me-1PI-GG}
  \end{subfigure}
  \begin{subfigure}[t]{0.24\textwidth}
    \centering
    \vskip -0.2cm
    \includegraphics[width=0.9\textwidth]{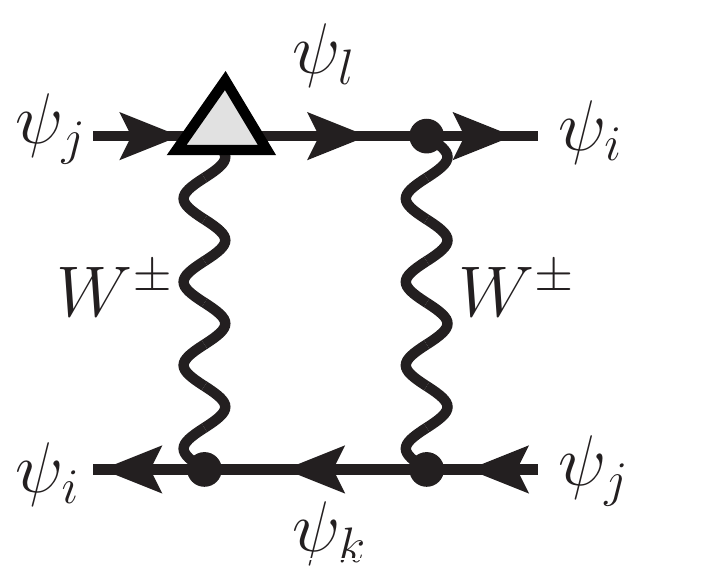}
    \caption{}
    \label{fig:psi2H2D-DF2-me-box}
  \end{subfigure}
\caption{\small  
  Classes of diagrams (up to permutations and insertions flipped to lower 
  fermion-line) contributing to $\Delta F = 2$ one-loop matrix elements of 
  the operators $\Op{Hd}$ and $\Op[(1,3)]{Hq}$. Operator insertions are 
  depicted by triangles and SM couplings by small dots. Diagrams
  (\ref{fig:psi2H2D-DF2-me-1HPR}) and (\ref{fig:psi2H2D-DF2-me-box}) 
  appear also with replacement of each $W^\pm \to G^\pm$ and further
  (\ref{fig:psi2H2D-DF2-me-1HPR}) with the $Z$-boson emitted from the virtual
  $\psi_k$. The operators $\Op{Hd}$ and $\Op[(1)]{Hq}$ generate diagrams
  (\ref{fig:psi2H2D-DF2-me-1HPR} -- \ref{fig:psi2H2D-DF2-me-1PI-GG}), 
  whereas operator $\Op[(3)]{Hq}$ generates diagrams (\ref{fig:psi2H2D-DF2-me-1HPR},
  \ref{fig:psi2H2D-DF2-me-1PI-GG}, \ref{fig:psi2H2D-DF2-me-box}).
}
  \label{fig:psi2H2D-DF2-me}
\end{figure}

The classes of diagrams we consider are shown in \reffig{fig:psi2H2D-DF2-me}.
Details on the Feynman rules of the operators $\Op{Hd}$ and $\Op[(1,3)]{Hq}$ are
provided in \refapp{app:SMEFT} in the mass eigenbasis for gauge bosons and
quarks after EWSB.  Among the 1-particle-irreducible~(1PI) diagrams only
the diagram (\ref{fig:psi2H2D-DF2-me-1PI-GG}) gives rise to divergences, which are
absorbed into counterterms and are known from the ADM calculation
\cite{Jenkins:2013wua}. The heavy-particle-reducible~(HPR)
diagram~(\ref{fig:psi2H2D-DF2-me-1HPR}) is well-known from the SM calculation
and gives rise to the gauge-dependent Inami-Lim function $C(x_t)$.  As usual, it
requires the inclusion of the counterterm of the flavour-off-diagonal
wave-function renormalization constant $\psi_j \to \psi_i$. The external momenta
and the up- and charm-quark masses are set to zero throughout. We do not use the
background-field method, such that top-, charm- and up-quark contributions are
not separately finite \cite{Bobeth:1999mk}, but their sum is finite after
GIM-summation, i.e. exploiting the unitarity of the CKM matrix $V$. The box
diagrams~(\ref{fig:psi2H2D-DF2-me-box}) are finite, but not
gauge-independent. We perform our calculation in general $R_\xi$ gauge for
electroweak gauge bosons in order to verify explicitly the gauge independence of
the final results after GIM-summation.

The result of the one-loop matching modifies the tree-level matching relations
\refeq{eq:GSM-DF2-matching} as
\begin{align}
  \label{eq:SMEFT-DF2-NLO-me}
  \Delta C_c (\muEW) & =  
    \sum_{a \in \psi^4} G^{(0,c)}_a \, \Wc[\psi^4]{a} (\muEW) 
  + \sum_{b \in \psi^2 H^2 D} G^{(1,c)}_b(x_t, \muEW)\,,
\end{align}
with $c = {\rm VLL},\, {\rm LR,\!1}$ and generation-indices omitted for 
brevity. In this equation
\begin{itemize}
\item $G^{(0,c)}_a$ is the tree-level matrix element of the
  $\psi^4$ operator $\Op{a}$ to the $\Delta F=2$--EFT operator $O_c$, 
  which can be read off from \refeq{eq:GSM-DF2-matching};
\item $G^{(1,c)}_b(x_t, \muEW)$ is the one-loop matrix element 
  of the $\psi^2 H^2 D$ operator $\Op{b}$ to $O_c$;
\item the $\muEW$ dependence of $G^{(1,c)}_b(x_t, \muEW)$ cancels
  the one present in $\Wc[\psi^4]{a} (\muEW)$ due to \refeq{eq:1stLLA-LR1} 
  and \refeq{eq:1stLLA-VLL}.
\end{itemize}
We find for $\Op{Hd}$
\begin{align}
  G^{(1,{\rm LR1})}_{Hd,ij}(x_t, \muEW) & 
  = v^2 \frac{\wc{Hd}{ij} (\muEW)}{\lambda_t^{ij}} x_t H_1(x_t, \muEW) \,  , 
\end{align}
and similarly for $\Op[(1)]{Hq}$
\begin{align}
  G^{(1,{\rm VLL})}_{Hq^{(1)},ij}(x, \muEW) & 
  = v^2 \frac{\wc[(1)]{Hq}{ij} (\muEW)}{\lambda_t^{ij}} x_t H_1(x_t, \muEW) .
\end{align}
In the case of $\Op[(3)]{Hq}$ 
\begin{equation}
\begin{aligned}
  \phantom{a}G^{(1,{\rm VLL})}_{Hq^{(3)},ij}(x, &\; \muEW) 
  =  - \frac{v^2}{\lambda_t^{ij}} \, x_t \Bigg[
       \wc[(3)]{Hq}{ij} (\muEW) \, H_2(x_t, \muEW) 
\\ &
  - \frac{2 S_0(x_t)}{x_t}
  \sum_m \left( \lambda_t^{im}\, \wc[(3)]{Hq}{mj} (\muEW)
              + \wc[(3)]{Hq}{im}(\muEW) \, \lambda_t^{mj} \right) \Bigg]\,,
\end{aligned}
\end{equation}
where the second term is due to the box diagrams in
\reffig{fig:psi2H2D-DF2-me-box}, proportional to the gauge-independent SM
Inami-Lim function $S_0(x_t)$ \refeq{eq:DF2:S0} and moreover involving not only
the coefficients $\wc[(3)]{Hq}{ij}$, but also those with $m \neq i$ and
$m \neq j$.

The expressions of the $\Delta F=2$ Wilson coefficients at NLO are then obtained
by inserting the 1stLLA results \refeq{eq:1stLLA-LR1} and \refeq{eq:1stLLA-VLL}
together with the NLO matrix elements $G^{(1,c)}_b(x_t, \muEW)$ into
\refeq{eq:SMEFT-DF2-NLO-me}. Furthermore, in $G^{(1,c)}_b(x_t, \muEW)$ we
neglect ``self-mixing'' and approximate $\Wc{\psi^2 H^2 D} (\muEW) \approx 
\Wc{\psi^2 H^2 D} (\muNP)$ -- see \refeq{eq:SMEFT-RGE} and 
\refeq{eq:psi2H2D-selfmixing} -- which is numerically small, such that finally
\begin{align}
  \label{eq:DF2-LR1-model-indep}
  \Delta C_{\rm LR,1}^{ij}(\muEW) & = 
  v^2 \frac{\wc{Hd}{ij}}{\lambda_t^{ij}} x_t
  \left\{ \ln \frac{\muNP}{\muEW} + H_1(x_t,\muEW) \right\} ,
\\[0.2cm] \nonumber
  \Delta C_{\rm VLL}^{ij} (\muEW) 
  = \frac{v^2}{\lambda_t^{ij}} x_t & \left\{
    [\Wc[(1)]{Hq} - \Wc[(3)]{Hq}]_{ij} \ln\frac{\muNP}{\muEW}
  + \wc[(1)]{Hq}{ij} H_1(x_t,\muEW) - \wc[(3)]{Hq}{ij} H_2(x_t,\muEW)
  \right.
\\ & \left.
  \label{eq:DF2-VLL-model-indep} 
  + \frac{2 S_0(x_t)}{x_t} \sum_m \left( \lambda_t^{im} \wc[(3)]{Hq}{mj} 
               + \wc[(3)]{Hq}{im} \lambda_t^{mj} \right) \right\}\,,
\end{align}
where we have omitted the argument $\muNP$ for all $\Wc{\psi^2 H^2 D}$
for brevity.

There are two gauge-independent functions\footnote{The function $H_1$ is
also present in different context in \cite{Aebischer:2015fzz}, Eq.(4.3).}
that depend on $\muEW$:
\begin{align}
  H_1(x, \muEW) & = 
  \ln\frac{\muEW}{M_W} - \frac{x - 7}{4 (x-1)} 
  - \frac{x^2 - 2 x + 4}{2 (x - 1)^2}\ln x ,
  \label{H1}
\\
  H_2(x, \muEW) & = 
  \ln\frac{\muEW}{M_W} + \frac{7 x - 25}{4 (x-1)} 
  - \frac{x^2 - 14 x + 4}{2 (x - 1)^2}\ln x .
  \label{H2}
\end{align}
The cancellation of the $\muEW$-dependence between the 1stLLA contribution
$\propto \ln(\muNP/\muEW)$ and the NLO functions $H_{1,2}(x_t,\muEW)$ can be easily
seen in \refeq{eq:DF2-LR1-model-indep} and \refeq{eq:DF2-VLL-model-indep}. 
However, we keep $\muEW$ on the l.h.s in (\ref{eq:DF2-LR1-model-indep}) and
(\ref{eq:DF2-VLL-model-indep}) to indicate the scale for the RG QCD evolution 
in the $\Delta F=2$--EFTs down to low energies. In this case the cancellation of
$\muEW$ dependence involves QCD effects and has been known \cite{Buchalla:1995vs}.

For convenience we give the composition of $H_{1,2}$ in
terms of the gauge-dependent Inami-Lim function $C(x, \xi_W)$ \cite{Inami:1980fz}
and some remainder functions $\widetilde{H}_{1,2}$ due to diagrams 
Figs. \ref{fig:psi2H2D-DF2-me-1PI-WG} -- \ref{fig:psi2H2D-DF2-me-box}:
\begin{align}
  H_i(x, \muEW) & =
  \frac{a_i}{x} C(x, \xi_W) + \widetilde{H}_i(x, \muEW, \xi_W),
\end{align}
with $a_1 = -8$ and $a_2 = 8$. The $\widetilde{H}_{1,2}$ can be easily calculated
from $H_{1,2}$ and the knowledge of $C(x,\xi_W)$ for every choice of $\xi_W$,
in particular
\begin{align}
  \label{eq:C-Inami-Lim}
  C(x, 1) & 
  = \frac{x}{8} \left(\frac{x-6}{x-1}+\frac{3x+2}{(x-1)^2} \;\ln x\right) ,
\end{align} 
with $C(x_t, 1) \approx 0.78$ for $x_t \approx 4$.

%
%
\subsection{Numerical Impact of NLO Contributions
  \label{sec:NLO-numeric}
}

\begin{figure}
\centering
  \includegraphics[width=0.48\textwidth]{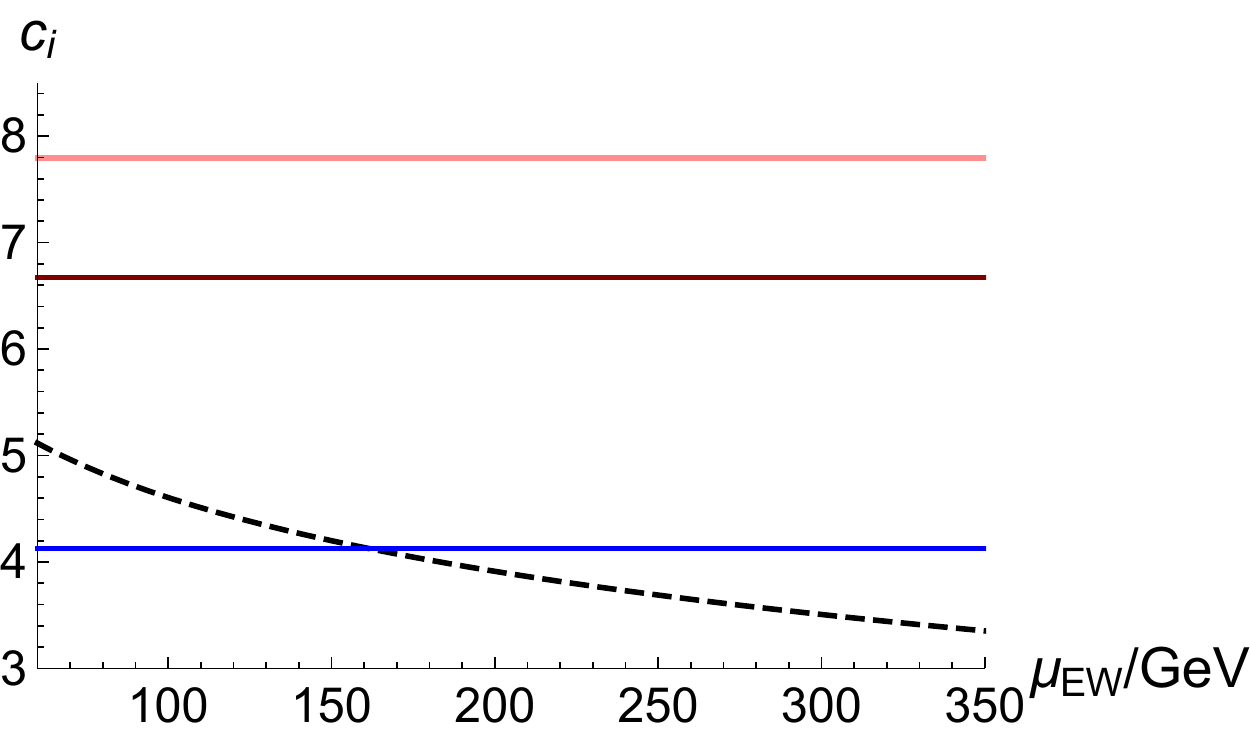}
  \includegraphics[width=0.48\textwidth]{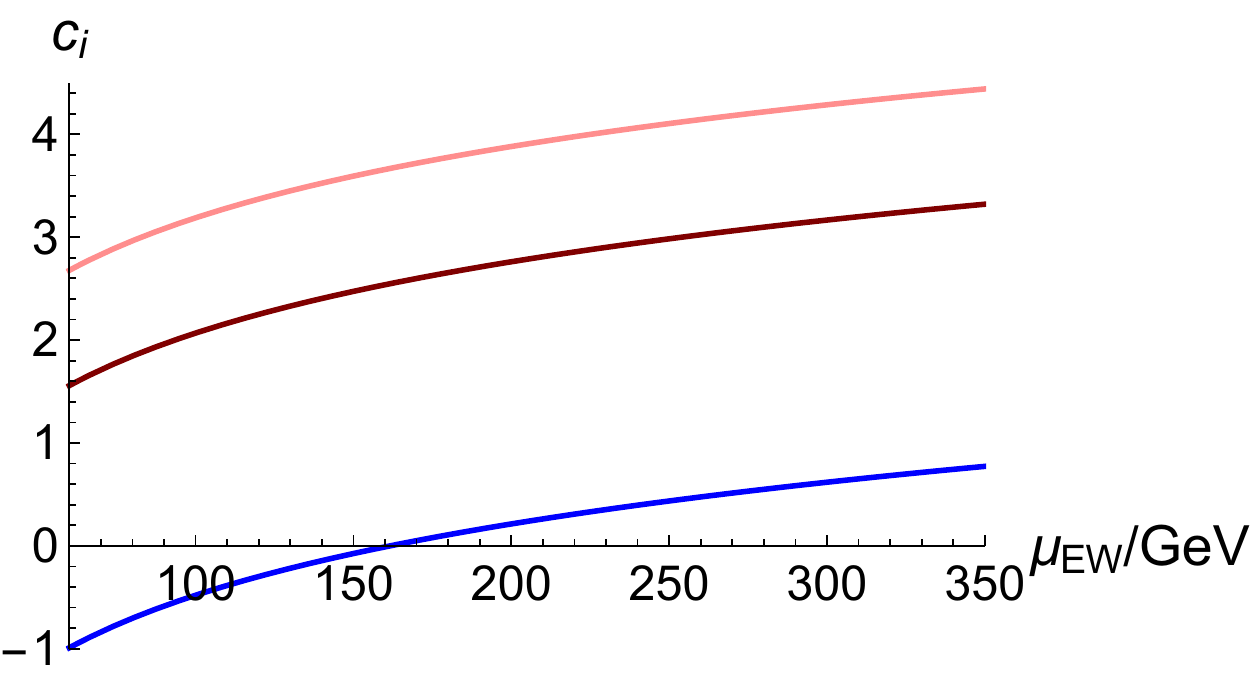}
\caption{\label{fig:scaledependence} \small 
  The $\muEW$-dependence of the coefficients ($c_i$) of $\Wc{Hd}$ in 
  $\Delta C_{\rm LR,1}(\muEW)$, as well $\Wc[(1)]{Hq}$ and $-\Wc[(3)]{Hq}$ in 
  $\Delta C_{\rm VLL}(\muEW)$ for LO and NLO, with $\muNP=10$~TeV.  The black 
  dashed curve in the left figure is the LO contribution, which is
  equal for all three coefficients. In the left (right) figure, the 
  blue line is the LO+NLO (NLO) contribution of $\Wc{Hd}{}$ in 
  $\Delta C_{\rm LR,1}$ as well as $\Wc[(1)]{Hq}{}$ in $\Delta C_{\rm VLL}$, 
  both flavour universal. The dark red line corresponds to LO+NLO (NLO)
  contribution of $\wc[(3)]{Hq}{ij}$ in $\Delta C_{\rm VLL}$ for $i=b$,
  $j=d,s$, while the light red line is for $ij=sd$. Note that the top-quark mass
  is kept fixed in this plot.}
\end{figure}

Fig.~\ref{fig:scaledependence} illustrates the cancellation of the scale
dependence present at LO: shown are the coefficients ($c_i$) of
$\mathcal C_{Hd}$ in $C_{\rm LR,1}(\muEW)$, as well $\mathcal C_{Hq}^{(1)}$ and
$\mathcal-C_{Hq}^{(3)}$ in $\Delta C_{\rm VLL}$ for LO and NLO. The absolute
size of the NLO contribution $H_1(x_t,\muEW)$ is close to zero for $\muEW=m_t$,
but cancels the logarithm which induces otherwise an uncertainty of about
$\sim 10\%$ related to the choice of this scale. The NLO correction involving
$H_2(x_t,\muEW)$ is sizable everywhere. At the order we are working, 
the renormalization scheme of the top quark mass is not specified as far as the
SMEFT contributions are concerned and hence in principle every scheme can be used. 
Here we use the value in the $\overline{\rm MS}$ scheme that enters also the SM 
contribution. The leading $\mu$ dependence of this top-quark mass is governed by 
QCD corrections that will be cancelled by the inclusion of NLO-QCD corrections 
to the SMEFT-contribution at the 2-loop level, being of higher order in this
context. Therefore we keep the scale of the top-quark mass fixed in 
\reffig{fig:scaledependence}. However, we include NLO QCD corrections 
to the SM contribution in the numerical evaluations.

The relative size of the NLO corrections w.r.t. the 1stLLA depends not only on
the scale $\muEW$ at which the contributions are evaluated, but also on the high
scale $\muNP$. For example from \refeq{eq:DF2-LR1-model-indep} follows with
$x_t \approx 4$
\begin{align}
  \label{eq:DF2-LR1-numeric}
  \Delta C_{\rm LR,1}^{ij}(\muEW) &
  =  v^2 \frac{\wc{Hd}{ij} (\muNP)}{\lambda_t^{ij}} x_t
  \left[ \left\{  \begin{array}{cl}
  2.5 & \mbox{for} \; \muNP=1\, \mbox{TeV} \\[1mm]
  4.8 & \mbox{for} \; \muNP=10\,\mbox{TeV}
  \end{array} \right\}
  \;\; - \;\; 0.7  \right] ,
\end{align}
where the values in braces correspond to the $\ln(\muNP/M_W)$ for the two
choices of $\muNP$ and the NLO contribution due to $H_1(x_t, M_W) = -0.7$
constitutes a destructive relative correction of 28\% to 15\% for $\muNP$ in the
range of 1~TeV to 10~TeV. The same holds for $\Delta C_{\rm VLL}^{ij}(\muEW)$
when generated by $\Wc[(1)]{Hq}$.

On the other hand, for $\Wc[(3)]{Hq}$ one part of the NLO matching contributions is
\begin{align}
  \ln \frac{\muNP}{M_W} + H_2(x_t, M_W) &
  = \left\{ \begin{array}{cl}
  2.5 & \mbox{for} \; \muNP=1\, \mbox{TeV} \\[1mm]
  4.8 & \mbox{for} \; \muNP=10\,\mbox{TeV}
  \end{array} \right\}
  \;\; + \;\; 3.0 \,,
\end{align}
that is, of order 100\% and constructive to the 1stLLA term. The second part due
to the box contribution proportional to 
\begin{align}
  \label{eq:DF2-S0-numeric-contr}
  \frac{2 S_0(x_t)}{x_t} & \quad \to \quad 1.1 
\end{align}
depends on $i$ and $j$ and an additional CKM suppression might occur. 
It is instructive to list the Cabibbo suppression, $\lambda_C \approx 0.2$, 
of each term in the sum of \refeq{eq:DF2-VLL-model-indep} for the three 
cases $ij = bs, bd, sd$
\begin{align}
  \nonumber
  ij & = bs : &
  {\cal O}(\lambda_C^2) & \cdot \wc[(3)]{Hq}{bb,\,ss}, &
  {\cal O}(1) & \cdot \wc[(3)]{Hq}{bs}, &
  {\cal O}(\lambda_C^3) & \cdot \wc[(3)]{Hq}{sd}, &
  {\cal O}(\lambda_C^5) & \cdot \wc[(3)]{Hq}{bd}; &
\\
  ij & = bd : &
  {\cal O}(\lambda_C^3) & \cdot \wc[(3)]{Hq}{bb,\,dd}, &
  {\cal O}(1) & \cdot \wc[(3)]{Hq}{bd}, &
  {\cal O}(\lambda_C^2) & \cdot \wc[(3)]{Hq}{sd}, &
  {\cal O}(\lambda_C^5) & \cdot \wc[(3)]{Hq}{bs}; &
\\ \nonumber
  ij & = sd : & 
  {\cal O}(\lambda_C^5) & \cdot \wc[(3)]{Hq}{ss,\,dd}, &
  {\cal O}(\lambda_C^2) & \cdot \wc[(3)]{Hq}{bd}, &
  {\cal O}(\lambda_C^3) & \cdot \wc[(3)]{Hq}{bs}, &
  {\cal O}(\lambda_C^4) & \cdot \wc[(3)]{Hq}{sd}.
\end{align}
It can be seen that for $B$-meson mixing, $ij = bd, bs$, there will be one term
$\lambda_t^{bb} \, \wc[(3)]{Hq}{bj} \sim {\cal O}(1) \times \wc[(3)]{Hq}{bj}$
without Cabibbo suppression for the $\wc[(3)]{Hq}{bj}$ itself that mediates the
process in 1stLLA, whereas all other contributions are suppressed by at least 
${\cal O}(\lambda_C^2)$. For Kaon-mixing, $ij=sd$, the largest CKM combination
will be $\lambda_t^{sb} \, \wc[(3)]{Hq}{bd} \sim \lambda_C^2\, \wc[(3)]{Hq}{bd}$ 
with quadratic Cabibbo suppression.  
Although at NLO for a given transition $ij$ the $\wc[(3)]{Hq}{mn}$ 
with $mn \neq ij$ are at least suppressed by ${\cal O}(\lambda_C^2)$ in
\refeq{eq:DF2-VLL-model-indep}, it cannot be excluded that the
Cabibbo suppression can be lifted in the case that some of the $\wc[(3)]{Hq}{mn}$ 
are  very hierarchical too, as already mentioned below \refeq{eq:corr-Bsmix-tcZ}, and 
thus might become even numerically leading contributions. In the remainder of this 
work we will always assume that this is not the case, and hence neglect
all Cabibbo-suppressed contributions, but for the most general situation a
global analysis would be required that puts simultaneous constraints on all
$\Wc[(3)]{Hq}$. The neglected contributions could be relevant for example if 
flavour-diagonal processes put significantly less severe bounds
on $\wc[(3)]{Hq}{kk}$ ($kk ={bb,\,ss,\,dd}$) than flavour-changing processes on the
flavour-off-diagonal couplings $\wc[(3)]{Hq}{ij}$ ($i\neq j$). The numerical effect
on $\Delta F=2$ observables is hard to predict without
the full analysis; however, for example flavour-diagonal contributions 
$\wc[(3)]{Hq}{ss,\, dd}$ in Kaon mixing are suppressed by ${\cal O}(\lambda_C^5)$
and hence much less likely to invalidate our assumption, compared to $B_s$ and $B_d$
mixing where the corresponding contributions are only suppressed by ${\cal O}(\lambda_C^2)$
and ${\cal O}(\lambda_C^3)$, respectively.
Note that these considerations do not affect most of our conclusions;
specifically, none of the plots presented for the right-handed scenario in 
\refsec{sec:RH-scenario} is changed. Concerning the left-handed scenario in 
\refsec{sec:LH-scenario}, the
constraints in \reffig{fig:constraintsL} derived from $\Delta F=2$ could be
changed, but not the ones from $\Delta F=1$ processes and similarly  for
\reffig{fig:constraintsLLONLO}.

%
%
\subsection{NLO Contributions in VLQ Models
  \label{sec:VLQ}
}

In this section we would like to illustrate the model dependence of NLO contributions 
in LH and RH scenarios in the context of vector-like quark (VLQ) models.
To this end we use the results for the coefficients $\Wc[(1,3)]{Hq}$ and $\Wc{Hd}$
evaluated in VLQ models \cite{delAguila:2000rc, Bobeth:2016llm} of one singlet,
one doublet and two triplets:
\begin{equation}
\begin{aligned}
  D(1, -1/3), & & \qquad
  Q_d(2, -5/6), & & \qquad
  T_d(3, -1/3), & & \qquad 
  T_u(3, +2/3),
\end{aligned}
\end{equation}
where the transformation properties are indicated as $(\SUtwoL, \UoneY)$ and
all VLQs are triplets under $\SUthreeC$. They interact with SM quarks ($q_L,\,
u_R,\, d_R$) and the Higgs doublet via Yukawa interactions
\begin{equation}
  \label{eq:Yuk:H}
\begin{aligned}
  - {\cal L}_{\rm Yuk}(H) & =  
  \left( \lambda_i^D \, H^\dagger \overline{D}_R 
       + \lambda_i^{T_d} \, H^\dagger \overline{T}_{dR}
       + \lambda_i^{T_u} \, \widetilde{H}^\dagger \overline{T}_{uR} \right) q_L^i
  + \bar{d}_R^i  \lambda_i^{Q_d} \, \widetilde{H}^\dagger Q_{dL} 
  + \mbox{h.c.} \, .
\end{aligned}
\end{equation}
The complex-valued Yukawa couplings $\lambda_i^{\rm VLQ}$ give rise to mixing
with the SM quarks and consequently to FC quark couplings of $Z$. The Wilson
coefficients $\Wc[(1,3)]{Hq}$ and $\Wc{Hd}$ are given in terms of Yukawa couplings 
$\lambda_i^{\rm VLQ}$ and the VLQ mass $M_{\rm VLQ}$. The Wilson coefficients are
\cite{delAguila:2000rc, Bobeth:2016llm}
\begin{equation}
  \label{eq:GSM:matching:SMEFT:psi2H2D}
\begin{aligned}
  D : &&
  \wc[(1)]{Hq}{ij}  = \wc[(3)]{Hq}{ij} &
  = - \frac{1}{4} \frac{\lambda_i^\ast \lambda_j}{M^2} , &
  \quad
  Q_d : &&
  \wc{Hd}{ij} &
  = - \frac{1}{2} \frac{\lambda_i^{} \lambda_j^\ast}{M^2} ,
\\
  T_d : &&
  \wc[(1)]{Hq}{ij}  = - 3\, \wc[(3)]{Hq}{ij} &
  = - \frac{3}{8} \frac{\lambda_i^\ast \lambda_j}{M^2} , &
  \quad
  T_u : &&
  \wc[(1)]{Hq}{ij}  = 3\, \wc[(3)]{Hq}{ij} &
  = + \frac{3}{8} \frac{\lambda_i^\ast \lambda_j}{M^2} ,
\end{aligned}
\end{equation}
at the high scale $\muNP \approx M_{\rm VLQ}$.

In the following we use these Wilson coefficients in \refeq{eq:DF2-VLL-model-indep}
in order to demonstrate the size of NLO corrections in specific models.
In the $\mbox{VLQ}=D$ one finds that the 1stLLA is vanishing such that 
the whole effect is first generated at NLO and lacks the enhancement by the
large logarithm:
\begin{equation}
\begin{aligned}
  \Delta C_{\rm VLL}^{ij} &
  = -\frac{x_t}{4} \frac{\lambda_i^* \lambda_j}{\lambda_t^{ij}} \frac{v^2}{M^2}
   \left[ H_1(x_t, M_W) - H_2(x_t, M_W) + \frac{2 S_0(x_t)}{x_t} + \ldots \right] 
\\ &
  = -\frac{x_t}{4} \frac{\lambda_i^* \lambda_j}{\lambda_t^{ij}} \frac{v^2}{M^2}
    \left[ -0.7 - 3.0 + 1.1 + \ldots \right]
  \approx -\frac{x_t}{4} \frac{\lambda_i^* \lambda_j}{\lambda_t^{ij}} \frac{v^2}{M^2}
    \times (-2.6 + \ldots)\,,
\end{aligned}
\end{equation}
where we have assumed that one of the indices $i,j = b$, see comments below
\refeq{eq:DF2-S0-numeric-contr}. For example in scenario $\mbox{VLQ} = T_d$ 
one finds (using $\muEW = M_W$)
\begin{equation}
\begin{aligned}
  \Delta C_{\rm VLL}^{ij} &
  = -\frac{3 x_t}{8} \frac{\lambda_i^* \lambda_j}{\lambda_t^{ij}} \frac{v^2}{M^2}
   \left[ \frac{4}{3} \ln\frac{\muNP}{M_W} + H_1(x_t, M_W)
        + \frac{H_2(x_t, M_W)}{3} - \frac{2 S_0(x_t)}{3 x_t} + \ldots \right] 
\\ &
  \approx -\frac{3 x_t}{8} \frac{\lambda_i^* \lambda_j}{\lambda_t^{ij}} \frac{v^2}{M^2}
  \left[  \left\{ \begin{array}{cl}
  3.3 & \mbox{for} \; \muNP=1\, \mbox{TeV} \\[1mm]
  6.4 & \mbox{for} \; \muNP=10\,\mbox{TeV}
  \end{array} \right\}
  \;\; - \;\; 0.1 + \ldots \right]
\end{aligned}
\end{equation}
and analogously for $\mbox{VLQ} = T_u$
\begin{equation}
\begin{aligned}
  \Delta C_{\rm VLL}^{ij} &
  = \frac{3 x_t}{8} \frac{\lambda_i^* \lambda_j}{\lambda_t^{ij}} \frac{v^2}{M^2}
   \left[ \frac{2}{3} \ln\frac{\muNP}{M_W} + H_1(x_t, M_W)
        - \frac{H_2(x_t,M_W)}{3} + \frac{2 S_0(x_t)}{3 x_t} + \ldots \right] 
\\ &
  \approx \frac{3 x_t}{8} \frac{\lambda_i^* \lambda_j}{\lambda_t^{ij}} \frac{v^2}{M^2}
  \left[  \left\{ \begin{array}{cl}
  1.7 & \mbox{for} \; \muNP=1\, \mbox{TeV} \\[1mm]
  3.2 & \mbox{for} \; \muNP=10\,\mbox{TeV}
  \end{array} \right\}
  \;\; - \;\; 1.3 + \ldots \right] .
\end{aligned}
\end{equation}
These results show that depending on the relative size of $\Wc[(1)]{Hq}$ w.r.t.
$\Wc[(3)]{Hq}$, NLO corrections can cancel or be comparable to the 1stLLA 
contributions. Moreover, the comparison of $D$ with $T_{u,d}$ shows that 
the NLO corrections by themselves indeed can be relevant even if the 
1stLLA contribution cancels in models with $\Wc[(1)]{Hq} - \Wc[(3)]{Hq} = 0$.%
 \footnote{Note that in full generality such a relation holds only at a specific
 scale, here $\muNP$, but self-mixing is a loop-suppressed correction in
 this context.}

For the RH scenario $Q_d$ there is only one coefficient $\Wc{Hd}$ such that the effect 
has been already discussed in \refeq{eq:DF2-LR1-numeric}.

%
%
%
\section{
  Comparison with the Literature
  \label{sec:COMPL}
}

This section is devoted to the comparison of the SMEFT approach to FC quark 
couplings of the $Z$ with previous studies of this NP scenario in the context of
rare Kaon and $B$-meson $\Delta F=1,2$ processes \cite{Buras:2012jb, 
Buras:2015yca, Buras:2015jaq, Endo:2016tnu}. We will focus in particular
on the parameterization of these effects given in \refeq{eq:Zcouplings}, 
to which we will refer in the following as ``simplified models''.

The SMEFT is well defined by the requirement that the low-energy field content
corresponds to the one of the SM, the imposition of the SM gauge group $\GSM$
and the presence of a mass gap between the electroweak scale and the new
dynamics $\muEW \ll \muNP$. Further, the RG equations yield the relations
between Wilson coefficients at both scales. In comparison, the simplified models
lack quantum-field-theoretical principles and constitute simply a postulation of
new FC quark couplings of the $Z$.  They leave questions open regarding for
instance the appropriate scale for the couplings and the implementation of gauge
invariance under the SM gauge group.  As a consequence their application beyond
tree-level seems problematic and one should assume the couplings
$\Delta^\psi_\chi$ ($\psi = u, d$ and $\chi = L,R$) in \refeq{eq:Zcouplings} to
be at $\muEW$.

\begin{figure}
  \centering
  \begin{subfigure}[t]{0.24\textwidth}
    \centering
    \includegraphics[width=0.9\textwidth]{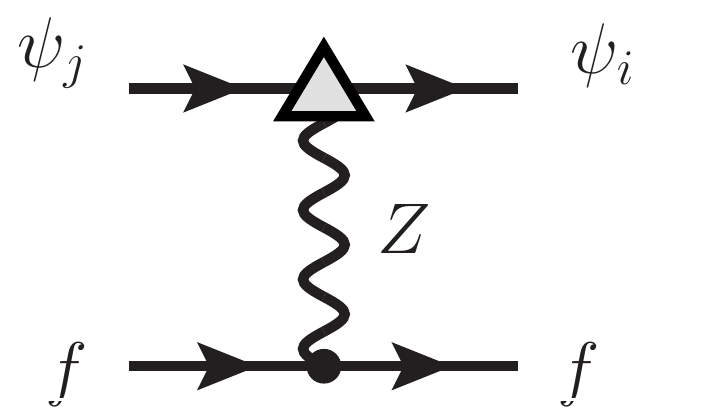}
    \caption{}
    \label{fig:EFT-DF1-Z-ff}
  \end{subfigure}
  \begin{subfigure}[t]{0.24\textwidth}
    \centering
    \includegraphics[width=0.9\textwidth]{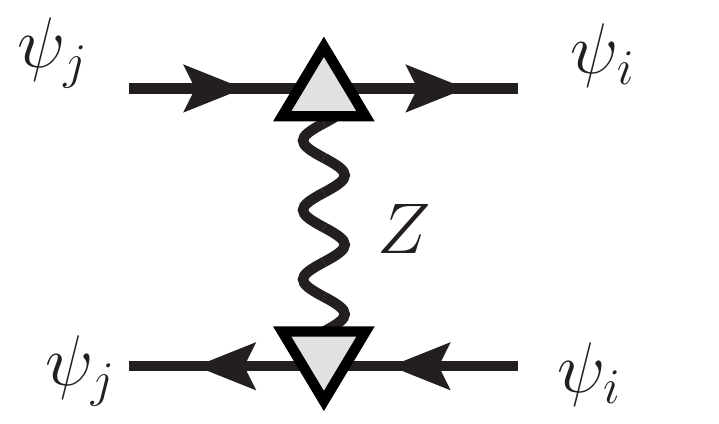}
    \caption{}
    \label{fig:EFT-DF2-Z}
  \end{subfigure}
\caption{\small  
  Tree-level mediated $\Delta F=1$ diagram (\ref{fig:EFT-DF1-Z-ff}) and 
  $\Delta F=2$ diagram (\ref{fig:EFT-DF2-Z}) processes as used in simplified
  models. Insertions of FC quark couplings of $Z$ of simplified models are
  depicted by the triangle and SM couplings by small dots.
}
  \label{fig:EFT-Z}
\end{figure}

Consider the example of $\Delta F=1$ decays $\psi_j \to \psi_i f\bar{f}$ (with
$\psi = u,d$ and $f=\nu, \ell, q$), mediated by the tree-level $Z$-exchange
depicted in \reffig{fig:EFT-DF1-Z-ff}. The FC coupling is due to either dim-6
SMEFT operators or $\Delta_\chi^\psi$ couplings in simplified models, whereas
the other coupling is the SM gauge coupling $Zf\bar{f} \propto g_Z$ that derives
from gauge invariance of the dim-4 SM Lagrangian under $\GSM$. In the SMEFT
case, we have neglected double insertions of dim-6 operators. We have also
approximated the flavour-diagonal $Z$ coupling in the simplified model by its SM
value. At this level, the results obtained in previous studies of $\Delta F=1$
transitions based on simplified models can be translated into constraints on the
SMEFT Wilson coefficients using \refeq{eq:Z-Deltas:dim-6-WC}.

$\Delta F=2$ processes provide important complementary constraints w.r.t
$\Delta F=1$, and depending on the presence of new LH and/or RH interactions the
correlations might change. In simplified models they are mediated via a
double-insertion, see \reffig{fig:EFT-DF2-Z}, and the amplitude scales as
$ (\Delta^\psi_{\chi})^2$. In SMEFT one has in general local
$(\Delta F=2)$-$\psi^4$ operators~\refeq{eq:SMEFT-DF2-NLO-me} with potential NLO
corrections from other classes of operators. In the special case of FC quark
couplings of the $Z$ these $\psi^4$ operators are generated via Yukawa-enhanced
RG evolution from $\psi^2 H^2 D$ operators -- see \refeq{eq:1stLLA-LR1} and
\refeq{eq:1stLLA-VLL} -- and the $\Delta F=2$ amplitude scales as
$\Wc{\psi^2H^2D}$ and hence linearly in $\Delta^\psi_\chi$, see
\refeq{eq:Z-Deltas:dim-6-WC}. This linear dependence remains at NLO in SMEFT.

The quadratic dependence of $\Delta F=2$ amplitudes on $\Delta_\chi^\psi$
present in the results of simplified models in the literature is absent in SMEFT
at the level of single dim-6 operator insertions and arises in SMEFT when going
to the dim-8 level by inserting two dim-6 operators. The dim-8 contributions to
$\Delta F=2$ processes will become more important in SMEFT for smaller scales
$\muEW \lesssim \muNP$. In fact, since the dim-6 Yukawa-generated contributions
are one-loop suppressed one might ask at which scale the dim-8 contributions
start to have similar impact. Equating the naive dimensional scalings of dim-6
one-loop suppressed contributions, $v^2/\muNP^2 (4\pi)^{-2}$, with the ones of
dim-8 contributions, $v^4/\muNP^4$, yields a transition regime
$\muNP \sim 4 \pi v \approx 3$~TeV.

Lets consider first the SMEFT with $\muNP \gtrsim 4 \pi v$ and continue our
comparison with the simplified model. In this case, both approaches generate in
general different operators in the $\Delta F=2$--EFT \refeq{eq:DF2-hamiltonian}
below $\muEW$, as listed in \reftab{tab:DF2-generated-ops}.  A major difference
occurs here for RH interactions, where simplified models generate $O_{\rm VRR}$
and SMEFT $O_{\rm LR,1}$. The latter has a large enhancement under RG evolution
in QCD below $\muEW$ and chirally enhanced matrix elements compared to
$O_{\rm VRR}$, such that phenomenology completely changes, especially in the
Kaon sector -- see \refsec{sec:numeric-M12}.  But also for LH interactions
closer inspection shows that the involved couplings are the ones of up-type
quarks, $\Delta^u_L$, if one uses \refeq{eq:Z-Deltas:dim-6-WC} to relate SMEFT
Wilson coefficients to $\Delta^\psi_\chi$ couplings in the 1stLLA result
\refeq{eq:1stLLA-VLL}. Indeed, while in the simplified approach the LH-$Z$
couplings involved are just $\propto (\Wc[(1)]{Hq} + \Wc[(3)]{Hq})$, the leading
RG contribution in question is $\propto (\Wc[(1)]{Hq} - \Wc[(3)]{Hq})$ and
therefore proportional to up-quark couplings rather than down-quark couplings as
seen in \refeq{eq:Z-Deltas:dim-6-WC}.

\begin{table}
  \centering
\renewcommand{\arraystretch}{1.5}
\begin{tabular}{|c||c|c|}
\hline
& Simplified models
& SMEFT (dim-6)
\\
\hline\hline
  $\Delta F=2$--amplitude
& $\sim (\Delta_\chi^\psi)^2$
& $\sim \Delta_\chi^\psi$
\\
\hline
  LH
& $O_{\rm VLL} \sim (\Delta^d_L)^2$
& $O_{\rm VLL} \sim \Delta^u_L$
\\
\hline
  RH
& $O_{\rm VRR} \sim (\Delta^d_R)^2$ 
& $O_{\rm LR,1} \sim \Delta^d_R$
\\
\hline
  \multirow{2}{*}{LH$+$RH}
& $O_{\rm VLL} \sim (\Delta^d_L)^2$, $O_{\rm VRR} \sim (\Delta^d_R)^2$,  
& $O_{\rm VLL} \sim \Delta^u_L$
\\
& $O_{\rm LR,1} \sim (\Delta^d_L \Delta^d_R)$
& $O_{\rm LR,1} \sim \Delta^d_R$
\\
\hline
\end{tabular}
\renewcommand{\arraystretch}{1.0}
\caption{\small Comparison of simplified models and SMEFT (assuming $\muNP \gtrsim 
  4 \pi v$) for down-type quark $\Delta F=2$ processes. Scaling of the $\Delta F=2$
  amplitude in terms of couplings $\Delta^\psi_\chi$ ($\psi = u, d$ and $\chi = L,R$)
  from \refeq{eq:Zcouplings}. Operators generated in $\Delta F=2$--EFT below $\muEW$
  by LH, RH or LH$+$RH scenarios and their scaling with $\Delta^\psi_\chi$ (in 1stLLA
  for SMEFT, i.e. neglecting NLO corrections).
}
\label{tab:DF2-generated-ops}
\end{table}

If one considers the SMEFT with $\muNP \lesssim 4 \pi v$, double insertions of
dim-6 operators are expected to be of similar size or even dominate over loop
contributions with one dim-6 insertion. The quadratic dependence on
$\Delta_\chi^\psi$ via the dim-8 contributions is then present in the amplitude,
resembling the simplified model approach.

After having established the conditions for a correspondence of simplified
models and SMEFT, we will next compare our results with the ones in
\cite{Endo:2016tnu} on the issue of gauge dependence and renormalization 
scale dependence stressed in points 1. and 2. of \refsec{sec:intro}. 
These authors calculated first
the contributions in \reffig{fig:psi2H2D-DF2-me-1HPR} in simplified models for RH
and LH scenarios. Adding pure NP contributions from tree-level
exchange considered in \cite{Buras:2012jb} one finds then
\begin{align}
  \label{eq:HRR}
  {\cal N}_{ij} C_{\rm VRR}^{ij} & 
  = \frac{[\Delta_R^d]_{ij} [\Delta_R^d]_{ij}}{2 M_Z^2} , &
  {\cal N}_{ij} C_{\rm LR,1}^{ij} & 
  = \frac{[\Delta_L^d({\rm SM})]_{ij} [\Delta_R^d]_{ij}}{M_Z^2} ,  
\end{align}
\begin{align}
  \label{eq:HLL}
  {\cal N}_{ij} C_{\rm VLL}^{ij} & 
  = \frac{[\Delta_L^d]_{ij} [\Delta_L^d]_{ij}}{2 M_Z^2}
  + \frac{[\Delta_L^d({\rm SM})]_{ij} [\Delta_L^d]_{ij}}{M_Z^2} ,
\end{align}
for RH and LH scenarios, respectively. Here the FC $d_j \to d_i Z$ vertex of
the SM arises from the lower part of \reffig{fig:psi2H2D-DF2-me-1HPR},
\begin{align}
  \label{eq:DELTASM}
  [\Delta_L^d({\rm SM})]_{ij} & 
  = \lambda_t^{ij}\frac{g_2^3}{8\pi^ 2\cos\theta_W} C(x_t, \xi_W)\,,
\end{align}
where $\theta_W$ denotes the weak mixing angle and $C(x_t, \xi_W=1)$, given in
\refeq{eq:C-Inami-Lim}, is gauge-dependent.\footnote{Note that
  $[\Delta_L^d({\rm SM})]_{ij} \sim + g_2^3 C(x_t)$ in \refeq{eq:DELTASM}
  corresponds to the definition of the covariant derivative in
  \refeq{eq:def-cov-derivative}.} In the second version of their 
paper they included the diagrams (\ref{fig:psi2H2D-DF2-me-1PI-WG}) and 
(\ref{fig:psi2H2D-DF2-me-1PI-GG}) obtaining a gauge independent result for 
the NLO contributions to the coefficients $\Wc{Hd}$ and  $\Wc[(1)]{Hq}$,
which agrees with ours, but $\Wc[(3)]{Hq}$ has not been considered there. 
Even prior to the second version of this paper we have suggested how their 
original results could be corrected. Indeed using the relations in
\refeq{eq:Z-Deltas:dim-6-WC} we can cast our results into the ones of
\cite{Endo:2016tnu}. We find in the RH scenario the following replacement in
\refeq{eq:HRR}:
\begin{align}
  \label{eq:Deltaefftot}
  [\Delta_L^d({\rm SM})]_{ij} & 
  \;\;\; \rightarrow \;\;\;
  \frac{\lambda_t^{ij}\, g_2^3}{8\pi^ 2\cos\theta_W}
  \left[ C(x_t, \xi_W) - \frac{x_t}{8} \left(\ln\frac{\muNP}{\muEW}
                       + \widetilde{H}_1(x_t, \muEW, \xi_W) \right)\right],
\end{align}
with the latter coupling including LO and NLO corrections obtained in SMEFT. 
As discussed before, this combination of $C$ with $\widetilde{H}_1$ is 
gauge-independent and $\muEW$-independent. The numerical values
for $\xi_W=1$ and $x_t \approx 4$ are
\begin{align}
  C(x_t, \xi_W) - \frac{x_t}{8} \widetilde{H}_1(x_t, M_W, \xi_W) &
  \; \approx \; 0.78 \, - \,  0.41 \, ,
\end{align}
suggesting that a gauge-independent result would have been at least about a
factor of two smaller than the one used in the original version of 
\cite{Endo:2016tnu}. Further, the
logarithm in \refeq{eq:Deltaefftot} will typically dominate for reasonable
values of $\muNP$, flipping the sign of $[\Delta_L^d({\rm SM})]_{ij}$ compared
to~\cite{Endo:2016tnu}.  Although our comparison suggests that one can correct
the gauge dependence in the simplified model by the replacement
\refeq{eq:Deltaefftot}, conceptually this does not seem meaningful.

Thus the issue of gauge dependence of the NLO correction has been resolved.
Unfortunately, the present result in~\cite{Endo:2016tnu}, although gauge
independent, exhibits a very large renormalization scale dependence, simply
because the authors decided not to include the \emph{dominant} LO (1stLLA)
top-Yukawa RG effects above the electroweak scale, given in \refeq{eq:Deltaefftot} 
by the term $\propto \ln \muNP/\muEW$. Furthermore, the renormalization scale
$\mu$ in their paper, equivalent to the matching scale $\muEW$ in our paper,
has been set to $1\tev$. In this case the NLO correction by itself will be of
similar size and have the same sign as the (LO + NLO) result in 
\refeq{eq:Deltaefftot}. However, the choice $\muEW = 1$~TeV is clearly not
allowed, because at this matching scale one cannot integrate out $Z$,
$W$ and the top-quark, which have masses one order of magnitude smaller.
Moreover, neglecting the $\propto \ln \muNP/\muEW$ is not allowed from
the point of view of SMEFT, because it would imply that dimension
six operators are generated at $\muNP = \muEW$.

Indeed, returning to the right plot in \reffig{fig:scaledependence} (blue line 
for $\Wc{Hd}$), one can see that even varying $\muEW$ in an admissible range,
the neglect of RG Yukawa effects above the electroweak scale yields a very
strong $\muEW$ dependence. In particular, for scales close to $m_t$, the NLO
corrections considered in \cite{Endo:2016tnu} vanish. The fact that with a
particular choice of matching scale NLO corrections can often be absorbed
into the leading term is well known in the literature; however, if the leading
term is absent there is a serious problem. Thus a meaningful phenomenology
requires the inclusion of the LO RG effects. Doing so, the renormalization 
scale $\mu$ in \cite{Endo:2016tnu} becomes the scale of new physics, and the
results in that paper would correspond to $\muNP=1\tev$.\footnote{
Meanwhile the 
authors of  \cite{Endo:2016tnu} replaced the scale $\mu$ by $\muNP$ so 
that their final formula agrees with ours. However, we disagree with 
their statement that $\muNP$ comes from the diagram (c) in  
\reffig{fig:psi2H2D-DF2-me} as it is absent in this diagram and
can only come from LO RG effects as explained above.}

Next we would like to mention the analysis in \cite{Feruglio:2016gvd}. RG Yukawa
effects have been studied in a NP scenario in which the only operators with
non-vanishing coefficients at $\muNP$ are the third-generation (in the
interaction basis) semi-leptonic four-fermion operators ${\cal O}_{\ell q}^{(1)}
= (\bar \ell_L \gamma_{\mu} \ell_L) (\bar q_L \gamma^{\mu} q_L)$ and
${\cal O}_{\ell q}^{(3)} = (\bar \ell_L \gamma_{\mu} \sigma^{a} \ell_L) (\bar
q_L \gamma^{\mu} \sigma^{a} q_L)$
of the Warsaw basis \cite{Grzadkowski:2010es}.  This structure has been
motivated by the $B$-physics anomalies and can be met in models with vector
leptoquark mediators. It has been demonstrated that RG Yukawa effects modify
significantly the $Z$ couplings to leptons, ruling out a possible explanation of
the anomalies within this scenario because of the strong constraints from
$Z$-pole observables and lepton-flavour violating $\tau$ decays. In our models
these couplings are absent and are not generated by the RG Yukawa effects from
the operators in \refeq{eq:LH13} and \refeq{eq:RH1} considered by us.

Finally, in \cite{Cirigliano:2016yhc} the impact of RH charged currents on
$\epe$ in correlation with electric dipole moments (EDMs) has been analysed,
considering the operator $\Op{Hud}$ in (\ref{eq:O-Hud}). The phenomenology of
this model is very different from the one of our models as the main mediators
are $W^\pm$ with RH couplings and not the $Z$ boson. In this model contributions
to $\varepsilon_K$ are much smaller than in our models and it appears that it is
harder to explain the $\epe$ anomaly when other constraints, in particular from
EDMs, are taken into account. But the correlation of $\epe$ with EDMs pointed
out in this paper is clearly interesting.

%
%
%
\section{
  Implications for Flavour Observables
  \label{sec:implications}
}

\begin{figure}
  \includegraphics[width=\textwidth]{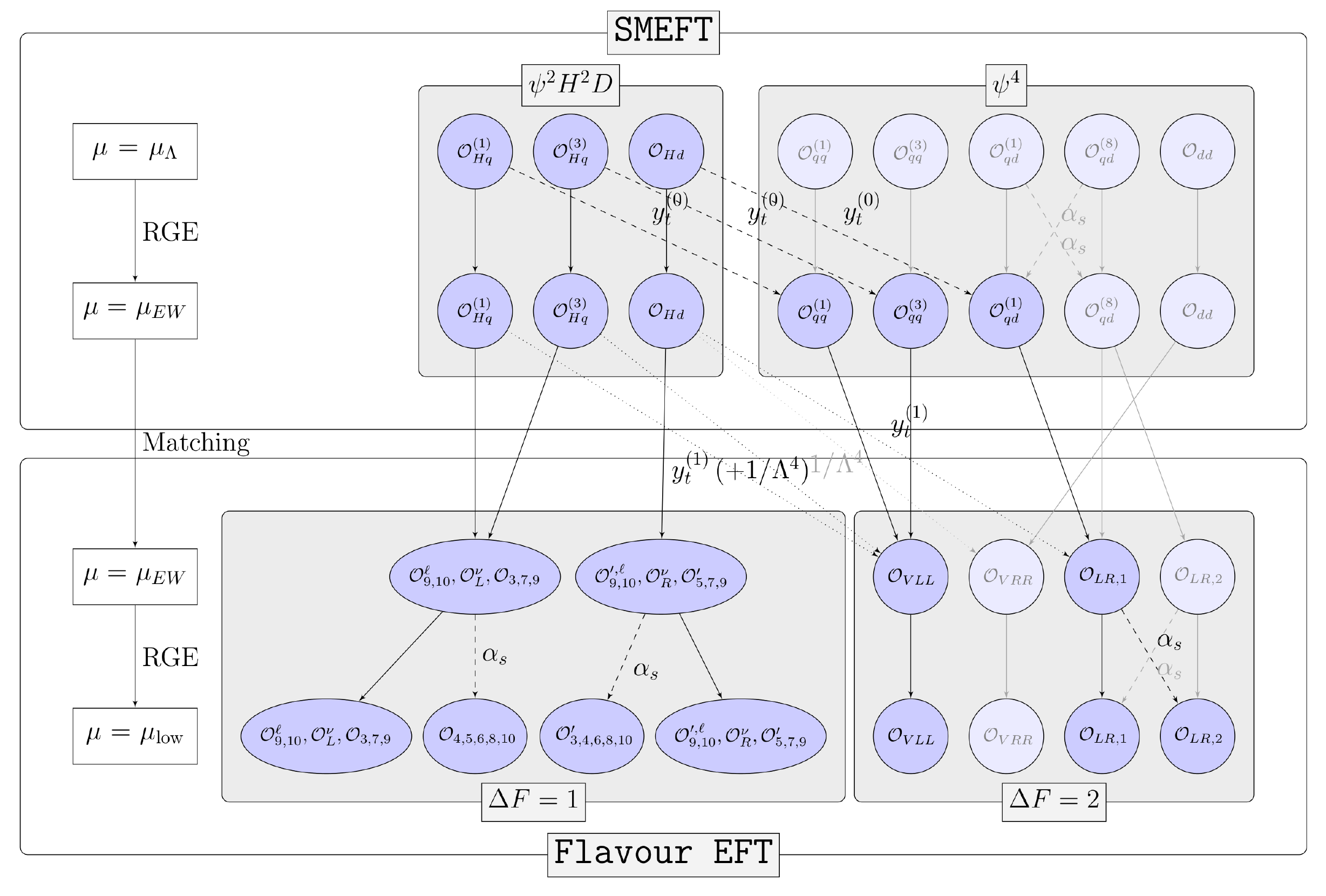}
\caption{\small \label{fig:visualization} 
  Visualization of matching and RG running for the operators under consideration
  in SMEFT and the $\Delta F=1,2$--EFTs. The darker nodes are those that are
  dominant within our approach. Solid lines indicate $\mathcal O(1)$ running/matching
  contributions, dashed lines RG mixing enhanced by a large logarithm and dotted
  lines NLO running/matching contributions.  Contributions via 1stLLA top-Yukawa
  RG mixing are denoted by $y_t^{(0)}$, 1-loop corrections to the matching at
  $\muEW$ by $y_t^{(1)}$. $1/\Lambda^4$ refers to contributions that appear at
  dimension eight, like double insertions of dim-6 operators.
}
\end{figure}

The matching and RG evolution in our setup is schematically shown in
\reffig{fig:visualization}. The darker nodes are included in the following
phenomenological analysis, while the lighter ones are contributions we do not
consider here, but that would appear in a general SMEFT analysis. Solid lines
represent direct matching and running contributions, while dashed lines are the
main contributions created via RG effects, either via Yukawa couplings
($\sim y_t^2$) or via QCD ($\sim \alpha_s$). Note that in some cases these
contributions can result in larger observable effects than the direct ones, due
to RG and chiral enhancement factors, \emph{e.g.} in the case of
$\mathcal O_{\rm LR,1}$. The dotted lines represent NLO matching effects.

The inclusion of new contributions from Yukawa RG evolution and the NLO
corrections calculated here have only direct impact on $\Delta F=2$ observables.
However, in a combined phenomenological analysis of $\Delta F=1,2$ processes
$\Delta F=2$ observables will restrict the available parameter space and hence
affect also predictions for $\Delta F=1$ processes, specifically for 
RH couplings of the $Z$. In the present paper we want to illustrate this impact
mainly in the latter case, as in this scenario the impact is very large.

Concentrating on RH couplings of the $Z$, we will first consider the correlation
between the ratio $\epe$, $\epsK$, $K_L\to\mu^+\mu^-$ and $K\to\pi\nu\bar\nu$ 
decays analyzed in \cite{Buras:2015jaq}, where the contributions in question have
not been taken into account. Subsequently we will consider the impact on the
correlation between $B_{s,d}^0-\bar B_{s,d}^0$ observables and rare
$b\to d,s + (\ell^+\ell^-,\, \nu\bar\nu)$ decays.

The experimental data and hadronic inputs are identical to
\cite{Bobeth:2016llm} with the exception that we include very recent
preliminary data for $B_{d,s}\to\mu^+\mu^-$ \cite{Aaij:2017vad},
combined with previous measurements in \cite{Altmannshofer:2017wqy}.

%
%
\subsection[Numerical impact on $M_{12}$]
{ \boldmath 
  Numerical impact on $M_{12}$
  \label{sec:numeric-M12}
}

The off-diagonal element of the mass matrix of neutral meson mixing including
the full set of $\Delta F=2$ operators \refeq{eq:DF2-hamiltonian} is given by
\begin{align}
  M_{12}^{ij\,\ast} & 
  = \frac{\langle \overline{M}^0 |{\cal H}_{\Delta  F=2}^{ij} | M^0 \rangle}{2 M_{M^0}}
  = \frac{G_F^2 M_W^2}{8\, \pi^2 M_{M^0}} (\lambda_t^{ij})^2
    \sum_{a} C_a^{ij}(\muLow) \; \langle \overline{M}^0 |O_a^{ij} | M^0 \rangle(\muLow)
\end{align}
in terms of Wilson coefficients and hadronic matrix elements of the operators,
$\langle O_a^{ij} \rangle \equiv \langle \overline{M}^0 |O_a^{ij} | M^0 \rangle$,
evaluated at the scale $\muLow$ relevant for the corresponding meson system
$ij = sd,\, bd,\, bs$. The hadronic matrix elements are provided by lattice
collaborations, who for historical reasons relate them usually to bag-factors,
thereby introducing additional dependences on the meson decay constant $F_{M}$
and the chirality-factor
\begin{align}
  r_\chi^{ij} &
  = \left[\frac{M_M}{m_i(\muLow) + m_j(\muLow)}\right]^2
\end{align}
that involves the $\overline{\rm MS}$ quark masses.

The NP contributions of FC quark couplings of the $Z$ in SMEFT require to
consider the operators $a= \rm VLL,\, LR1,\, LR2$, where $C_{\rm LR,2}$ enters
via QCD RG evolution.\footnote{Note that $C_{\rm VRR}$ can be included by
  $C_{\rm VLL}\to C_{\rm VLL} + C_{\rm VRR}$.} For the Kaon system, we adapt the
results of bag factors from RBC-UKQCD \cite{Garron:2016mva}. For the $B_{d,s}$
systems, we use the products of decay constants and bag factors,
$F_{B_j}^2 B_a^{ij}$, from FNAL/MILC \cite{Bazavov:2016nty}. Both
sets of coefficients are tabulated in
\reftab{tab:M12:input}. The relations between bag factors and matrix
elements at $\muLow$ for the choice of operator basis by RBC-UKQCD
\cite{Garron:2016mva} are
\begin{align}
  \langle O_{\rm VLL}^{sd} \rangle &
  = \frac{2}{3} M_K^2 F_K^2 B_1^{sd},
&
  \langle O_a^{sd} \rangle &
  = N_a^{sd} r_\chi^{sd} M_K^2 F_K^2 B_a^{sd}\,,
\end{align}
with $N_a^{sd} = (-1/3,\, 1/2)$ for $a = (\rm LR1,\, LR2)$, and for FNAL/MILC 
\cite{Bazavov:2016nty}
\begin{align}
  \langle O_{\rm VLL}^{bj} \rangle &
  = \frac{2}{3} M_{B_j}^2 F_{B_j}^2 B_1^{bj},
&
  \langle O_a^{bj} \rangle &
  = N_a^{bj} \left(r_\chi^{bj} + d_a^{bj}\right) M_{B_j}^2 F_{B_j}^2 B_a^{bj}\,,  
\end{align}
with $N_a^{bj} = (-1/3,\, 1/2)$ and $d_a^{bj} = (3/2,\, 1/6)$.

\begin{table}
  \centering
\renewcommand{\arraystretch}{1.5}
\begin{tabular}{|c||cc|c|ccc|}
\hline
  $ij$  & $\muLow$ [GeV] & $N_f$ & $r_\chi$ & $B_1^{ij}$ & $B_4^{ij}$ & $B_5^{ij}$
\\
\hline\hline
  $sd$  & $3.0$          & 3     & $30.8$   & 0.525(16)      & 0.920(20)      & 0.707(45)
\\
\hline\hline
  &  &  &  & $F_{B_j}^2 B_1^{ij}$ & $F_{B_j}^2 B_4^{ij}$ & $F_{B_j}^2 B_5^{ij}$
\\
\hline
  $bd$  & $4.18$         & 5     & $1.6$    & 0.0342(30)      & 0.0390(29)    & 0.0361(36)
\\
  $bs$  & $4.18$         & 5     & $1.6$    & 0.0498(32)      & 0.0534(32)    & 0.0493(37)
\\
\hline
\end{tabular}
\renewcommand{\arraystretch}{1.0}
\caption{\small \label{tab:M12:input}
  Scale settings and number of flavours, $N_f$, as well as numerical
  inputs of bag factors entering $M_{12}^{ij}$, see \cite{Bazavov:2016nty}
  and \cite{Garron:2016mva} for correlations. For the Kaon system threshold
  crossings to $N_f=4$ and $N_f = 3$ have been chosen as $4.18$~GeV and $1.4$~GeV.
}
\end{table}

In order to illustrate the RG and chiral enhancement of NP contributions in
$\Delta C_{\rm LR,1}$~\refeq{eq:DF2-LR1-model-indep} compared to
$\Delta C_{\rm VLL}$ \refeq{eq:DF2-VLL-model-indep}, we express in $M_{12}^{ij}$
the $C_i(\muLow)$ in terms of the $C_i(\muEW)$, using QCD RG evolution at NLO
\cite{Buras:2000if, Buras:2001ra}.  We keep the hadronic input unevaluated and
use $\muEW \approx 163$~GeV in order to be able to adapt the SM calculations of
$M_{12}^{sd}$ at NNLO \cite{Buras:1990fn, Brod:2010mj, Brod:2011ty} and of
$M_{12}^{bj}$ at NLO. The semi-numerical result in terms of $C_i(\muEW)$ is
\begin{align}
  \frac{M_{12}^{sd\,\ast}}{{\cal F}_{sd}} &
  = \Big[168.7 + {\rm i}\, 194.1 
  + 0.8\, \Delta C_{\rm VLL}^{sd} \Big] B_1^{sd}
  - \Delta C_{\rm LR,1}^{sd} \left(25.9\, B_4^{sd} + 14.1\, B_5^{sd} \right) ,
\\
  \frac{M_{12}^{bj\,\ast}}{\widetilde{{\cal F}}_{bj}} & =
  \Big[1.95 
  + 0.84\, \Delta C_{\rm VLL}^{bj} \Big] F_{B_j}^2 B_1^{bj}
  - \Delta C_{\rm LR,1}^{bj}\, F_{B_j}^2 \left(1.18\, B_4^{bj} + 1.42\, B_5^{bj} \right) .
\end{align}
The SM contribution is given by the first numbers in brackets $\propto B_1$
and the normalization factors read
\begin{align}
  {\cal F}_{ij} & 
  = \widetilde{{\cal F}}_{ij} F_M^2
  = \frac{G_F^2 M_W^2}{12\, \pi^2} (\lambda_t^{ij})^2 M_M F_M^2 . 
\end{align}
The huge enhancement of $\Delta C_{\rm LR,1}^{sd}$ w.r.t. $\Delta C_{\rm VLL}^{sd}$
in Kaon mixing stems in large part from $r_\chi^{sd}$, whereas the effect is a
factor of three in $B$-meson mixing. The SMEFT contributions 
\refeq{eq:DF2-LR1-model-indep} and \refeq{eq:DF2-VLL-model-indep} can be inserted
into both equations to obtain numerical predictions for $M_{12}^{ij}$. There are
two experimental constraints in each sector on $M_{12}^{ij}$,
\begin{align}
  ij & = sd :
& \Delta M_K    & =       2\, \mbox{Re} \left(M_{12}^{sd}\right), 
& \varepsilon_K & \propto \mbox{Im} \left(M_{12}^{sd}\right) ,
\\
  ij & = bj :
& \Delta M_{B_j} & = 2 \big|M_{12}^{bj}\big|, 
& \phi_j   & = \mbox{Arg} \left(M_{12}^{bj} \right)\,,
\end{align}
where we have assumed that SM QCD penguin pollution and new physics in
$b\to s c\bar{c}$ processes are negligible, see \cite{Jung:2012mp,
DeBruyn:2014oga, Frings:2015eva,Ligeti:2015yma} for recent works.

%
%
\subsection[Semileptonic $\Delta F=1$ Processes]
{ \boldmath 
  Semileptonic $\Delta F=1$ Processes
  \label{sec:DF1-semileptonic} 
}

The $\Delta F=1$ semileptonic processes $d_j\to d_i + (\ell^+\ell^-,\, \nu\bar\nu)$
are highly sensitive to FC quark couplings of the $Z$. The $\psi^2 H^2 D$
operators modify them at tree-level via \reffig{fig:EFT-DF1-Z-ff} in SMEFT. 
The relevant parts of the $\Delta F=1$--EFTs,
\begin{align}
  {\cal H} & =
  -\frac{4 G_F}{\sqrt{2}} \lambda^{ij}_t \frac{\alpha_e}{4 \pi} 
   \sum_a C_a^{ij} O_a^{ij}\,,
\end{align}
involve the six semileptonic operators
\begin{align}
  O_{9(9')}^{ij} & 
  = [\bar{d}_i \gamma_\mu P_{L(R)} d_j] [\bar{\ell} \gamma^\mu \ell] ,
&
  O_{10(10')}^{ij} & 
  = [\bar{d}_i \gamma_\mu P_{L(R)} d_j] [\bar{\ell} \gamma^\mu \gamma_5 \ell] ,
\end{align}
\begin{align}
  O_{L(R)}^{ij} & 
  = [\bar{d}_i \gamma_\mu P_{L(R)} d_j] [\bar{\nu} \gamma^\mu (1-\gamma_5) \nu] .
\end{align}
The Wilson coefficients of the LH operators at $\muEW$ read as follows
\cite{Alonso:2014csa, Buras:2014fpa}:
\begin{align}
  C_9^{ij}     & 
  = \frac{Y(x_t)}{s_W^2} - 4 Z(x_t) 
  - \frac{\pi}{\alpha_e}\frac{v^2}{\lambda_t^{ij}} 
    (1 - 4 s_W^2)\left[\Wc[(1)]{Hq} + \Wc[(3)]{Hq}\right]_{ij} + \ldots,
\\
  C_{10}^{ij}  & 
  = - \frac{Y(x_t)}{s_W^2} 
  + \frac{\pi}{\alpha_e}\frac{v^2}{\lambda_t^{ij}}
    \left[\Wc[(1)]{Hq} + \Wc[(3)]{Hq}\right]_{ij} + \ldots \, ,
\\
  C_L^{ij}     &
  = - \frac{X(x_t)}{s_W^2} 
  + \frac{\pi}{\alpha_e}\frac{v^2}{\lambda_t^{ij}} 
    \left[\Wc[(1)]{Hq} + \Wc[(3)]{Hq}\right]_{ij} + \ldots,
\end{align}
where the SM contributions are given by the gauge-independent functions
$X(x_t)$, $Y(x_t)$ and $Z(x_t)$~\cite{Buchalla:1990qz}. The tree-level
matching of SMEFT gives rise to the dependence on the sum
$\Wc[(1)]{Hq}+\Wc[(3)]{Hq}$ and the dots indicate potential additional
contributions, for instance from $\psi^4$ operators. The chirality-flipped
Wilson coefficients
\begin{align}
  C_{9'}^{ij}  & 
  = - (1 - 4 s_W^2) \frac{\pi}{\alpha_e}\frac{v^2}{\lambda_t^{ij}} \wc{Hd}{ij} + \ldots \, ,
&
  C_{10'}^{ij} & 
  = \frac{\pi}{\alpha_e}\frac{v^2}{\lambda_t^{ij}} \wc{Hd}{ij} + \ldots \, ,
\\ & 
&
  C_R^{ij}     & 
  = \frac{\pi}{\alpha_e}\frac{v^2}{\lambda_t^{ij}} \wc{Hd}{ij} + \ldots \, ,  
\end{align}
depend on $\Wc{Hd}$. $C_{9,9'}$ depend on the same combination of
coefficients $\Wc{\psi^2 H^2 D}$ as $C_{10,10'}$, but with an additional
suppression factor $1-4 s_W^2 \approx 0.08$. There is also a strict relation
$C_{10'}^{ij} = C_R^{ij}$, which holds equivalently for the NP parts of
$C_{10}$ and $C_L$. Therefore all semileptonic $\Delta F=1$ processes depend
on only one left-handed and one right-handed combination of Wilson
coefficients.

Whenever only one of these combinations is present, strong correlations are
present in each sector $ij=sd,bd,bs$. For instance the semileptonic decays
with neutrinos and a pseudoscalar meson in the final state, \emph{e.g.}
$K^+ \to \pi^+ \nu\bar\nu$, $K_L\to \pi^0 \nu\bar\nu$, $B\to K\nu\bar\nu$ and
the leptonic decays $M_{ij} \to \ell^+\ell^-$ ($K_L\to \mu^+\mu^-$,
$B_q\to \ell^+\ell^-$) depend only on $C_{10,10'}$ and $C_{L,R}$,
respectively. The dependence of the corresponding branching ratios on
$\Wc{\psi^2 H^2 D}$ Wilson coefficients reads
\begin{align}\label{eq:BRMPnunu}
  \Br(M_j\to P_i \nu\bar\nu) & 
  \propto \left|- \frac{X(x_t)}{s_W^2} + \frac{\pi}{\alpha_e}\frac{v^2}{\lambda_t^{ij}} 
   \left[ \Wc[(1)]{Hq} + \Wc[(3)]{Hq} + \Wc{Hd} \right]_{ij} \right|^2 ,
\\\label{eq:BRMmumu}
  \Br(M_{ij} \to \ell^+\ell^-) & 
  \propto \left|- \frac{Y(x_t)}{s_W^2} \,+ \frac{\pi}{\alpha_e}\frac{v^2}{\lambda_t^{ij}} 
   \left[ \Wc[(1)]{Hq} + \Wc[(3)]{Hq} - \Wc{Hd} \right]_{ij} \right|^2  \,.
\end{align}
These observables are clearly correlated if only LH or RH couplings are
present, but are independent as soon as both couplings are finite.  The full
set of semileptonic decays $d_j\to d_i + (\ell^+\ell^-,\, \nu\bar\nu)$ includes
further observables that depend also on $C_{9,9'}$, and allow to put further
constraints on these Wilson coefficients.

Additional correlations exist between $\Delta F=1$ and $\Delta F=2$ processes,
at least for the RH scenario, which will be discussed in the next section.

%
%
\subsection[Correlations between $\Delta F=2$ and $\Delta F=1$ Processes]
{\boldmath 
  Correlations between $\Delta F=2$ and $\Delta F=1$ Processes
  \label{sec:4B} 
}

The dependence on $\Wc{Hd}$ of the $\Delta F=2$ contribution
\refeq{eq:DF2-LR1-model-indep} implies a strong correlation of the
aforementioned semileptonic decays with $M_{12}^{ij}$ for RH interactions,
to be discussed below. As already mentioned in \refsec{sec:COMPL}, such a
correlation is not present for NP LH $Z$ couplings. This is due to the
presence of two Wilson coefficients, conveniently written as the combinations
\begin{align}
  \Wc[(\pm)]{Hq} & \equiv\Wc[(1)]{Hq} \pm \Wc[(3)]{Hq}\,,
\end{align}
which appear in different combinations in $\Delta F=1$ and $\Delta F=2$:
$\Delta F=1$ processes depend only on $\Wc[(+)]{Hq}$, whereas
$\Delta F=2$ processes depend on $\Wc[(-)]{Hq}$ when restricting to the 1stLLA
term. As can be seen from \refeq{eq:DF2-VLL-model-indep}, the latter changes
once the NLO corrections are included:
\begin{equation}\label{eq:CVLLnum}
\begin{aligned}
  \Delta C_{\rm VLL}^{ij} (\muEW) 
  = \frac{v^2}{\lambda_t^{ij}} x_t & \Bigg[
    \wc[(-)]{Hq}{ij} \Big( \ln\frac{\muNP}{M_W} + 
    \left\{ \begin{array}{c} 0.6 \\[1mm] 1.2 \end{array} \right\} \Big)
  - \wc[(+)]{Hq}{ij} 
    \left\{ \begin{array}{c} 1.3 \\[1mm] 1.9 \end{array} \right\} \Bigg]
\\ 
  = \frac{v^2}{\lambda_t^{ij}} x_t & \Bigg[
    \wc[(-)]{Hq}{ij} 
  \left\{ \begin{array}{c} 5.4 \\[1mm] 6.0 \end{array} \right\}
  - \wc[(+)]{Hq}{ij} 
    \left\{ \begin{array}{c} 1.3 \\[1mm] 1.9 \end{array} \right\} \Bigg]
  \quad\mbox{for}\quad 
  \left\{ \begin{array}{l} ij = bd, bs \\[1mm] ij = sd \end{array} \right.\,.
\end{aligned}
\end{equation}
Here we have used $\muNP = 10$~TeV and numerical results presented in
\refsec{sec:NLO-numeric} for $H_1(x_t, M_W) = -0.7$, $H_2(x_t, M_W) = +3.0$ and
\refeq{eq:DF2-S0-numeric-contr} for $ij=bd, bs$. The dependence of $\Delta F=2$
on $\Wc[(+)]{Hq}$ is by about a factor three weaker compared to one of
$\Wc[(-)]{Hq}$. Despite the presence of $\Wc[(+)]{Hq}$, $\Delta F=2$
constraints will not constrain $\Delta F=1$ observables, as long as the
$\Wc[(\pm)]{Hq}$ are arbitrary. However, in specific models they can be related,
yielding again correlations. The influence of $\Delta F=2$ remains weaker than
in the RH case though, given the absence of chiral and RG enhancements.

In addition to $\Delta F=2$ and semileptonic $\Delta F=1$ processes we consider
in the Kaon sector also one non-leptonic $\Delta F=1$ observable, namely 
$\epe$. We parameterize NP effects in this quantity as \cite{Buras:2015jaq}
\begin{align}
  \label{GENERAL} 
  \frac{\varepsilon'}{\varepsilon} & =
  \left(\frac{\varepsilon'}{\varepsilon}\right)^{\rm SM} +
  \left(\frac{\varepsilon'}{\varepsilon}\right)^{\rm NP} =
  \left(\frac{\varepsilon'}{\varepsilon}\right)^{\rm SM}+\kepe\cdot 10^{-3} 
\end{align}
and use the expressions given in \cite{Bobeth:2016llm} to express
$\kappa_{\epsilon'}$ as a linear function of $\wc{Hd}{sd}$ and
$\wc[(+)]{Hq}{sd}$, see also \cite{Buras:2015jaq}. These expressions are
unaffected by the new contributions calculated in this work.  As in
\cite{Bobeth:2016llm} we use the very conservative bound
$\kappa_{\epsilon'}\in[0,2]$, reflecting the fact that the experimental world
average from the NA48 \cite{Batley:2002gn} and KTeV
\cite{AlaviHarati:2002ye,Abouzaid:2010ny} collaborations,
\begin{align}
  \label{eq:epe:EXP}
  (\epe)_\text{exp} & = (16.6 \pm 2.3) \times 10^{-4} ,
\end{align}
is larger than recent theoretical estimates \cite{Blum:2015ywa, Bai:2015nea,
Buras:2015yba, Buras:2015xba, Buras:2016fys, Kitahara:2016nld}.

\subsubsection{\boldmath 
  Correlations for RH $Z$ couplings
  \label{sec:RH-scenario}
}

\begin{figure}
  \includegraphics[width=5.1cm,height=5.1cm]{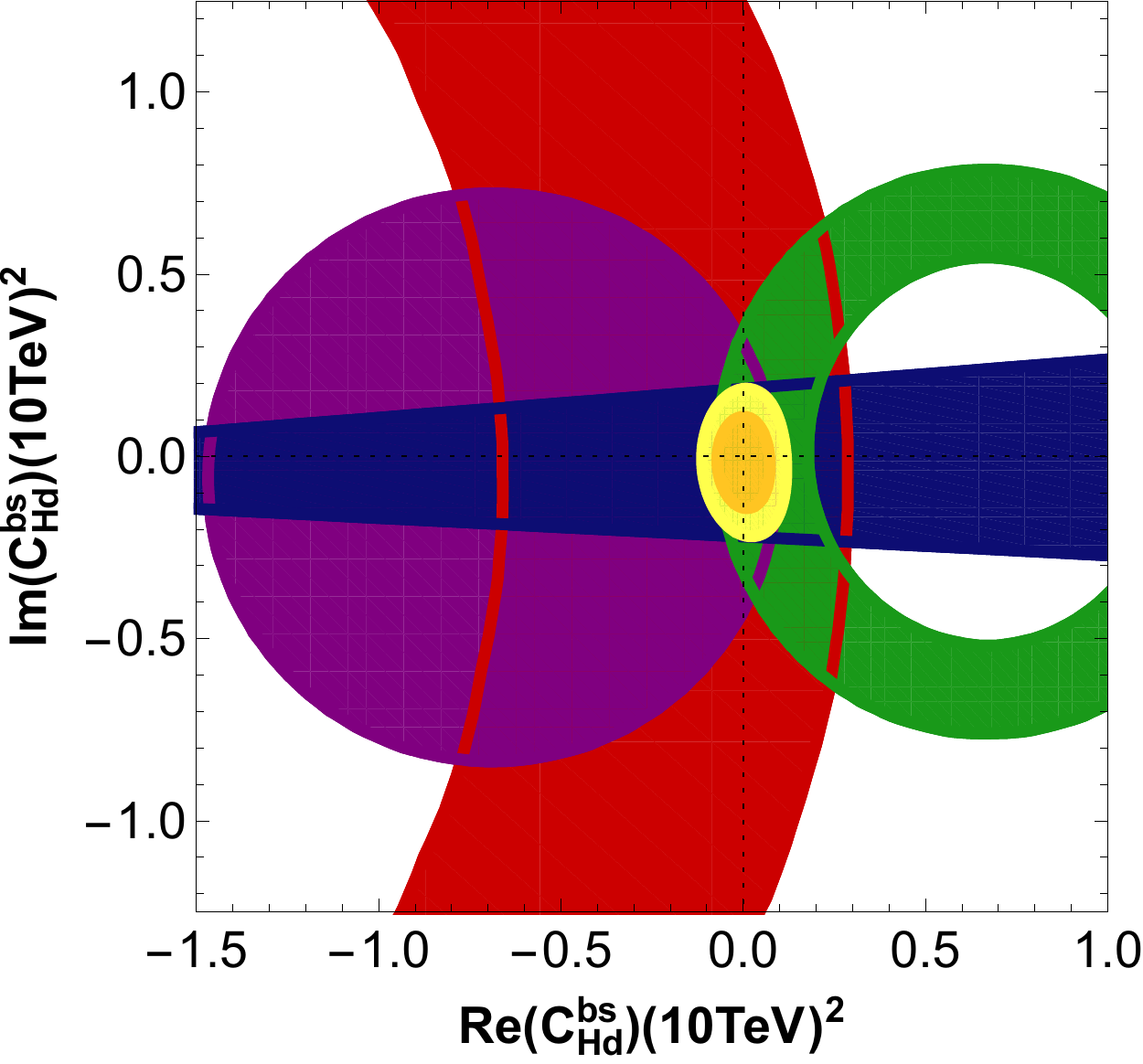}
~
  \includegraphics[width=5.1cm,height=5.1cm]{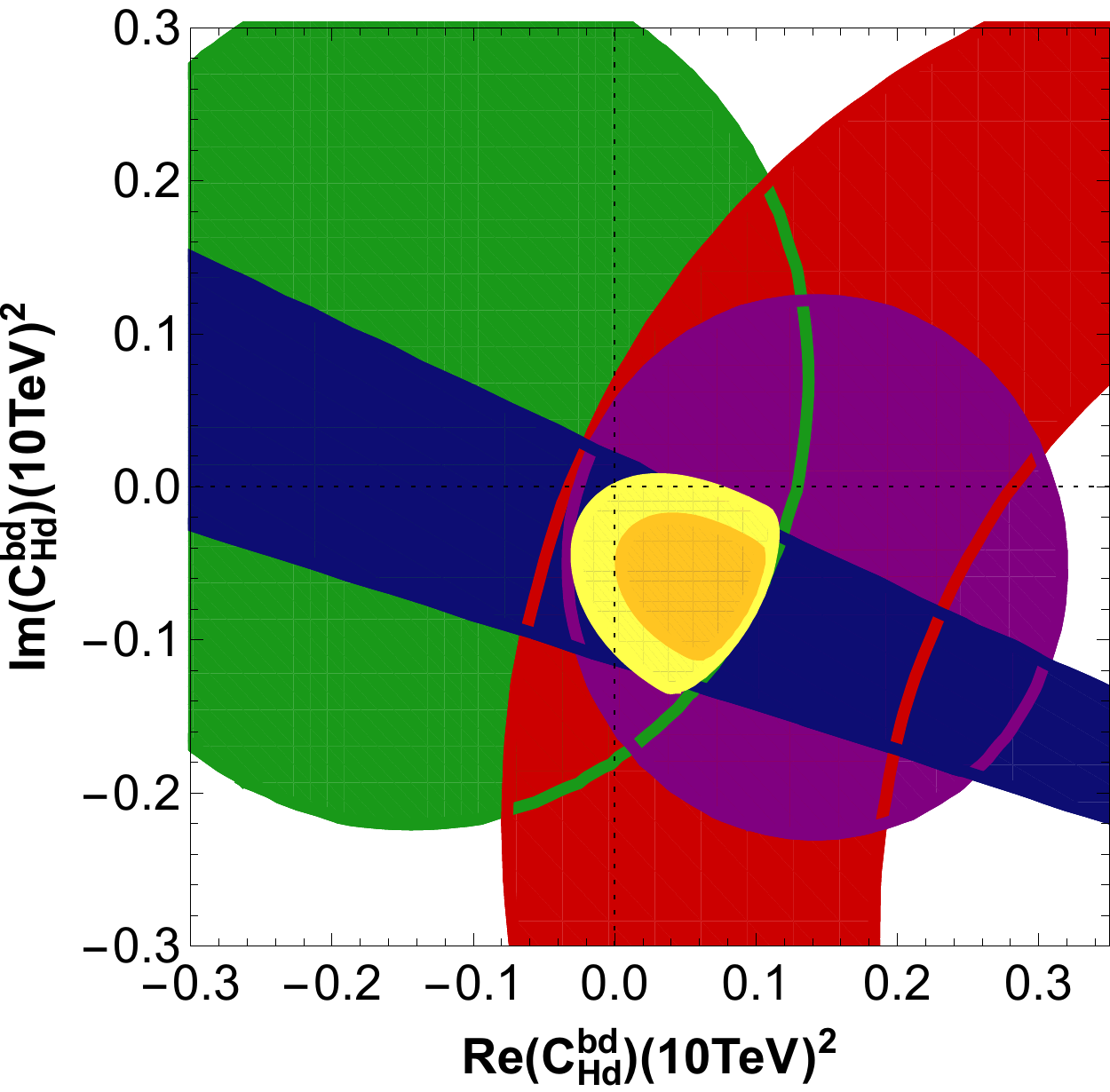}
~
  \includegraphics[width=5.1cm,height=5.1cm]{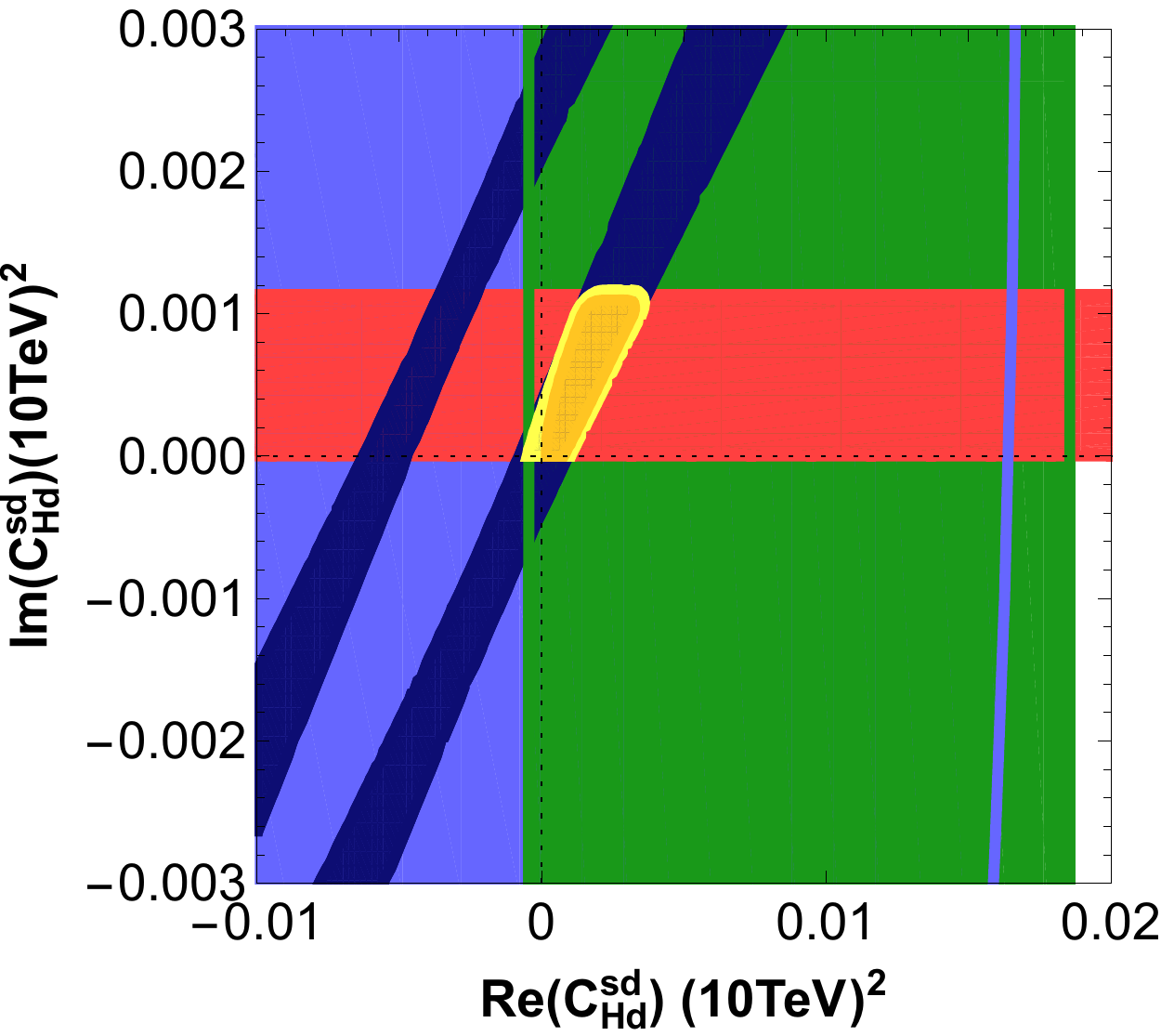}
\caption{\small \label{fig:constraints} 
  Constraints on the couplings $\wc{Hd}{ij}$ from $b\to s$ (left), $b\to d$
  (middle) and $s\to d$ (right) observables at $\muNP=10$~TeV, assuming these
  are the only couplings present at $\muNP$. The constraints shown correspond to
  the observables $\Delta m_s$ (dark red), $\phi_s$ (dark blue),
  $\Br(B_s\to\mu^+\mu^-)$ (green) and $\Br(B^+\to K^+\mu^+\mu^-)_{[15,22]}$
  (purple) for $b\to s$, to $\Delta m_d$ (dark red), $\sin2\beta$ (dark blue),
  $\Br(B_d\to\mu^+\mu^-)$ (green) and $\Br(B^+\to \pi^+\mu^+\mu^-)_{[15,22]}$
  (purple) for $b\to d$, and $\varepsilon_K$ (dark blue), $\epsilon'/\epsilon$
  (light red), $\Br(K^+\to\pi^+\nu\bar\nu)$ (light blue) and
  $\Br(K_L\to\mu^+\mu^-)$ (green) for $s\to d$ transitions.  The global fit to
  each sector is shown in yellow. All coloured areas correspond to $95\%$~CL,
  only the dark yellow area to $68\%$.
}
\end{figure}

We start by assuming the presence of only RH NP $Z$ couplings, \emph{i.e.}
$\Wc{Hd}\neq 0$. In this case, in principle two observables per sector
$ij=sd,bd,bs$ are sufficient to determine both real and imaginary part of this
coefficient. The fits for the three sectors are shown in
\reffig{fig:constraints}. We have chosen $\muNP = 10$~TeV in order
to guarantee sufficient suppression of potential dimension eight contributions
as explained in \refsec{sec:COMPL}. In all three sectors a consistent combined fit is
possible, restricting $\Wc{Hd}$ to lie in a range close to the SM point
$\Wc{Hd}=0$: the obtained ranges are
\begin{align}
  |\Wc[bs]{Hd}| & \lesssim \frac{0.25}{(10\,{\rm TeV})^2} , & 
  |\Wc[bd]{Hd}| & \lesssim \frac{0.15}{(10\,{\rm TeV})^2} , &
  |\Wc[sd]{Hd}| & \lesssim \frac{0.004}{(10\,{\rm TeV})^2} . 
\end{align}
The hierarchy in these results follows roughly that of the corresponding CKM combinations
$\lambda_{ij}^t$. It is seen how the combined fit is determined in all three
sectors by observables from both $\Delta F=1$ and $\Delta F=2$: $\phi_s$,
$\Br(B^+\to K^+\mu^+\mu^-)$ and $\Br(B_s\to\mu^+\mu^-)$ for $b\to s$,
$\sin2\beta$, $\Br(B^+\to \pi^+\mu^+\mu^-)$ and $\Br(B_d\to\mu^+\mu^-)$ for
$b\to d$, and $\varepsilon_K$, $\epsilon'/\epsilon$, $K^+\to\pi^+\nu\bar\nu$ and
$\Br(K_L\to \mu^+\mu^-)$ in $s\to d$. The increased importance of $\Delta F=2$
observables in this context compared to earlier works is due to the new
contributions calculated above. Especially $\varepsilon_K$, fully dominated by
the new contribution from $O_{\rm LR,1}$, is now the most constraining
observable for $s\to d$ together with $\epsilon'/\epsilon$.

\begin{figure}
  \includegraphics[width=5.1cm,height=5.1cm]{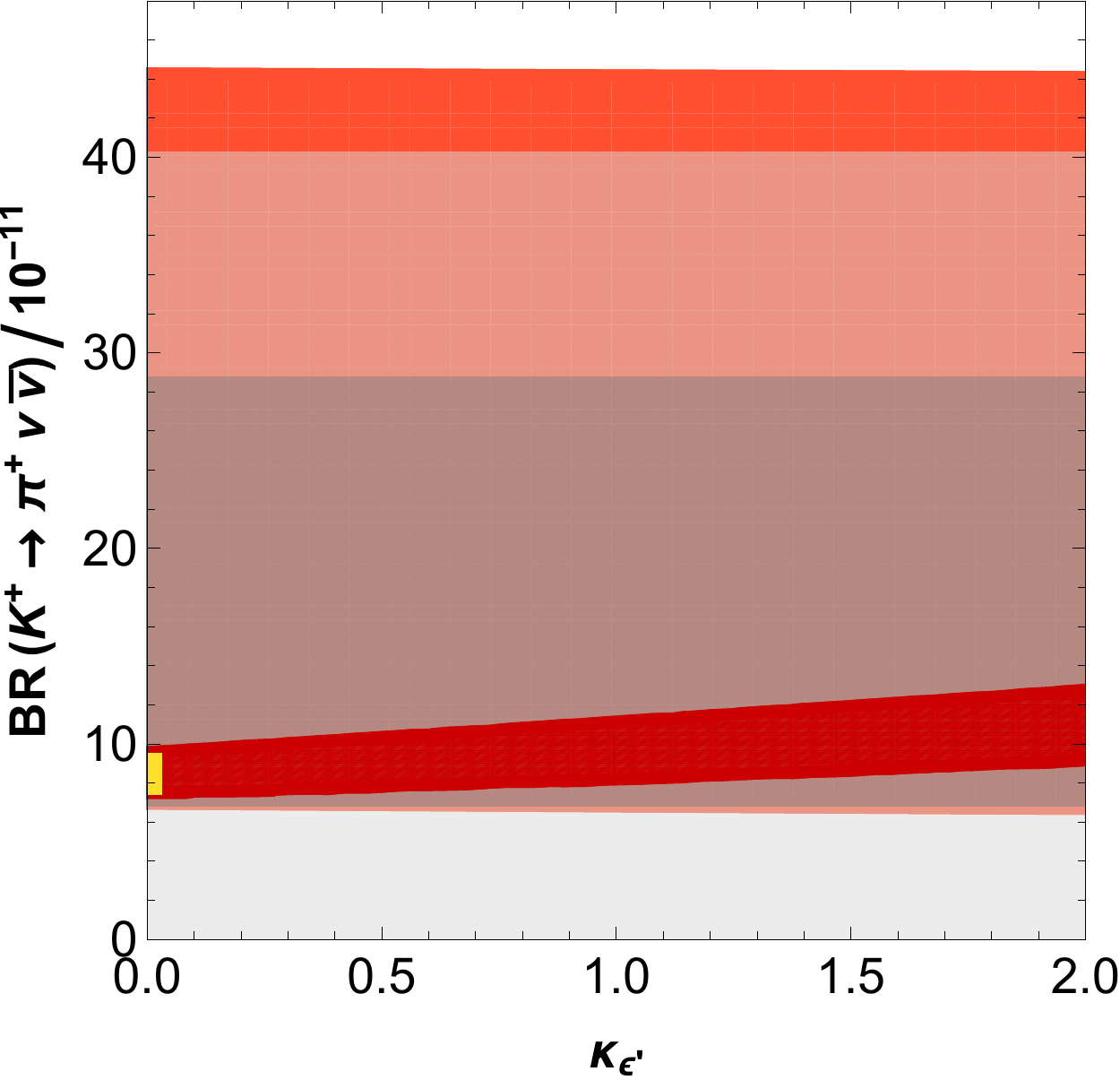}
~
  \includegraphics[width=5.1cm,height=5.1cm]{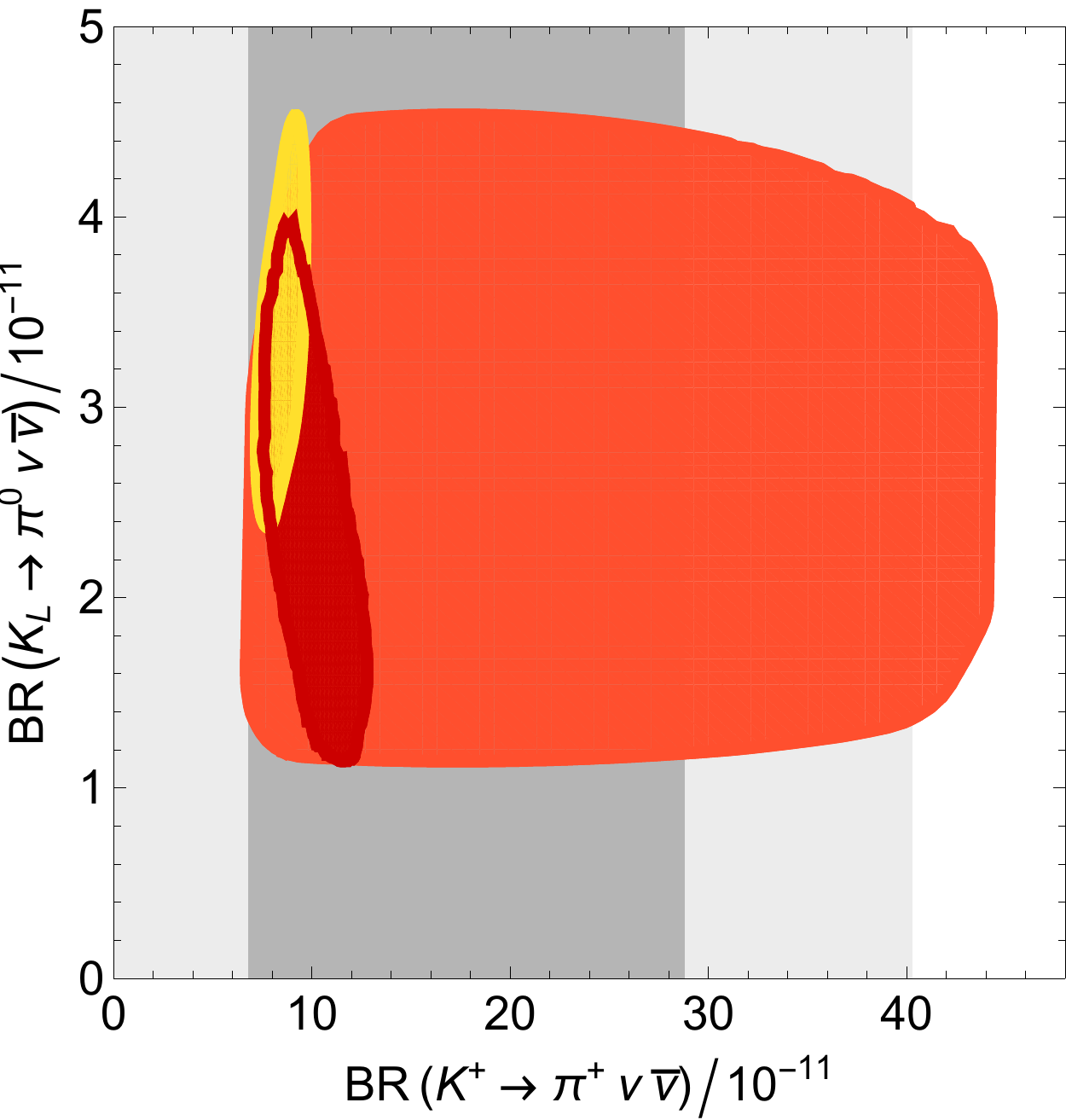}
~
  \includegraphics[width=5.1cm,height=5.1cm]{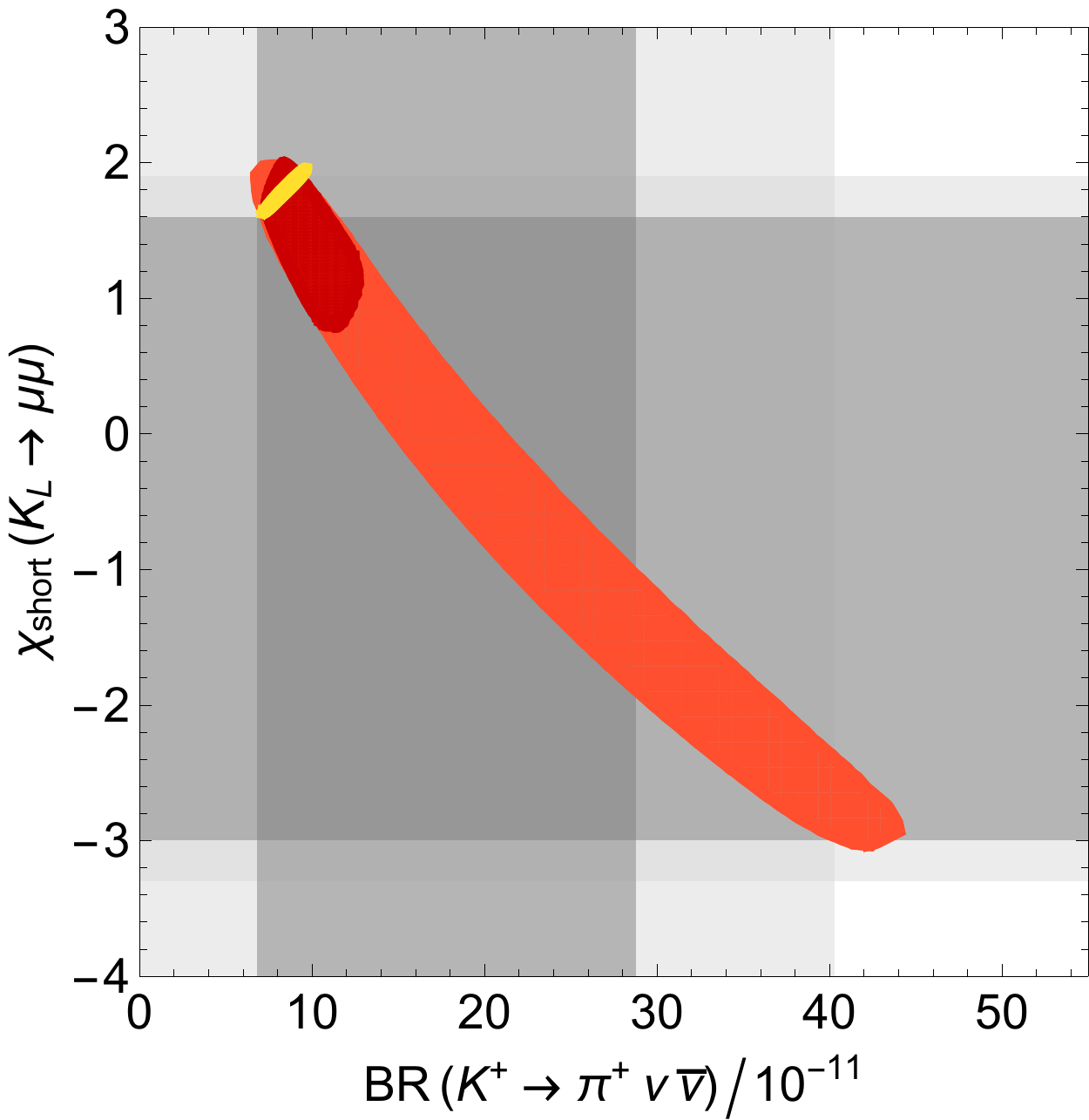}
\caption{\label{fig:predictionsK} \small
  Correlations between $s\to d$ observables and $\kappa_{\epsilon'}$ in the
  presence of right-handed NP FC $Z$ couplings, only, including (darker
  colours) and excluding (lighter colours) constraints from $\Delta F=2$:
  $\Br(K^+\to\pi^+\nu\bar\nu)$ vs. $\kappa_{\epsilon'}$ (left),
  $\Br(K_L\to\pi^0\nu\bar\nu)$ vs.  $\Br(K^+\to\pi^+\nu\bar\nu)$ (middle), and
  $\chi_{\rm short}(K_L\to\mu^+\mu^-)$ vs. $\Br(K^+\to\pi^+\nu\bar\nu)$ (right).
  All coloured areas correspond to $95\%$~CL, the yellow areas are
  the SM predictions. The dark and light grey areas indicate the
  $1$- and $2\sigma$ experimental ranges.
}
\end{figure}

To illustrate the influence of $\Delta F=2$ observables further, we show in
\reffig{fig:predictionsK} the resulting correlations between observables in the
Kaon sector with and without taking the $\Delta F=2$ constraint from
$\varepsilon_K$ into account. Taking only $\Delta F=1$ into account, in many
cases the present upper limits for observables like $\Br(K^+\to\pi^+\nu\bar\nu)$
can be reached, \emph{i.e.} enhancements compared to the SM of up to a factor
5. On the other hand, the resulting predictions for rare decays when including
$\Delta F=2$ are rather close to the SM; specifically, $\Br(\kpn)$ is
  predicted to be enhanced, but only up to $50\%$ of the SM value.

\begin{figure}
  \includegraphics[width=6.5cm,height=6.5cm]{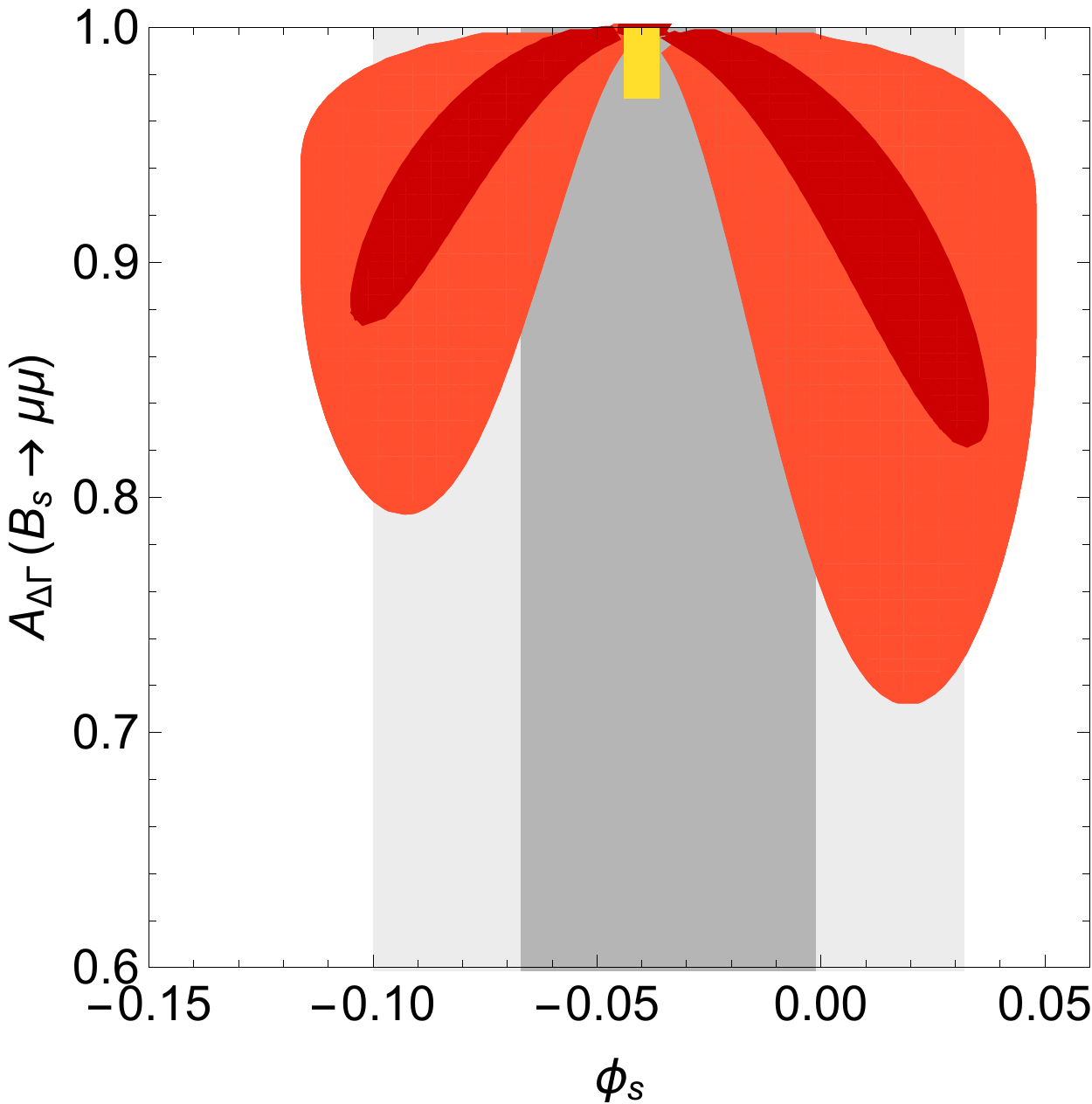}
\hfill
  \includegraphics[width=6.5cm,height=6.5cm]{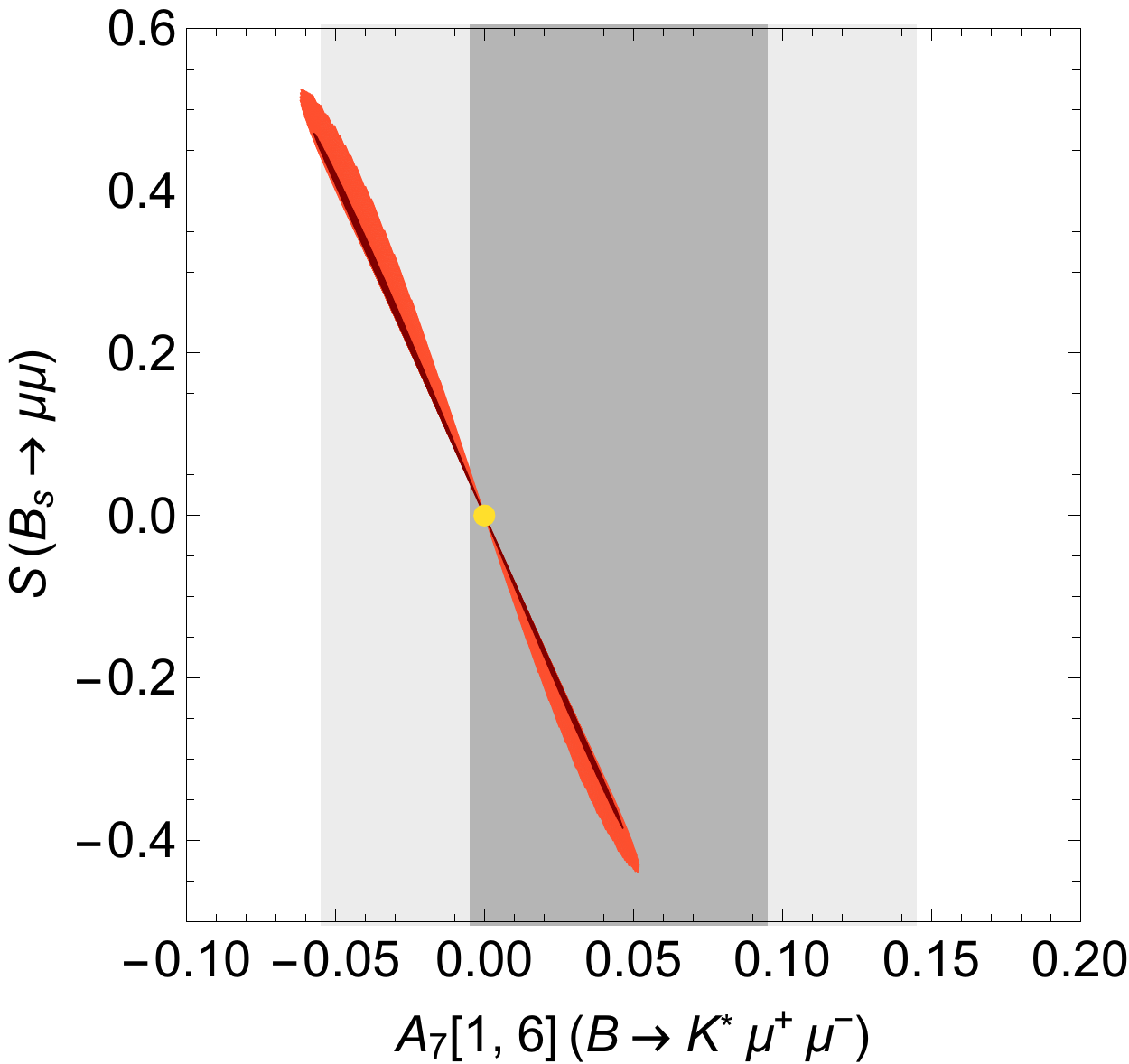}
\caption{\small \label{fig:predictionsbs}
  Correlations between $b\to s$ observables in the presence of right-handed NP
  FC $Z$ couplings, only, including (darker colours) and excluding (lighter
  colours) constraints from $\Delta F=1$: $A_{\Delta\Gamma}(B_s\to\mu^+\mu^-)$
  vs. $\phi_s$ (left) and $S(B_s\to\mu^+\mu^-)$ vs. $A_7[1,6] (B\to K^*\mu^+\mu^-)$ 
  (right).  All coloured areas correspond to $95\%$~CL, the yellow areas are the SM 
  predictions. The dark and light grey areas indicate the $1$- and $2\sigma$
  experimental ranges.  
}
\end{figure}

\begin{figure}
  \includegraphics[width=6.5cm,height=6.5cm]{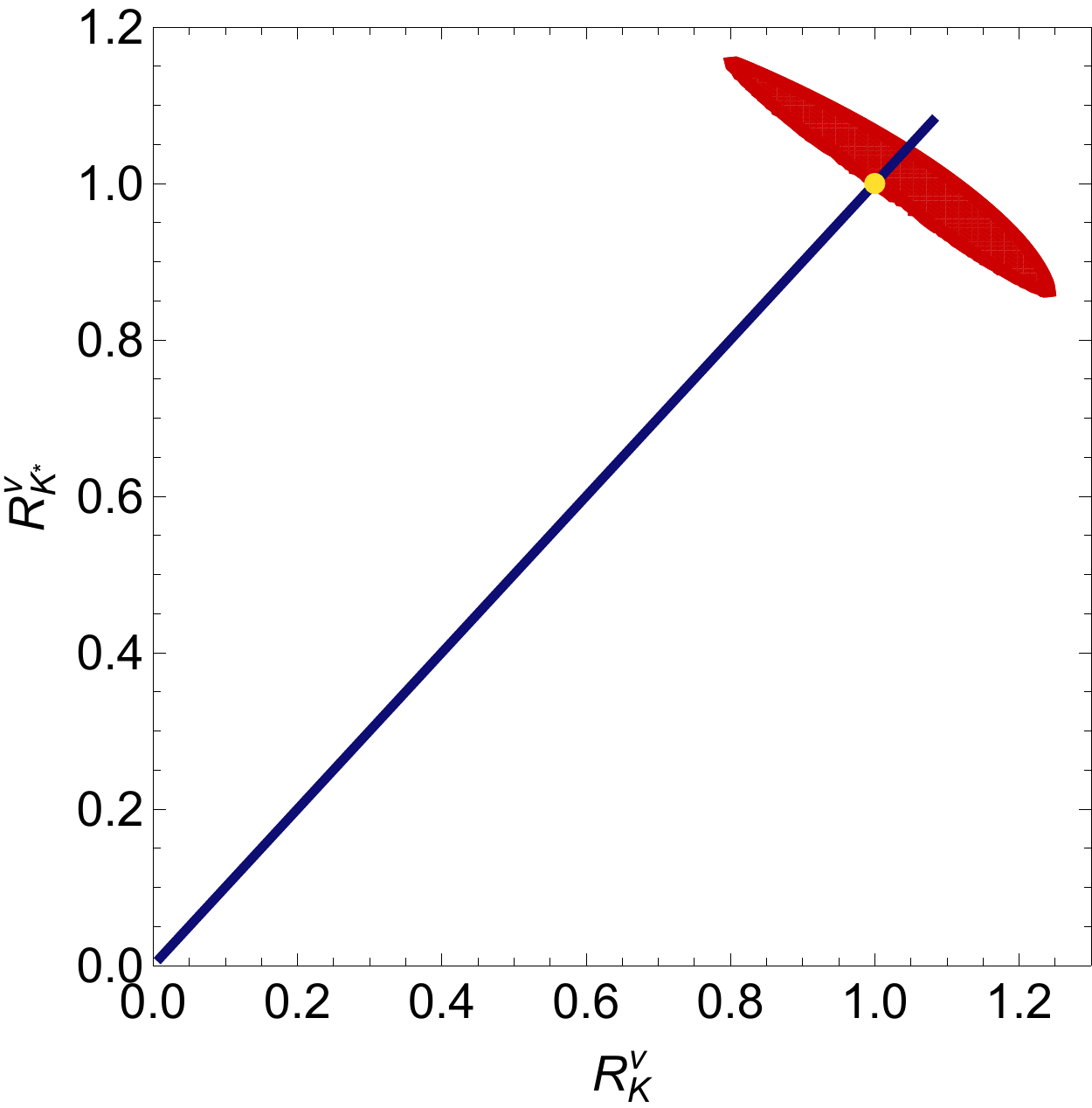}
\hfill
  \includegraphics[width=6.5cm,height=6.5cm]{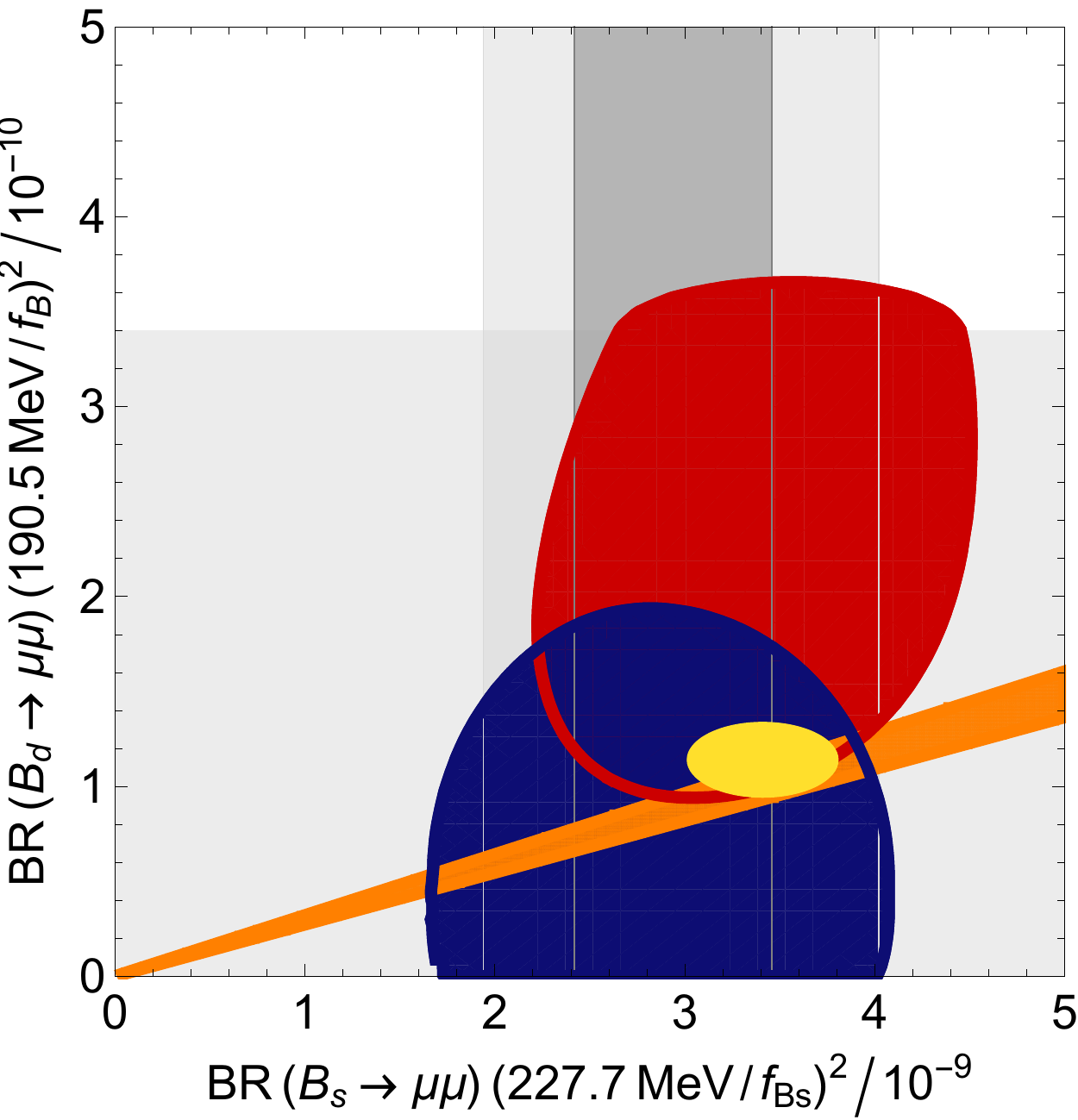}
\caption{\small \label{fig:RLcomp}
  Correlations between $R_{K^*}^\nu$ vs. $R_K^\nu$ (left) and 
  $\Br(B_d\to \mu^+\mu^-)$ vs. 
  $Br(B_s\to \mu^+\mu^-)$ (right) in purely LH (blue) or RH (red) scenarios. 
  All coloured areas correspond to $95\%$~CL, the yellow areas are the SM predictions.  
  The dark and light grey areas indicate the $1$- and $2\sigma$ experimental ranges.
  The orange band in the right plot corresponds to a scenario of 
  constrained minimal flavour violation (CMFV) \cite{Buras:2014fpa}.
}
\end{figure}

In contrast to $s\to d$, we show in \reffig{fig:predictionsbs} correlations with
and without $\Delta F=1$ constraints for $b\to s$ transitions. Since from
\reffig{fig:constraints} it can be seen that $\Delta F=1$ dominate the global
fit, it does not surprise that the allowed ranges again become much larger when
excluding the corresponding observables; clearly only the combination of
$\Delta F=1$ and $\Delta F=2$ constraints paints the full picture. There is
a strong correlation in the RH scenario between the mass-eigenstate rate
asymmetry $A_{\Delta\Gamma}(B_s\to\mu^+\mu^-)$ and $\phi_s$ that can be tested
in the near future at LHCb. We show also the strong correlation between the
mixing-induced CP asymmetry $S(B_s \to \mu^+\mu^-)$ and one of the T-odd
CP asymmetries in $B\to K^*\mu^+\mu^-$, where apart from the shown $A_7$,
also $A_8$ and $A_9$ are subject of improving measurements at LHCb.  Note
that to very high accuracy $(A_{\Delta\Gamma})^2 + (S)^2 = 1$ in
$B_s\to \mu^+\mu^-$ due to a vanishing direct CP-asymmetry. In RH scenarios
there is also a strong correlation between $A_{7,8,9}$.

Finally, in \reffig{fig:RLcomp} we directly compare models with only RH and only
LH NP $Z$ couplings by showing the correlations between $R_K^\nu$ and
$R_{K^*}^\nu$ \cite{Buras:2014fpa},
\begin{align}
  R_{K^{(*)}}^\nu & 
  = \frac{\Br(B\to K^{(*)}\nu\bar\nu)}{\Br(B\to K^{(*)}\nu\bar\nu)|_{\rm SM}} ,
\end{align} 
as well as $\Br (B_s\to\mu^+\mu^-)$ and $\Br (B_d\to\mu^+\mu^-)$. For RH models, 
shown in dark red, we observe a strong anti-correlation between the two modes
with neutrinos in the final state, each allowed to deviate up to $\sim 20\%$ 
from its SM value. Furthermore, $\Br (B_s\to\mu^+\mu^-)$ is slightly pulled to
larger values by $\Br(B^+\to K^+ \mu^+\mu^-)$, see \reffig{fig:constraints}, and
$\Br (B_d\to\mu^+\mu^-)$ is predicted to be at least as large as the SM value,
with values allowed up to the present experimental upper limit; lower values are
in tension with $\Br(B^+\to \pi^+ \mu^+\mu^-)$ as well as $\Delta m_d$, see
again \reffig{fig:constraints}.

\subsubsection{\boldmath 
  Correlations for LH $Z$ couplings
  \label{sec:LH-scenario}
}

\begin{figure}
  \includegraphics[width=5.1cm,height=5.1cm]{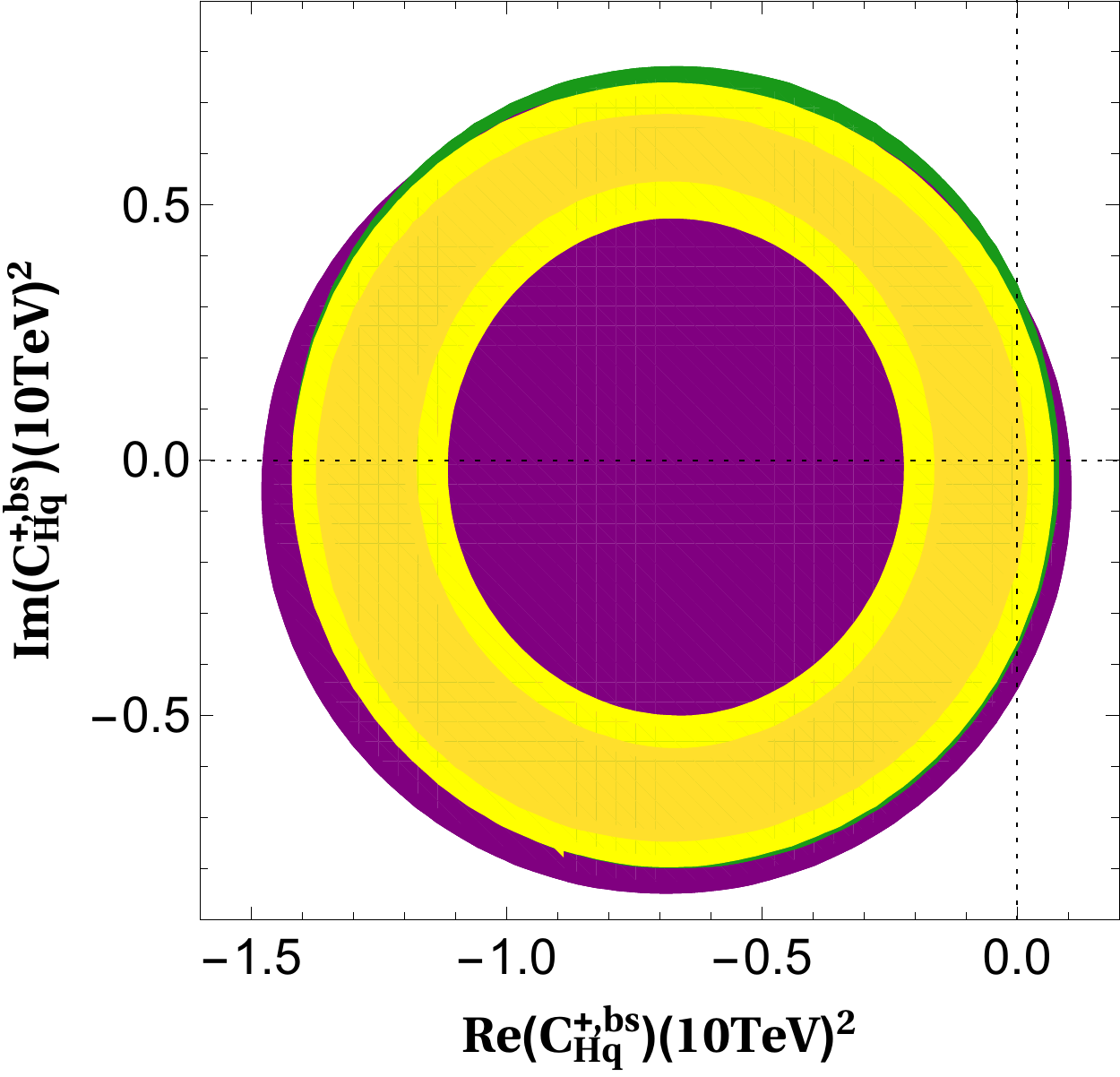}
~
  \includegraphics[width=5.1cm,height=5.1cm]{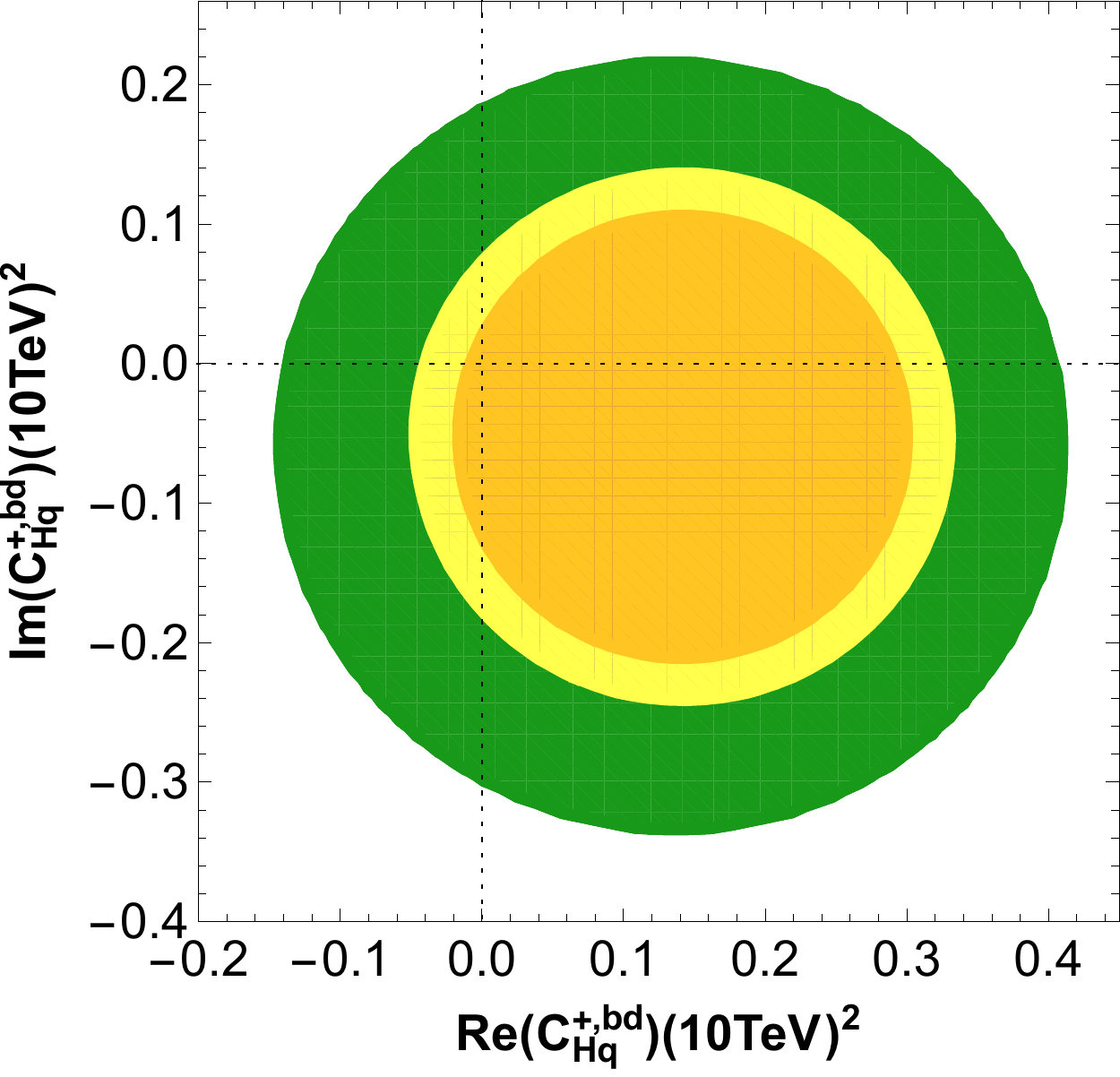}
~
  \includegraphics[width=5.1cm,height=5.1cm]{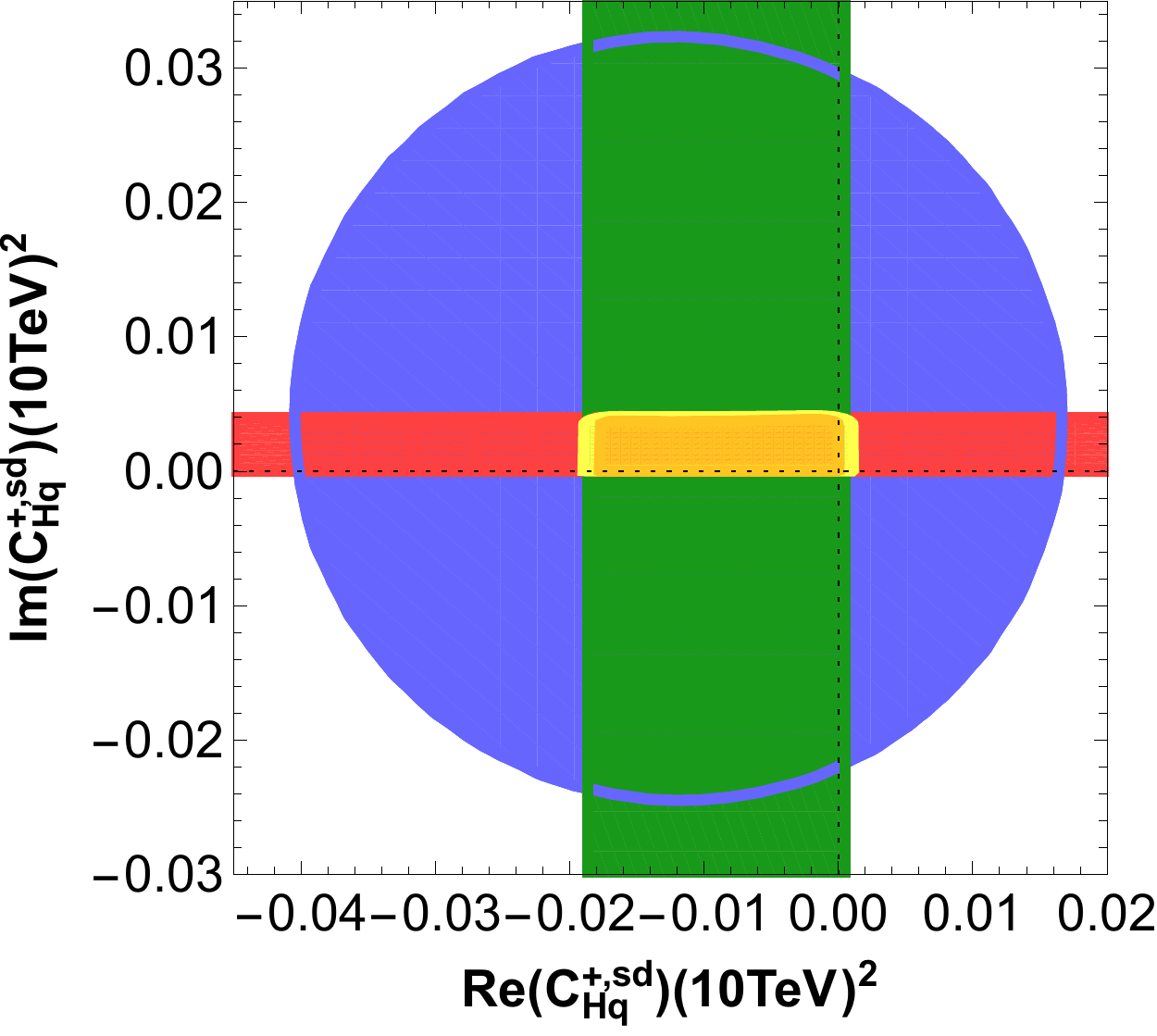}\newline
    \includegraphics[width=5.1cm,height=5.1cm]{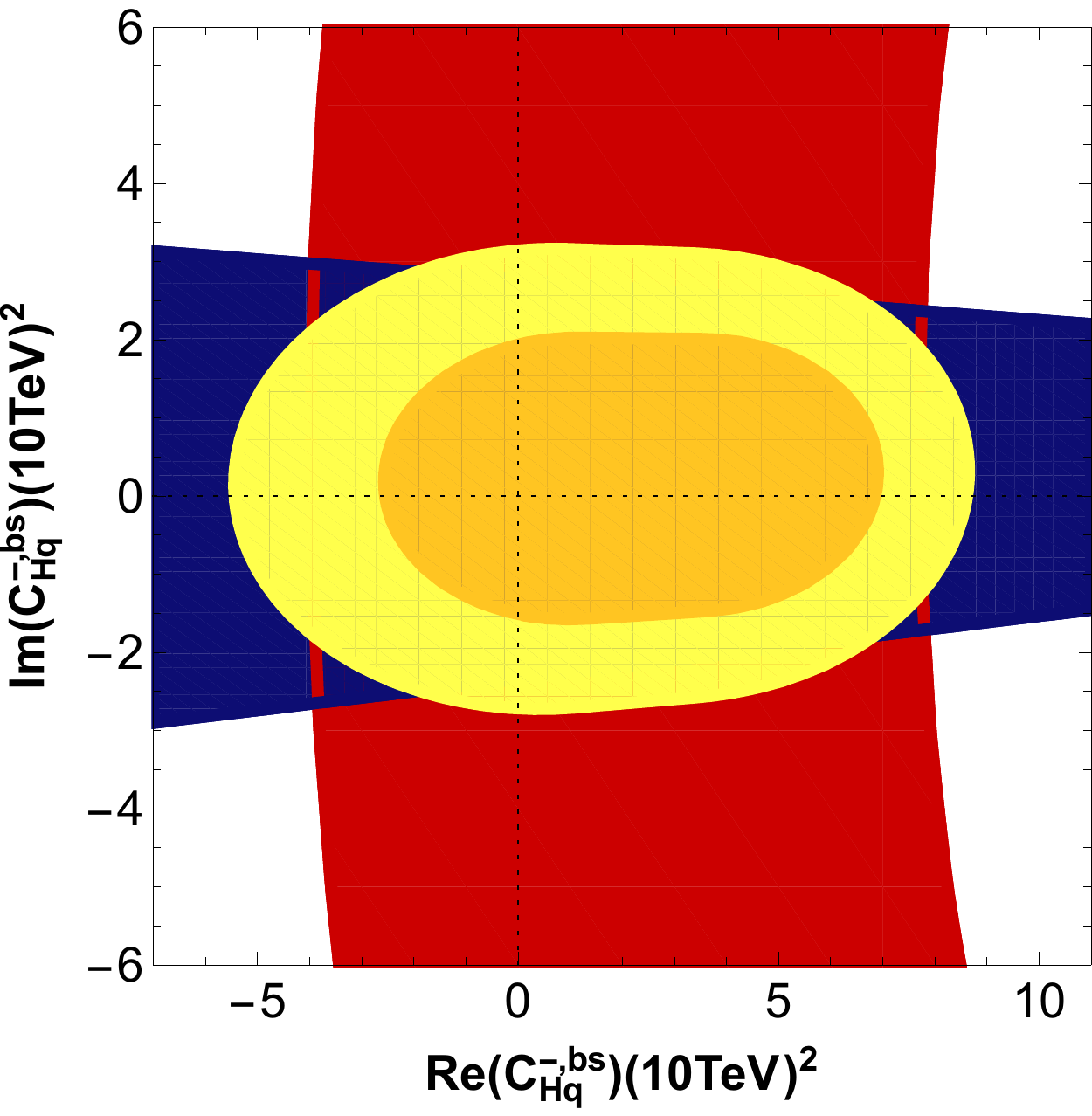}
~
  \includegraphics[width=5.1cm,height=5.1cm]{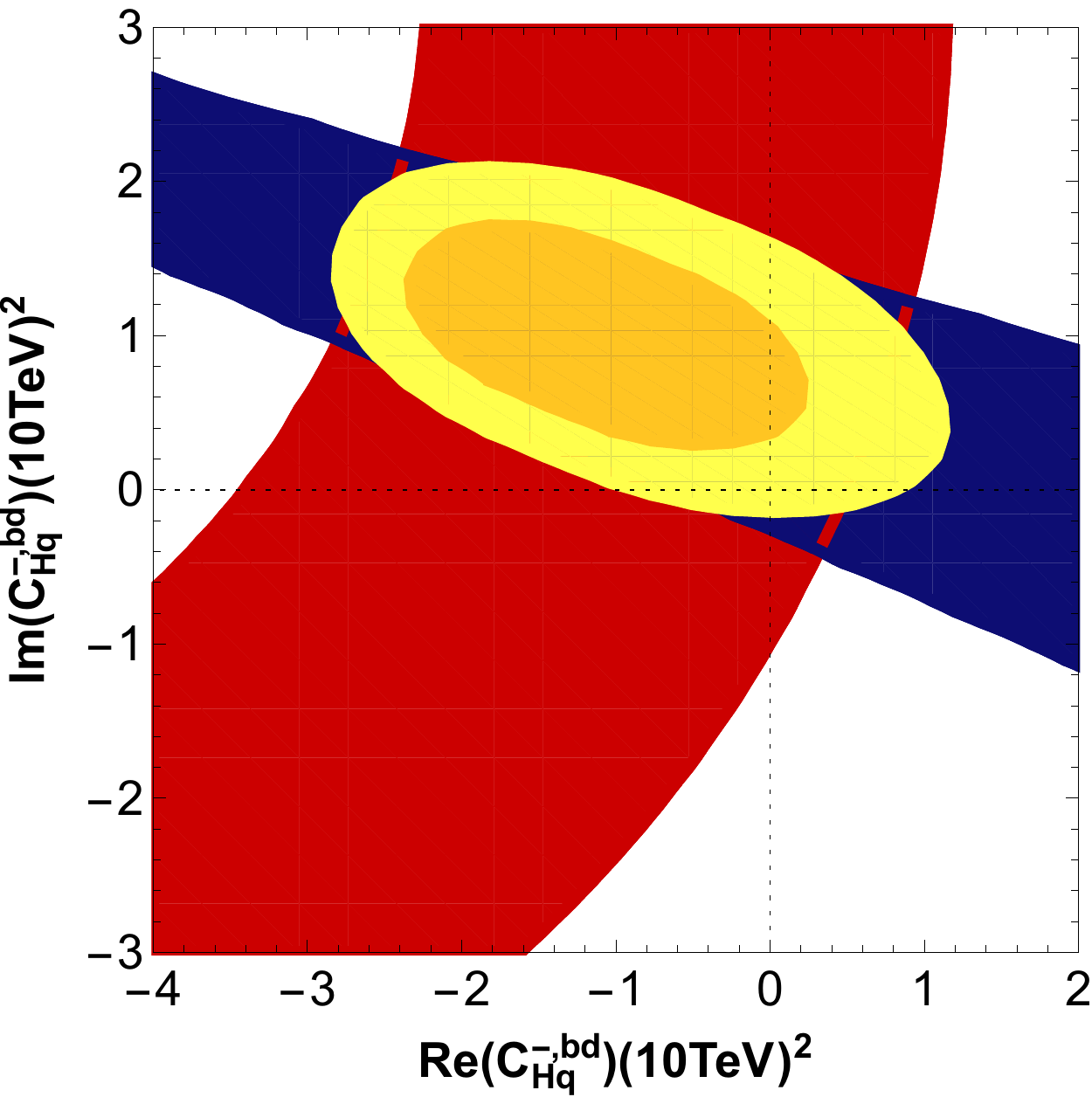}
~
  \includegraphics[width=5.1cm,height=5.1cm]{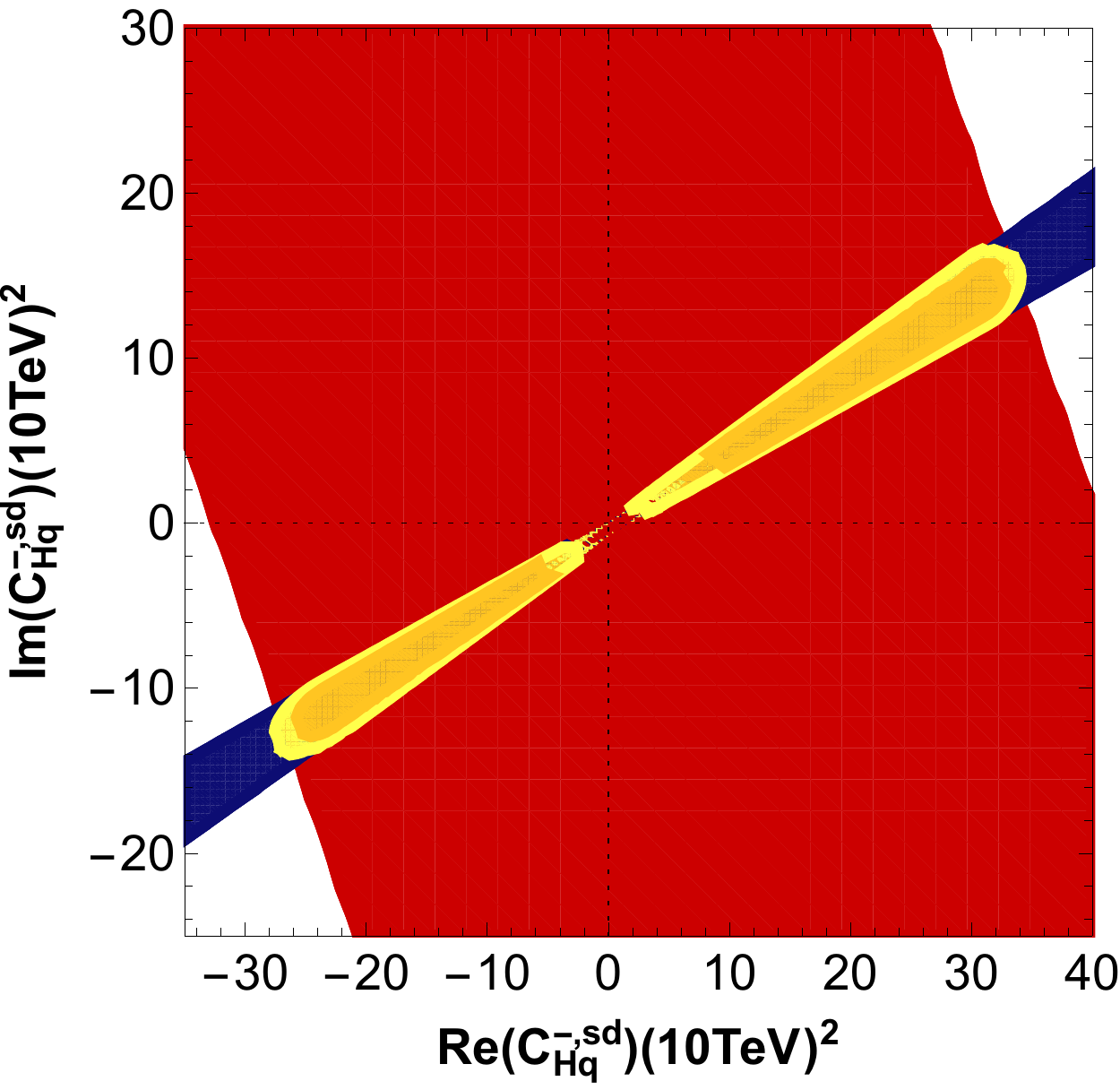}
\caption{\small \label{fig:constraintsL} 
  Constraints on the couplings $\wc[(\pm)]{Hq}{ij}$ from $b\to s$ (left),
  $b\to d$ (middle) and $s\to d$ (right) observables at $\muNP=10$~TeV, assuming
  these are the only couplings present at $\muNP$. The $\Delta F=2$ constraints
  shown in the $\wc[(-)]{Hq}{ij}$ planes in the lower row are at LO. The
  constraints shown correspond to the observables $\Delta m_s$ (dark red),
  $\phi_s$ (dark blue), $\Br(B_s\to\mu^+\mu^-)$ (green) and
  $\Br(B^+\to K^+\mu^+\mu^-)_{[15,22]}$ (purple) for $b\to s$, to $\Delta m_d$
  (dark red), $\sin2\beta$ (dark blue), $\Br(B_d\to\mu^+\mu^-)$ (green) and
  $\Br(B^+\to \pi^+\mu^+\mu^-)_{[15,22]}$ (purple) for $b\to d$, and
  $\varepsilon_K$ (dark blue), $\epsilon'/\epsilon$ (light red),
  $\Br(K^+\to\pi^+\nu\bar\nu)$ (light blue) and $\Br(K_L\to\mu^+\mu^-)$ (green)
  for $s\to d$ transitions. The global fit to each
  sector is shown in yellow.  All coloured areas correspond to $95\%$~CL, only
  the dark yellow area to $68\%$.
}
\end{figure}

Making the assumption that only LH NP couplings of the $Z$ are non-vanishing,
$\Wc[(1,3)]{Hq}\neq0$, changes the picture qualitatively. For each sector there
are now two complex coefficients; in order to obtain plots similar to
\reffig{fig:constraints}, we show in \reffig{fig:constraintsL} the constraints
from $\Delta F=1$ in the $\Wc[(+)]{Hq}$ plane and from $\Delta F=2$ in the
$\Wc[(-)]{Hq}$ plane; the latter constraints are shown at LO, \emph{i.e.} based
on \refeq{eq:1stLLA-LR1} and \refeq{eq:1stLLA-VLL}, to have only this
coefficient appear.  These coefficients are both much weaker constrained than in
the RH case. The reasons for that are not only the absence of chiral and RG
enhancements for LH contributions and the presence of two coefficients, but also
the different interference pattern in $\Delta F=1$: while for the RH the two
main constraints intersect only in a small area, they essentially lie on top of
each other for LH couplings.  This is due to a relative sign for LH and RH
contributions analogous to the one between \refeq{eq:BRMPnunu} and
\refeq{eq:BRMmumu}.  Finally, in the case of $\wc[(-)]{Hq}{sd}$ the fact that
the long-distance contribution to $\Delta m_K$ has large uncertainties renders
this constraint extremely weak; here progress on the lattice is necessary to
make this a useful constraint.

\begin{figure}
  \includegraphics[width=6.5cm,height=6.5cm]{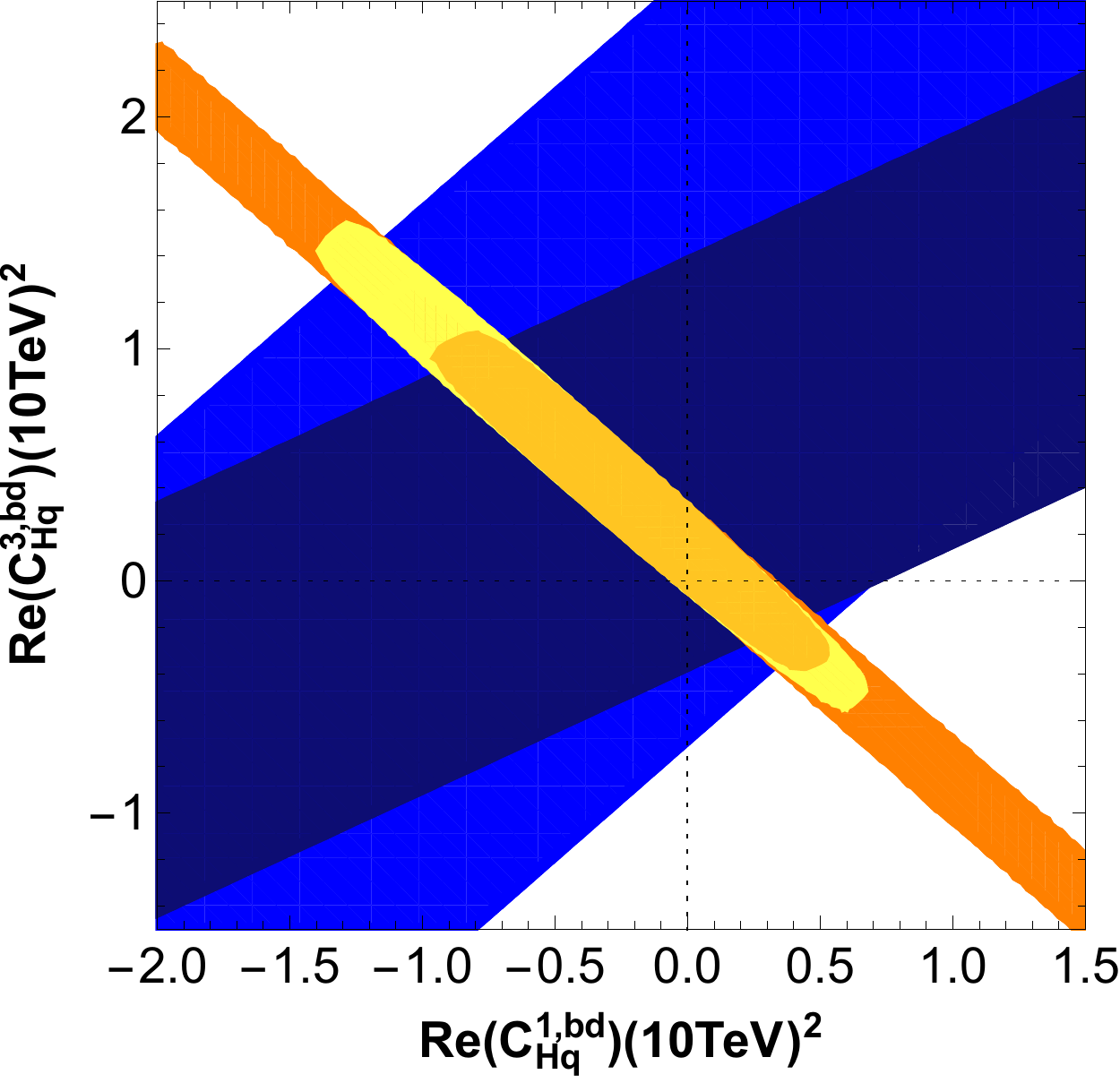}
\hfill
  \includegraphics[width=6.5cm,height=6.5cm]{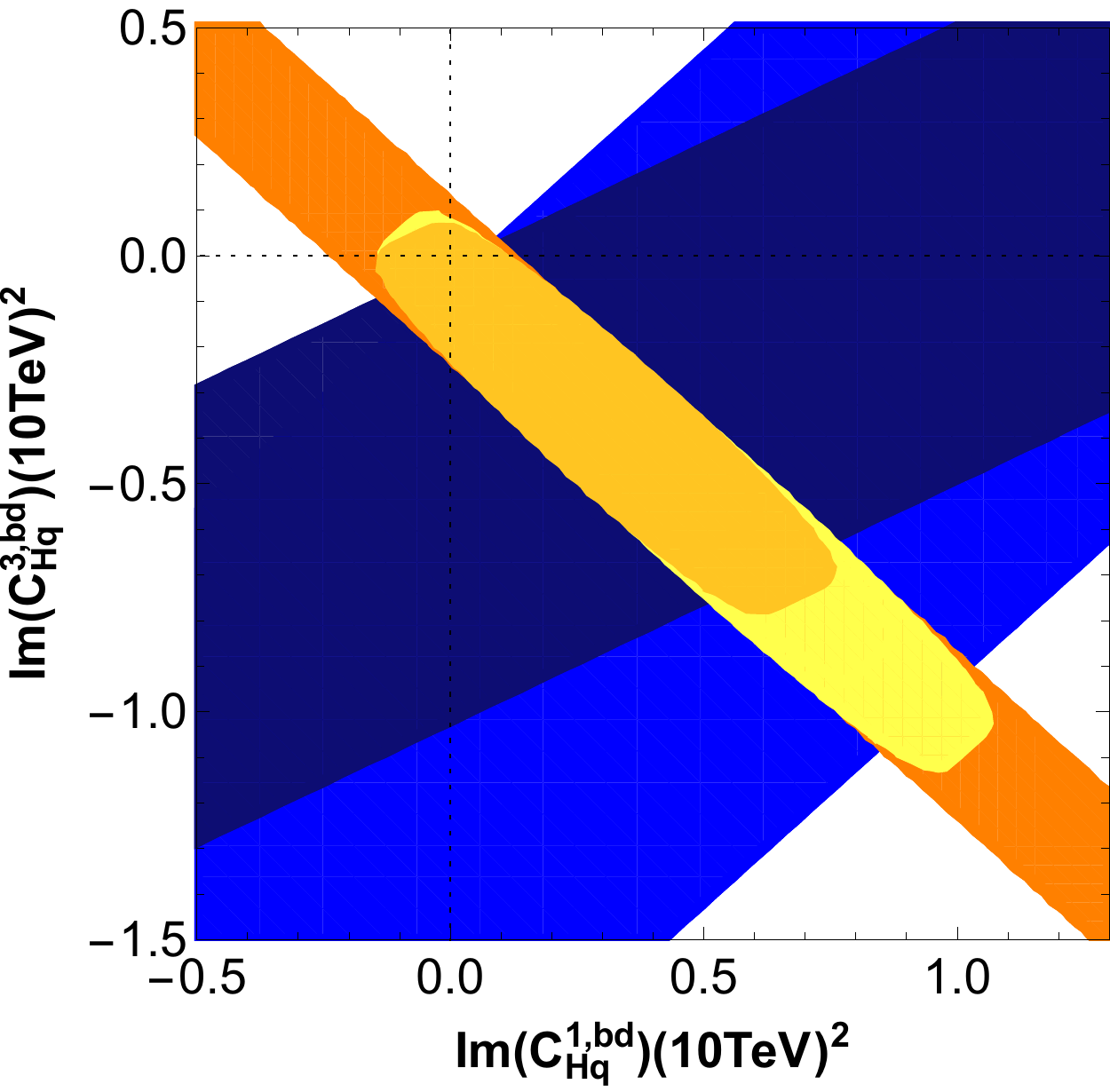}
\caption{\small \label{fig:constraintsLLONLO} 
  Constraints on the couplings $\wc[(1,3)]{Hq}{bd}$ at $\muNP=10$~TeV, assuming
  these are the only couplings present at $\muNP$. The combined $\Delta F=1$
  constraints are shown in orange, the $\Delta F=2$ constraints at LO in blue
  and at NLO in dark blue. The combined fit is shown at LO in yellow and at NLO
  in dark yellow.  All coloured areas correspond to $95\%$~CL.  }
\end{figure}

In order to demonstrate the influence of our NLO calculation, we show in
\reffig{fig:constraintsLLONLO} additionally the combined fits for $b\to d$ in
the planes of the real- and imaginary parts of $\Wc[(1,3)]{Hq}$ at LO and
NLO. At NLO the allowed regions shrink due to the larger coefficients in
\refeq{eq:CVLLnum}; additionally the $\Delta F=2$ constraint is rotated in the
$\Wc[(1)]{Hq}-\Wc[(3)]{Hq}$-plane, due to the additional contribution from
$\Wc[(+)]{Hq}$, see \refeq{eq:CVLLnum}.

\begin{figure}
  \includegraphics[width=6.5cm,height=6.5cm]{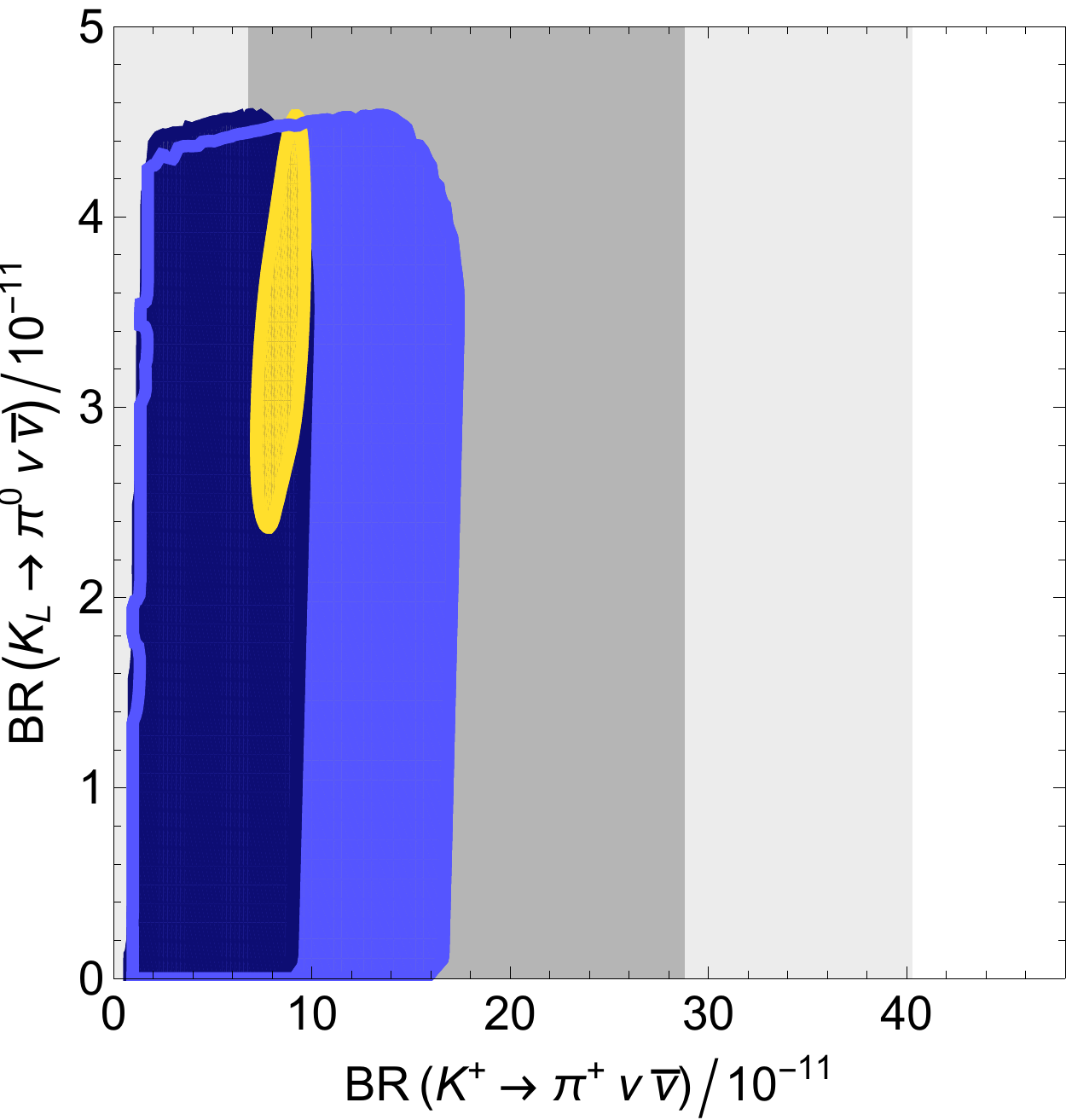}
\hfill
  \includegraphics[width=6.5cm,height=6.5cm]{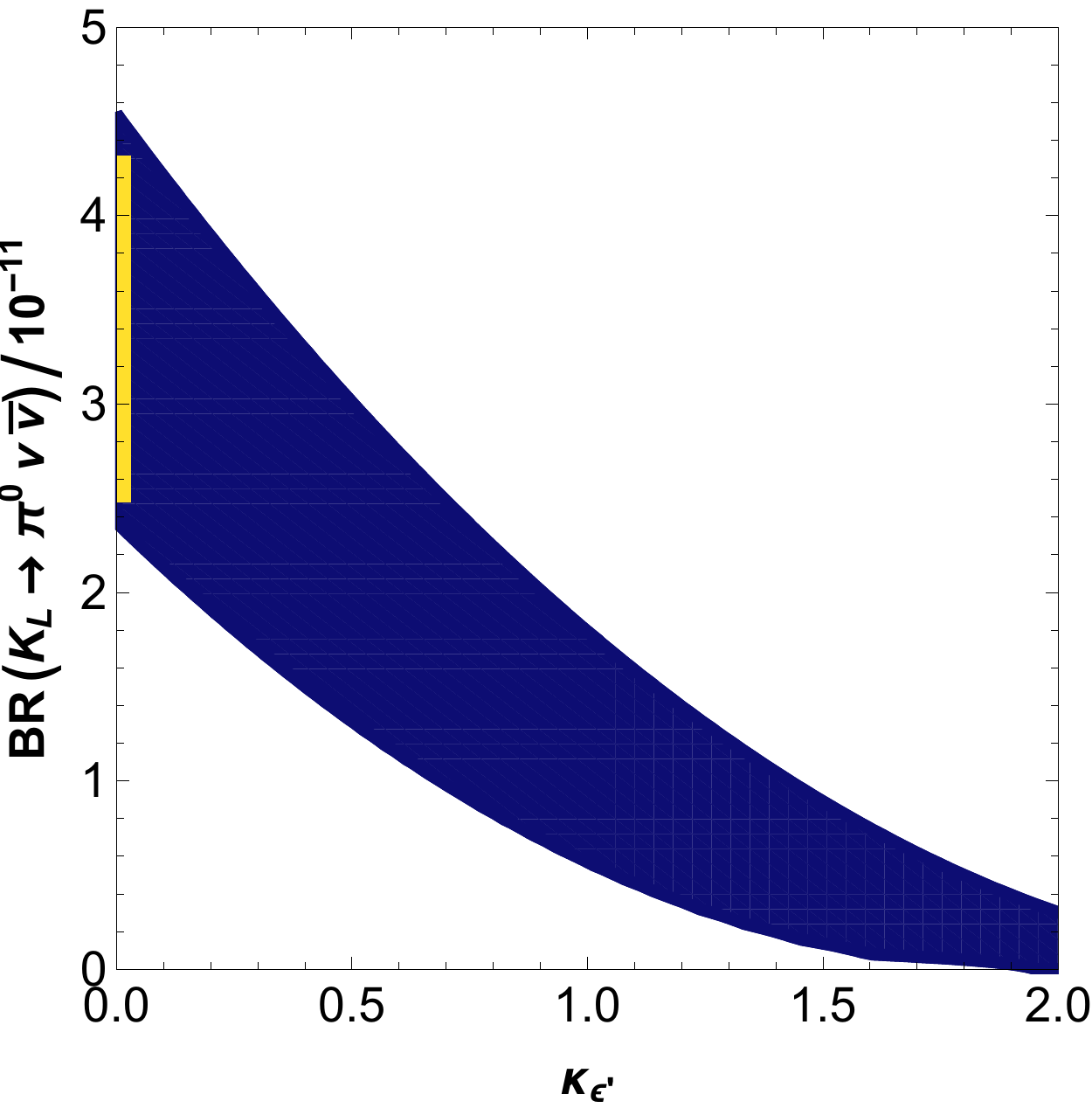}
\caption{\small \label{fig:predictionsL}
  Correlations between $s\to d$ observables in the presence of left-handed NP FC
  $Z$ couplings, only.  $\Br(K_L\to\pi^0\nu\bar\nu)$
  vs. $\Br(K^+\to\pi^+\nu\bar\nu)$ (left) and $\Br(K_L\to\pi^0\nu\bar\nu)$
  vs. $\kappa_{\epsilon'}$ (right). The dark blue area in the left plot
  corresponds to using the assumption on the phase in $K_L\to\mu^+\mu^-$, see
  text, the light blue area to not making this assumption. All coloured areas
  correspond to $95\%$~CL, the yellow line is the SM prediction.  The dark and
  light grey areas indicate the $1$- and $2\sigma$ experimental ranges.  }
\end{figure}

While there are in general no correlations between $\Delta F=1$ and
$\Delta F=2$, the ones within the $\Delta F=1$ sector remain. We illustrate
these correlations in \reffig{fig:predictionsL}. Importantly, we observe that
also in this case no sizable enhancement of $\Br(K^+\to\pi^+\nu\bar\nu)$ is
possible. This is related to our treatment of $K_L\to\mu^+\mu^-$, which utilizes
the approach \cite{Isidori:2003ts} that derived bounds on the short-distance
part $\chi_{\rm SD}$ of its decay amplitude.  Based on the quite general
assumptions stated in~\cite{Isidori:2003ts, D'Ambrosio:1997jp} the sign of the
interference between long- and short-distance contributions can be predicted,
leading to a stronger bound on the short-distance part,
$ -3.1 \leq \chi_{\rm SD}\leq 1.7$, used in our fits. In
\reffig{fig:predictionsL} on the left we show the correlation between
$\Br(K^+\to\pi^+\nu\bar\nu)$ and $\Br(K_L\to\pi^0\nu\bar\nu)$ with (dark blue)
and without (light blue) this assumption. Without this assumption positive
values for the real part of $\Wc[(+)]{Hq}$ become allowed, which in turn allows
for an enhancement of $\Br(K^+\to\pi^+\nu\bar\nu)$ of up to a factor of two
compared to the SM, but still this branching ratio is in our scenarios stronger
constrained from other modes than from the direct measurement. The correlation
between $\epsilon'/\epsilon$ and $\Br(K_L\to\pi^0\nu\bar\nu)$, shown in the same
figure on the right, is not affected by the assumption on $K_L\to \mu^+\mu^-$,
since these constraints are related to the imaginary part of $\Wc[(+)]{Hq}$,
only. These observables are anti-correlated, so that a large value for
$\kappa_{\epsilon'}$ would imply a strong suppression of
$\Br(K_L\to\pi^0\nu\bar\nu)$ \cite{Blanke:2015wba, Buras:2015jaq}.

The correlations between $R_K^\nu$ and $R_{K^*}^\nu$ and between
$\Br (B_s\to\mu^+\mu^-)$ and $\Br (B_d\to\mu^+\mu^-)$ allow to distinguish LH
and RH scenarios for a large part of the parameter space, as shown in
\reffig{fig:RLcomp} (LH in dark blue): since in the LH scenario the interference
with the SM is the same in $B\to K\nu\bar\nu)$ and $B\to K^*\nu\bar\nu)$, there
is a very strict prediction $R_K^\nu/R_{K^*}^\nu\equiv 1$ \cite{Buras:2014fpa}. An
enhancement of each of the ratios is possible only up to $\sim 10\%$ in this
scenario, but a strong suppression is possible, in contrast to RH models. For
$\Br(B_d\to\mu^+\mu^-)$ a moderate enhancement up to $\sim 2\times 10^{-10}$ is
possible, but again a strong suppression, in contrast with the RH case.

%
%
%
\section{
  Summary and Conclusions
  \label{sec:summary}
}

In this paper we have addressed the $Z$-mediated contributions to
$\Delta F=2$ observables from the point of view of the SMEFT. Such an analysis
goes beyond the simplified framework presented in \cite{Buras:2012jb}, in which
only pure BSM contributions involving two vertices generated by NP have been
included and the Yukawa renormalization group effects not been taken into
account.  Once the latter effects are included at LO, their sizable
unphysical scale dependence requires the calculation of NLO corrections. Both
effects have been calculated in the present paper for the first time.

Among the new findings, listed as points 1.-5. in the Introduction, the most
important is the generation of large LR operator contributions through RG Yukawa
evolution in models with flavour-changing RH neutral currents. We have
calculated these effects at LO using the results of \cite{Jenkins:2013wua}.

At NLO we have addressed the contributions, pointed out recently in
\cite{Endo:2016tnu}, in which one of the BSM vertices in $Z$ exchange is
replaced by the SM Z-penguin vertex. We have pointed out that the latter
contributions are by themselves gauge dependent and, using SMEFT, calculated the
remaining contributions that cancel this gauge dependence. This NLO
calculation has significant impact on the original results presented in
\cite{Endo:2016tnu}, where these contributions have not been included;
however, the dominant new effect comes from the RG Yukawa evolution,
which is included in this work for the first time. The comparison 
with the published version of \cite{Endo:2016tnu} is given in \refsec{sec:COMPL}.

In the course of our analysis we have found two new gauge-independent functions
$H_1(x_t)$ and $H_2(x_t)$, given in \refeq{H1} and \refeq{H2}, respectively,
that enter the phenomenology of these new contributions together with FC quark
couplings of the $Z$ generated by NP, given in \refeq{eq:SMEFT-wilson-coeffs}, 
and the Yukawa RG effects mentioned above.

The impact of these new effects has been illustrated model-independently by
considering the correlations between $\Delta F=1$ and $\Delta F=2$
observables in the down-quark sector. These are strongest in the presence of
only RH NP $Z$ couplings: the new effects strengthen the constraints from
$\Delta F=2$, especially in the Kaon sector.  For instance, $\varepsilon_K$
now restricts the coupling $\wc{Hd}{sd}$ in such a way that only small
enhancements of $\Br(\kpn)$ remain allowed, about $50\%$. In the $b\to s$
sector and to less extent also in the $b\to d$ sector, $\Delta F=1$ 
constraints remain dominant, but allow still for sizable NP contributions, 
\emph{e.g.} in $B_{d,s}\to\mu^+\mu^-$. In particular $B_d\to\mu^+\mu^-$
can be enhanced to the present upper bound. Nevertheless, the strong
correlations in this scenario will allow for distinguishing it from other NP
models with coming data from the LHC (LHCb, CMS, ATLAS) and Belle~II.

For NP models that yield only LH FC $Z$ couplings, contributions to
$\Delta F=1$ and $\Delta F=2$ are in general completely decoupled, since two
Wilson coefficients are present -- $\Wc[(1)]{Hq}$ and $\Wc[(3)]{Hq}$ -- and
enter in different combinations in $\Delta F=1$ and $\Delta F=2$. Furthermore
the RG and chiral enhancements present for the RH $\Delta F=2$ contributions
are absent, such that large NP effects remain allowed in this sector,
especially in $\Delta m_K$, where the SM prediction suffers from large
long-distance effects.  The correlations for $\Delta F=1$ processes remain,
however, since they are all sensitive to the same combination of Wilson
coefficients, $\Wc[(+)]{Hq}$. We find that $\Br(\kpn)$ is limited by its SM
value in this case, which is related to our treatment of the constraint from
$\Br(K_L\to\mu^+\mu^-)$ \cite{Isidori:2003ts, D'Ambrosio:1997jp}; should the
corresponding assumptions be violated, an enhancement up to a factor of two
w.r.t. the SM is possible, a bound that is still much stronger than the
present experimental limit. Measuring a significant enhancement of this mode
could therefore exclude both scenarios; this is also true for $\Br(\klpn)$
which can only be suppressed compared to its SM value, which is due to the
constraint from $\epe$, in accordance with \cite{Buras:2015jaq}. 

In LH scenarios the $\SUtwoL$ invariance of SMEFT implies that
the two Wilson coefficients $\Wc[(1)]{Hq}$ and $\Wc[(3)]{Hq}$ enter
also up-type quark $\Delta F=1$ FCNC processes with the same linear 
combination as in $\Delta F = 2$ down-type mixing. Therefore there are in
principle also correlations among down-type $\Delta F=2$ mixing and up-type 
$\Delta F=1$ FCNC processes. They can be quite strong -- see \refeq{eq:corr-Bsmix-tcZ}
-- when certain conditions are met, \emph{i.e.} the hierarchy of CKM elements 
remains as extracted in the SM and is not overcompensated by a
hierarchy in $\wc[(1,3)]{Hq}{ij}$.

One of the important messages from our paper is that while $\epe$ can easily be
enhanced in the LH and RH scenarios considered by us, $\Br(\kpn)$ can only be
suppressed (enhanced up to a factor of two) in the LH case if the stricter
(conservative) bound on $K_L\to\mu\bar\mu$ is used and enhanced by at most $50\%$ 
in the RH case. If the future results from the NA62 experiment will find much 
larger enhancement of $\Br(\kpn)$
and later KOTO will also find an enhanced $\Br(\klpn)$, the only solution in
the context of the $Z$ scenario would be to consider the operators $\Op[(1)]{Hq}$,
$\Op[(3)]{Hq}$ and $\Op{Hd}$ simultaneously \cite{Buras:2015jaq}. Alternatively
other contributions, like the ones from four-fermion operators generated by
$Z^\prime$ exchanges or exchanges of other heavy particles will be required.

Our analysis did not specify the origin of FC $Z$-boson couplings.  The
inclusion of these new effects in VLQ models in which concrete dynamics
generates such couplings is discussed in \cite{Bobeth:2016llm}. 

%
%
%

\section*{Acknowledgements}
The research Ch.B, A.J.B and M.J was supported by the DFG cluster of excellence
``Origin and Structure of the Universe''. The work of A.C. is supported by the
Alexander von Humboldt Foundation. This work is also supported in part by the
DFG SFB/TR~110 ``Symmetries and the Emergence of Structure in QCD".

%
%
%

\appendix

\section{\boldmath SMEFT}
\label{app:SMEFT}

The covariant derivative in our conventions is 
\begin{align}
  \label{eq:def-cov-derivative}
  {\cal D}_\mu & =
  \partial_\mu - i g_2 \frac{\sigma^a}{2} W^a_\mu - i g_1 Y B_\mu 
\end{align}
with the $\SUtwoL$ and $\UoneY$ gauge couplings $g_{2,1}$ and $\sigma^a$ denoting 
the Pauli matrices. The $\UoneY$-hyper charge of the Higgs doublet is $Y_H=1/2$.
We define the SM Yukawa couplings of quarks as in \cite{Grzadkowski:2010es},
\begin{align}
  \label{eq:SM-dim-4-Yuk}
  - {\cal L}_{\rm Yuk} & 
  = \bar q_L \, Y_{d} \,H \,d_R  + \bar q_L \, Y_{u}\, \widetilde{H}\, u_R    
    + \mbox{h.c.} \,.
\end{align}
The Higgs doublet $H$ is parameterized in $R_\xi$-gauge as
\begin{align}
  H & 
  = \begin{pmatrix} H^+ \\ H^0 \end{pmatrix}
  = \begin{pmatrix} G^+ \\ \left(v+h^0+i G^0\right)/\sqrt{2} \end{pmatrix}\,,
\end{align}
with $G^+$ and $G^0$ denoting the would-be-Goldstone bosons and $h^0$ the 
SM Higgs. In the absence of dim-6 effects $v = (\sqrt{2} G_F)^{-1/2}$, 
however, in SMEFT this equality is not guaranteed anymore and changed by the dim-6
contribution of the $H^6$-operator \cite{Alonso:2013hga} that modifies
the Higgs potential.

The derivatives in $\psi^2 H^2 D$ operators \refeq{eq:LH13} and \refeq{eq:RH1} 
are defined in a Hermitian way \cite{Grzadkowski:2010es},
\begin{equation}
\begin{aligned}
  H^\dagger i \overleftrightarrow{\cal D}_{\!\!\!\mu} H &
  \equiv i \left[ H^\dagger ({\cal D}_{\!\mu} H)
       - ({\cal D}_{\!\mu} H)^\dagger H \right] ,
\\
  H^\dagger i \overleftrightarrow{\cal D}^a_{\!\!\!\mu} H & 
  \equiv i \left[ H^\dagger\sigma^a ({\cal D}_{\!\mu} H)
       - ({\cal D}_{\!\mu} H)^\dagger \sigma^a H \right] .
\end{aligned}
\end{equation}
After EWSB the $\psi^2 H^2 D$ operators take rather lengthy forms in the
mass eigenbasis:
\begin{align}
  \wc{Hd}{ij} \op{Hd}{ij} & 
  = \wc{Hd}{ij} (H^\dagger i \overleftrightarrow{\cal D}_{\!\!\!\mu} H)
    [\bar{d}^i \gamma^\mu P_R d^j] \,,
\\[2mm]
  \wc[(1)]{Hq}{ij} \op[(1)]{Hq}{ij} & 
  = \wc[(1)]{Hq}{ij} (H^\dagger i \overleftrightarrow{\cal D}_{\!\!\!\mu} H)
    \left(V_{mi}^{} V_{nj}^* [\bar{u}^m \gamma^\mu P_L u^n] 
          + [\bar{d}^i \gamma^\mu P_L d^j] \right) ,
\\[2mm]
  \wc[(3)]{Hq}{ij} \op[(3)]{Hq}{ij} & \nonumber
  =  \left( \wc[(3)]{Hq}{ij} 
    (H^\dagger i \overleftrightarrow{\cal D}^{1}_{\!\!\!\mu} H
    -  \, i H^\dagger i \overleftrightarrow{\cal D}^{2}_{\!\!\!\mu} H)
    V_{mi}^{} [\bar{u}^m \gamma^\mu P_L d^j] + \mbox{h.c.} \right)
\\ &   
  + \wc[(3)]{Hq}{ij} (H^\dagger i \overleftrightarrow{\cal D}^3_{\!\!\!\mu} H)
    \left(V_{mi}^{} V_{nj}^* [\bar{u}^m \gamma^\mu P_L u^n] 
          - [\bar{d}^i \gamma^\mu P_L d^j] \right) .          
\end{align}
Note that CKM elements appear in the mass eigenbasis only whenever left-handed
up-type quarks are involved, due to the choice explained in \refsec{sec:SMEFT}.
The terms with covariant derivatives contain gauge, Higgs and 
would-be-Goldstone interactions that are equal in $\Op{Hd}$ and 
the singlet-operator $\Op[(1)]{Hq}$,
\begin{align}
  \nonumber
  H^\dagger i \overleftrightarrow{\cal D}_{\!\!\!\mu} H 
  & =
  - (v + h^0) (\partial_\mu G^0) + G^0 (\partial_\mu h^0)
  + i G^- \overleftrightarrow{\partial}_{\!\!\!\mu} G^+
\\ & \nonumber
  + g_2 \left[ \left(v + h^0 - i G^0 \right) G^+ W^-_\mu  + \mbox{h.c.} \right]
  + 2 e G^- G^+ A_\mu
\\ & \label{eq:FR-Dmu}
  + \frac{g_Z}{2} \left[ 2 (c_W^2 - s_W^2) G^- G^+ - v^2 
      - 2 v h^0 - (h^0)^2 - (G^0)^2\right] Z_\mu ,
\intertext{but differ for the triplet operator $\Op[(3)]{Hq}$,}  
  \nonumber
  H^\dagger i \overleftrightarrow{\cal D}^3_{\!\!\!\mu} H & =
  + (v + h^0) (\partial_\mu G^0) - G^0 (\partial_\mu h^0)
  + i G^- \overleftrightarrow{\partial}_{\!\!\!\mu} G^+
  +  2 e G^- G^+ A_\mu
\\ & \label{eq:FR-D3mu}
  + \frac{g_Z}{2} \left[ 2 (c_W^2 - s_W^2) G^- G^+ + v^2 
       + 2 v h^0 + (h^0)^2 + (G^0)^2\right] Z_\mu ,
\\[2mm]
  \nonumber
 \frac{1}{\sqrt{2}}  (H^\dagger i \overleftrightarrow{\cal D}^{1}_{\!\!\!\mu} H
 - & \, i H^\dagger i \overleftrightarrow{\cal D}^{2}_{\!\!\!\mu} H) =  
     v i \partial_\mu G^+ + h^0 i \overleftrightarrow{\partial_\mu} G^+
  + G^0  \overleftrightarrow{\partial_\mu} G^+
\\ & \nonumber
  + \frac{g_2}{2} \left[ v^2 + 2 v h^0 + (h^0)^2 + (G^0)^2 + 2 G^- G^+ \right] W^+_\mu
\\ &   \label{eq:FR-D1p2mu}
  + (v + h^0 - i G^0) G^+ \left[ e A_\mu - g_Z s_W^2 Z_\mu \right] .
\end{align}
Here $g_Z\equiv \sqrt{g_1^2 + g_2^2}$, $s_W \equiv \sin\theta_W$ and $c_W\equiv
\cos\theta_W$, where $\theta_W$ denotes the weak mixing angle, which again differs
by dim-6 contributions from its SM analogue. The partial derivatives act only on
fields within parentheses. Eventually only a few terms are required for the Feynman 
rules that enter the calculation of the diagrams in \reffig{fig:psi2H2D-DF2-me}.

The $\psi^2 H^2 D$ operators \refeq{eq:LH13}-\refeq{eq:O-Hud} undergo also mixing
among themselves. Here we list for completeness the Yukawa-enhanced contributions 
\cite{Jenkins:2013wua}:
\begin{equation}
  \label{eq:psi2H2D-selfmixing}
\begin{aligned}
  \dotWc[(1)]{Hq} = & 
  \; 6\, \mbox{Tr} \big[\Yuk{u} \YukD{u}] \Wc[(1)]{Hq}
  + 2 \left( \Yuk{u} \YukD{u} \Wc[(1)]{Hq} + \Wc[(1)]{Hq} \Yuk{u} \YukD{u} \right)
, \\ &
  - \frac{9}{2} \left( \Yuk{u} \YukD{u} \Wc[(3)]{Hq} 
                     + \Wc[(3)]{Hq} \Yuk{u} \YukD{u} \right)
  - \Yuk{u} \Wc{Hu} \YukD{u} ,
\\
  \dotWc[(3)]{Hq} = &
  \; 6\, \mbox{Tr} \big[\Yuk{u} \YukD{u}] \Wc[(3)]{Hq}
  + \Yuk{u} \YukD{u} \Wc[(3)]{Hq} + \Wc[(3)]{Hq} \Yuk{u} \YukD{u}
  - \frac{3}{2} \left( \Yuk{u} \YukD{u} \Wc[(1)]{Hq} 
                     + \Wc[(1)]{Hq} \Yuk{u} \YukD{u} \right) ,
\\
  \dotWc{Hu} = &
  - 2 \YukD{u} \Wc[(1)]{Hq} \Yuk{u}
  + 6\, \mbox{Tr}[\Yuk{u} \YukD{u}] \Wc{Hu} 
  + 4 \left(\YukD{u} \Yuk{u} \Wc{Hu} + \Wc{Hu} \YukD{u} \Yuk{u} \right) , 
\\
  \dotWc{Hd} = & \; 6\, \mbox{Tr} \big[\Yuk{u} \YukD{u}] \Wc{Hd} ,
\\
  \dotWc{Hud} = &
  \; 6\; \mbox{Tr}[\Yuk{u} \YukD{u}] \Wc{Hud}
  + 3 \YukD{u} \Yuk{u} \Wc{Hud} \,.
\end{aligned}
\end{equation}
These ADMs show that in SMEFT the LH-$Z$ interactions $\Wc[(1,3)]{Hq}$ do not
generate RH-$Z$ interactions in the down-type sector ($\Wc{Hd}$) and vice versa.
However, there \emph{is} mixing of the LH-$Z$ interaction $\Wc[(1)]{Hq}$ into
the RH-$Z$ interactions of up-type sector ($\Wc{Hu}$) and vice versa. In order
to draw conclusions on the phenomenological impact, the explicit flavour
structure should be worked out though, see for example \refeq{eq:ADM-basis-change}.
In the main part of our work we assume a scenario where $\psi^2 H^2 D$
operators are the dominant ones at $\muNP$ and generate $\Delta F=2$--$\psi^4$ 
operators at $\muEW$. For this purpose we have neglected the mixing among the 
various $\psi^2 H^2 D$ operators in the evolution from $\muNP$ to $\muEW$ in 
\refeq{eq:SMEFT-RGE}, which enters loop-suppressed in $\Delta F=2$ processes.

%
%
%

\renewcommand{\refname}{R\lowercase{eferences}}

\addcontentsline{toc}{section}{References}

\bibliographystyle{JHEP}
\bibliography{Bookallrefs}

\end{document}